\documentclass[11pt,a4paper]{article}

\usepackage[utf8]{inputenc}
\usepackage[T1]{fontenc}
\usepackage{lmodern}                 
\usepackage{microtype}               
\usepackage{amsmath,amssymb,amsfonts}

\usepackage[a4paper,
            top=0.95in, bottom=1.0in,
            left=0.85in, right=0.85in,
            footskip=0.45in]{geometry}

\usepackage{graphicx}
\usepackage{subcaption}
\usepackage{booktabs}
\usepackage[font=small,            
            labelfont=bf,          
            justification=justified,
            singlelinecheck=false,
            skip=6pt,
            margin=8pt]{caption}

\usepackage{xcolor}
\usepackage{url}
\usepackage[colorlinks=true,
            linkcolor={blue!55!black},
            citecolor={blue!55!black},
            urlcolor={blue!55!black},
            breaklinks=true]{hyperref}
\usepackage{tcolorbox}

\usepackage{titlesec}
\titlespacing*{\section}     {0pt}{2.0ex plus 0.6ex minus 0.4ex}{1.0ex plus 0.3ex}
\titlespacing*{\subsection}  {0pt}{1.6ex plus 0.5ex minus 0.3ex}{0.8ex plus 0.2ex}
\titlespacing*{\subsubsection}{0pt}{1.3ex plus 0.4ex minus 0.3ex}{0.6ex plus 0.2ex}

\newtcolorbox{theorembox}[1][]{colback=green!5,colframe=green!50!black,title=\textbf{Theorem (Mathematical Consequence)},#1}
\newtcolorbox{assumptionbox}[1][]{colback=blue!5,colframe=blue!50!black,title=\textbf{Physically Motivated Assumption},#1}
\newtcolorbox{occambox}[1][]{colback=orange!5,colframe=orange!50!black,title=\textbf{Minimal Complexity Choice},#1}

\title{Covariant Two-Field Perturbations Through a\\Curvature-Driven Bounce in Closed FRW Cosmology}

\author{Oleksandr Kravchenko}
\date{%
    \small \textit{UponCode LLC, Research Division, Cheyenne, WY, USA}\\%
    \textit{OkMath Research Initiative}\\[2ex]%
    \footnotesize \texttt{cosmology@okmath.org}%
}

\begin{document}

\providecommand{\aMinBounce}{1.73254\times 10^{5}}
\providecommand{\NpostBounce}{398.0}
\providecommand{\HBounce}{0}
\providecommand{\gDeviationMax}{3.1\times 10^{-9}}
\providecommand{\OmegaKCutoffLogTen}{-345.69}
\providecommand{\OmegaKDecouplingLogTen}{-295.69}
\providecommand{\OmegaKCutoffUpperExponent}{-345}
\providecommand{\OmegaKDecouplingUpperExponent}{-295}
\providecommand{\nsAnalytical}{0.9667}
\providecommand{\rAnalytical}{0.0033}
\providecommand{\nAlphaPoints}{11}
\providecommand{\nsAlphaMean}{0.9672}
\providecommand{\phiCMBmeasured}{5.4290}
\providecommand{\NtotalInflationAlphaScan}{86.44}
\providecommand{\nAlphaPointsNT}{11}
\providecommand{\nsAlphaMeanNT}{0.96699}
\providecommand{\nsAlphaStdNT}{2.55\times 10^{-6}}
\providecommand{\KchiMaxNT}{11.6}
\providecommand{\gCMBminNT}{0.748}
\providecommand{\gCMBmaxNT}{1.000}
\providecommand{\fNLCMBdeltaN}{+0.0133}
\providecommand{\fNLCMBconsistency}{+0.0134}
\providecommand{\fNLCMBconsistencyLeading}{+0.0139}
\providecommand{\fNLCMBdelta}{1.54\times 10^{-4}}
\providecommand{\rValClosedFRW}{1.732\times 10^{-6}}
\providecommand{\rLythStewartReference}{0.003333}
\providecommand{\PhSixtyClosedFRW}{6.749\times 10^{-12}}
\providecommand{\PhSixtyFlatFRWReference}{6.749\times 10^{-12}}
\providecommand{\PhSameExitAgreementPct}{0.0033}
\providecommand{\PhSixtyClosedFRWDetailed}{6.7492924\times 10^{-12}}
\providecommand{\PhSixtyFlatFRWDetailed}{6.7490665\times 10^{-12}}
\providecommand{\HSixtyTensorExit}{5.7710751\times 10^{-6}}
\providecommand{\PRSixtyClosedFRW}{3.896\times 10^{-6}}
\providecommand{\PRSixtyClosedFRWStateMin}{3.896\times 10^{-6}}
\providecommand{\PRSixtyClosedFRWStateMax}{4.162\times 10^{-6}}
\providecommand{\rValClosedFRWStateMin}{1.622\times 10^{-6}}
\providecommand{\rValClosedFRWStateMax}{1.732\times 10^{-6}}
\providecommand{\ExactScalarCompactResidual}{6.053\times 10^{-16}}
\providecommand{\ExactScalarBounceLimitResidual}{4.032\times 10^{-16}}
\providecommand{\ExactScalarWronskianResidual}{5.309\times 10^{-11}}
\providecommand{\ExactScalarAmplitudePhaseResidual}{3.584\times 10^{-8}}
\providecommand{\ExactScalarCoordinateResidual}{1.199\times 10^{-8}}
\providecommand{\ExactScalarOrthonormalResidual}{1.199\times 10^{-8}}
\providecommand{\ExactScalarMethodResidual}{2.186\times 10^{-9}}
\providecommand{\ExactScalarInitialMapResidual}{3.682\times 10^{-16}}
\providecommand{\ExactScalarHamiltonianResidual}{5.612\times 10^{-18}}
\providecommand{\ExactScalarMomentumResidual}{8.78\times 10^{-11}}
\providecommand{\ExactScalarBounceKappaOverMPlSq}{2.21\times 10^{-16}}
\providecommand{\ExactScalarMinAOverTwoMPlSq}{1.093\times 10^{-7}}
\providecommand{\ExactScalarMinBOverTwoMPlSq}{8.743\times 10^{-8}}
\providecommand{\ExactScalarMinCSq}{0.7998}
\providecommand{\ExactScalarBounceCSqMin}{3.015\times 10^{5}}
\providecommand{\ExactScalarBounceCSqMax}{3.015\times 10^{5}}
\providecommand{\ExactScalarStateSpanEllSixtyPct}{6.39}
\providecommand{\TwoTwoOriginalThreshold}{10^{-3}}
\providecommand{\TwoTwoMaxResidual}{5.587\times 10^{-10}}
\providecommand{\TwoTwoIntegratedCells}{64}
\providecommand{\TwoTwoStrictCells}{64}
\providecommand{\TwoTwoControlCells}{32}
\providecommand{\TwoTwoTotalCells}{96}
\providecommand{\BounceWindowHalfWidth}{0.0566755}
\providecommand{\BounceWindowMethodHalfWidth}{0.0566755}
\providecommand{\BounceWindowMaxCH}{1.925\times 10^{-12}}
\providecommand{\BounceWindowMaxCMPct}{1.628\times 10^{-8}}
\providecommand{\BounceWindowMaxCHTerm}{8.238\times 10^{-8}}
\providecommand{\BounceWindowMaxCMTermPct}{1.628\times 10^{-8}}
\providecommand{\BounceWindowMaxRefinement}{1.152\times 10^{-11}}
\providecommand{\BounceWindowMaxMethod}{1.411\times 10^{-12}}
\providecommand{\BounceWindowRadauMaxCH}{1.925\times 10^{-12}}
\providecommand{\BounceWindowRadauMaxCMPct}{1.628\times 10^{-8}}
\providecommand{\BounceWindowMethodCells}{24}
\providecommand{\BounceWindowLegCells}{128}
\providecommand{\BounceWindowBackgroundCells}{64}
\providecommand{\BounceWindowSamples}{2001}
\providecommand{\SfRMaxResidual}{2.407\times 10^{-9}}
\providecommand{\SfPowerDriftMax}{5.858\times 10^{-12}}
\providecommand{\SfHamiltonianMedianMin}{5.695\times 10^{-19}}
\providecommand{\SfHamiltonianMedianMax}{1.225\times 10^{-18}}
\providecommand{\SfHamiltonianTermMedianMaxPct}{2.816\times 10^{-7}}
\providecommand{\SfMomentumMedianEllTwoPct}{2.111\times 10^{-7}}
\providecommand{\SfMomentumMedianEllThreePct}{2.206\times 10^{-7}}
\providecommand{\SfMomentumMedianEllFivePct}{2.813\times 10^{-7}}
\providecommand{\SfMomentumGatePct}{1.5}
\providecommand{\FriedmannAbsoluteResidualMax}{1.024\times 10^{-25}}
\providecommand{\FriedmannScaledResidualMax}{3.072\times 10^{-15}}
\providecommand{\FriedmannAbsoluteResidualGate}{10^{-12}}
\providecommand{\FriedmannScaledResidualGate}{10^{-6}}
\providecommand{\AttractorMedianDeviation}{1.374\times 10^{-4}}
\providecommand{\AttractorMedianGate}{5.000\times 10^{-3}}
\providecommand{\AttractorPointCount}{12{,}506}
\providecommand{\AttractorPhiRef}{9.90}
\providecommand{\nsCurvatureScale}{4.172}
\providecommand{\nsCurvatureScaleLS}{1.000}
\providecommand{\nsCurvatureScaleStateMin}{0.862}
\providecommand{\nsCurvatureScaleStateMax}{4.172}
\providecommand{\nsCurvatureScaleInstantaneous}{0.862}
\providecommand{\ISRAtFiducialMchi}{0.983}
\providecommand{\ISRSupSampledGrid}{2.434}
\providecommand{\ISRAtFiducialMchiStateMin}{0.896}
\providecommand{\ISRAtFiducialMchiStateMax}{0.983}
\providecommand{\ISRSupSampledGridStateMin}{1.193}
\providecommand{\ISRSupSampledGridStateMax}{2.434}
\providecommand{\ISRAtMchiEightMinus}{0.995}
\providecommand{\ISRAtMchiEightMinusStateMin}{0.907}
\providecommand{\ISRAtMchiEightMinusStateMax}{0.995}
\providecommand{\ISRAtMchiSevenMinus}{0.995}
\providecommand{\ISRAtMchiSevenMinusStateMin}{0.907}
\providecommand{\ISRAtMchiSevenMinusStateMax}{0.995}
\providecommand{\ISRAtMchiSixMinus}{0.983}
\providecommand{\ISRAtMchiFiveMinus}{0.383}
\providecommand{\ISRAtMchiFiveMinusStateMin}{0.349}
\providecommand{\ISRAtMchiFiveMinusStateMax}{0.383}
\providecommand{\ISRAtMchiFourMinus}{0.0291}
\providecommand{\ISRAtMchiFourMinusStateMin}{0.0265}
\providecommand{\ISRAtMchiFourMinusStateMax}{0.0291}
\providecommand{\ISRLoglogSlope}{-0.35}
\providecommand{\ISRAtEllTwo}{2.434}
\providecommand{\ISRAtEllTwoStateMin}{0.948}
\providecommand{\ISRAtEllTwoStateMax}{2.434}
\providecommand{\ISRAtEllFive}{0.748}
\providecommand{\ISRAtEllFiveStateMin}{0.748}
\providecommand{\ISRAtEllFiveStateMax}{1.193}
\providecommand{\ISRAtEllSixty}{0.983}
\providecommand{\ISRAtEllSixtyStateMin}{0.896}
\providecommand{\ISRAtEllSixtyStateMax}{0.983}
\providecommand{\ISRAtEllHundred}{0.984}
\providecommand{\ISRAtEllHundredStateMin}{0.984}
\providecommand{\ISRAtEllHundredStateMax}{1.009}
\providecommand{\ISRAtEllTwoHundred}{0.988}
\providecommand{\ISRAtEllTwoHundredStateMin}{0.962}
\providecommand{\ISRAtEllTwoHundredStateMax}{0.988}
\providecommand{\ISRAtEllFiveHundred}{0.988}
\providecommand{\ISRAtEllFiveHundredStateMin}{0.988}
\providecommand{\ISRAtEllFiveHundredStateMax}{0.994}
\providecommand{\ISRAtEllThousand}{0.988}
\providecommand{\ISRAtEllThousandStateMin}{0.988}
\providecommand{\ISRAtEllThousandStateMax}{0.993}
\providecommand{\PRExitEllSixty}{7.755\times 10^{-6}}
\providecommand{\PSSExitEllSixty}{7.626\times 10^{-6}}
\providecommand{\mchiFiducial}{10^{-6}}
\providecommand{\mchiOverHinfFiducial}{0.17}

\IfFileExists{results_macros.tex}{

\def\aMinBounce{1.73254\times 10^{5}}  
\def\NpostBounce{398.0}  
\def\HBounce{0}  
\def\gDeviationMax{3.1\times 10^{-9}}  
\def\OmegaKCutoffLogTen{-345.69}  
\def\OmegaKDecouplingLogTen{-295.69}  
\def\OmegaKCutoffUpperExponent{-345}  
\def\OmegaKDecouplingUpperExponent{-295}  

\def\ExactScalarCompactResidual{6.053\times 10^{-16}}  
\def\ExactScalarBounceLimitResidual{4.032\times 10^{-16}}  
\def\ExactScalarWronskianResidual{5.309\times 10^{-11}}  
\def\ExactScalarAmplitudePhaseResidual{3.584\times 10^{-8}}  
\def\ExactScalarCoordinateResidual{1.199\times 10^{-8}}  
\def\ExactScalarOrthonormalResidual{1.199\times 10^{-8}}  
\def\ExactScalarMethodResidual{2.186\times 10^{-9}}  
\def\ExactScalarInitialMapResidual{3.682\times 10^{-16}}  
\def\ExactScalarHamiltonianResidual{5.612\times 10^{-18}}  
\def\ExactScalarMomentumResidual{8.78\times 10^{-11}}  
\def\ExactScalarBounceKappaOverMPlSq{2.21\times 10^{-16}}  
\def\ExactScalarMinAOverTwoMPlSq{1.093\times 10^{-7}}  
\def\ExactScalarMinBOverTwoMPlSq{8.743\times 10^{-8}}  
\def\ExactScalarMinCSq{0.7998}  
\def\ExactScalarBounceCSqMin{3.015\times 10^{5}}  
\def\ExactScalarBounceCSqMax{3.015\times 10^{5}}  
\def\ExactScalarStateSpanEllSixtyPct{6.39}  

\def\rValClosedFRW{1.732\times 10^{-6}}  
\def\rLythStewartReference{0.003333}  
\def\PhSixtyClosedFRW{6.749\times 10^{-12}}  
\def\PhSixtyFlatFRWReference{6.749\times 10^{-12}}  
\def\PhSameExitAgreementPct{0.0033}  
\def\PhSixtyClosedFRWDetailed{6.7492924\times 10^{-12}}  
\def\PhSixtyFlatFRWDetailed{6.7490665\times 10^{-12}}  
\def\HSixtyTensorExit{5.7710751\times 10^{-6}}  
\def\PRSixtyClosedFRW{3.896\times 10^{-6}}  
\def\PRSixtyClosedFRWStateMin{3.896\times 10^{-6}}  
\def\PRSixtyClosedFRWStateMax{4.162\times 10^{-6}}  
\def\rValClosedFRWStateMin{1.622\times 10^{-6}}  
\def\rValClosedFRWStateMax{1.732\times 10^{-6}}  

\def\ISRSupSampledGrid{2.434}  
\def\ISRAtFiducialMchi{0.983}  
\def\ISRAtMchiSixMinus{0.983}  
\def\PRExitEllSixty{7.755\times 10^{-6}}  
\def\ISRAtFiducialMchiStateMin{0.896}  
\def\ISRAtFiducialMchiStateMax{0.983}  
\def\ISRSupSampledGridStateMin{1.193}  
\def\ISRSupSampledGridStateMax{2.434}  
\def\PSSExitEllSixty{7.626\times 10^{-6}}  
\def\ISRAtEllTwo{2.434}  
\def\ISRAtEllTwoStateMin{0.948}  
\def\ISRAtEllTwoStateMax{2.434}  
\def\ISRAtEllFive{0.748}  
\def\ISRAtEllFiveStateMin{0.748}  
\def\ISRAtEllFiveStateMax{1.193}  
\def\ISRAtEllSixty{0.983}  
\def\ISRAtEllSixtyStateMin{0.896}  
\def\ISRAtEllSixtyStateMax{0.983}  
\def\ISRAtEllHundred{0.984}  
\def\ISRAtEllHundredStateMin{0.984}  
\def\ISRAtEllHundredStateMax{1.009}  
\def\ISRAtEllTwoHundred{0.988}  
\def\ISRAtEllTwoHundredStateMin{0.962}  
\def\ISRAtEllTwoHundredStateMax{0.988}  
\def\ISRAtEllFiveHundred{0.988}  
\def\ISRAtEllFiveHundredStateMin{0.988}  
\def\ISRAtEllFiveHundredStateMax{0.994}  
\def\ISRAtEllThousand{0.988}  
\def\ISRAtEllThousandStateMin{0.988}  
\def\ISRAtEllThousandStateMax{0.993}  

\def\ISRAtMchiEightMinus{0.995}  
\def\ISRAtMchiEightMinusStateMin{0.907}  
\def\ISRAtMchiEightMinusStateMax{0.995}  
\def\ISRAtMchiSevenMinus{0.995}  
\def\ISRAtMchiSevenMinusStateMin{0.907}  
\def\ISRAtMchiSevenMinusStateMax{0.995}  
\def\ISRAtMchiFiveMinus{0.383}  
\def\ISRAtMchiFiveMinusStateMin{0.349}  
\def\ISRAtMchiFiveMinusStateMax{0.383}  
\def\ISRAtMchiFourMinus{0.0291}  
\def\ISRAtMchiFourMinusStateMin{0.0265}  
\def\ISRAtMchiFourMinusStateMax{0.0291}  
\def\ISRLoglogSlope{-0.35}  

\def\FriedmannAbsoluteResidualMax{1.024\times 10^{-25}}  
\def\FriedmannScaledResidualMax{3.072\times 10^{-15}}  
\def\FriedmannAbsoluteResidualGate{10^{-12}}  
\def\FriedmannScaledResidualGate{10^{-6}}  
\def\AttractorMedianDeviation{1.374\times 10^{-4}}  
\def\AttractorMedianGate{5.000\times 10^{-3}}  
\def\AttractorPointCount{12{,}506}  
\def\AttractorPhiRef{9.90}  

\def\TwoTwoOriginalThreshold{10^{-3}}  
\def\TwoTwoMaxResidual{5.587\times 10^{-10}}  
\def\TwoTwoIntegratedCells{64}  
\def\TwoTwoStrictCells{64}  
\def\TwoTwoControlCells{32}  
\def\TwoTwoTotalCells{96}  

\def\BounceWindowHalfWidth{0.0566755}  
\def\BounceWindowMethodHalfWidth{0.0566755}  
\def\BounceWindowMaxCH{1.925\times 10^{-12}}  
\def\BounceWindowMaxCMPct{1.628\times 10^{-8}}  
\def\BounceWindowMaxCHTerm{8.238\times 10^{-8}}  
\def\BounceWindowMaxCMTermPct{1.628\times 10^{-8}}  
\def\BounceWindowMaxRefinement{1.152\times 10^{-11}}  
\def\BounceWindowMaxMethod{1.411\times 10^{-12}}  
\def\BounceWindowRadauMaxCH{1.925\times 10^{-12}}  
\def\BounceWindowRadauMaxCMPct{1.628\times 10^{-8}}  
\def\BounceWindowMethodCells{24}  
\def\BounceWindowLegCells{128}  
\def\BounceWindowBackgroundCells{64}  
\def\BounceWindowSamples{2001}  

\def\SfRMaxResidual{2.407\times 10^{-9}}  
\def\SfPowerDriftMax{5.858\times 10^{-12}}  
\def\SfHamiltonianMedianMin{5.695\times 10^{-19}}  
\def\SfHamiltonianMedianMax{1.225\times 10^{-18}}  
\def\SfHamiltonianTermMedianMaxPct{2.816\times 10^{-7}}  
\def\SfMomentumMedianEllTwoPct{2.111\times 10^{-7}}  
\def\SfMomentumMedianEllThreePct{2.206\times 10^{-7}}  
\def\SfMomentumMedianEllFivePct{2.813\times 10^{-7}}  
\def\SfMomentumGatePct{1.5}  

\def\nsCurvatureScale{4.172}  
\def\nsCurvatureScaleLS{1.000}  
\def\nsCurvatureScaleStateMin{0.862}  
\def\nsCurvatureScaleStateMax{4.172}  
\def\nsCurvatureScaleInstantaneous{0.862}  

\def\mchiFiducial{10^{-6}}  
\def\mchiOverHinfFiducial{0.17}  

\def\nsAnalytical{0.9667}  
\def\rAnalytical{0.0033}  
\def\fNLCMBdeltaN{+0.0133}  
\def\fNLCMBconsistency{+0.0134}  
\def\fNLCMBconsistencyLeading{+0.0139}  
\def\fNLCMBdelta{1.54\times 10^{-4}}  

\def\nsAlphaMean{0.9672}  
\def\nAlphaPoints{11}  
\def\phiCMBmeasured{5.4290}  
\def\NtotalInflationAlphaScan{86.44}  
\def\nsAlphaMeanNT{0.96699}  
\def\nsAlphaStdNT{2.55\times 10^{-6}}  
\def\KchiMaxNT{11.6}  
\def\gCMBminNT{0.748}  
\def\gCMBmaxNT{1.000}  
\def\nAlphaPointsNT{11}  
}{}

\maketitle

\begin{abstract}
We study perturbations through a locally non-singular bounce in a closed ($K=+1$) Friedmann--Robertson--Walker universe with a regularised two-field sigma model. Positive curvature drives it in General Relativity without Null-Energy-Condition violation. In the exact fiducial reduced-scalar calculation, kinetic and gradient coefficients remain positive on the declared harmonic--time grid and adaptive evaluations; tensor characteristics are luminal. A wall-free homogeneous-shear scan establishes only sampled robustness. The expanding branch approaches Starobinsky inflation and yields standard leading slow-roll observables at $N=60$.

We integrate the scalar sector in two field-space representations of the gauge-invariant comoving curvature perturbation $\mathcal R$, both in Newtonian gauge: a coordinate-field system and an orthonormal adiabatic--entropy system. An algebraic Gordon--Wands projection supplies a third reconstruction; the complete evolved systems are stated in Appendix~\ref{app:integrated_scalar_systems}. Across $\TwoTwoIntegratedCells$ turning cells, the maximum cross-representation residual is $\TwoTwoMaxResidual$. Across $H=0$, the Einstein-constraint maxima are $\mathcal H_k=\BounceWindowMaxCH$ and $100\mathcal C_M=\BounceWindowMaxCMPct\%$; tolerance refinement gives $\BounceWindowMaxRefinement$, while a local DOP853--Radau comparison gives $\BounceWindowMaxMethod$. These are finite-grid statements.

On the fiducial no-turn trajectory we derive and integrate the exact finite-harmonic closed-$S^3$ reduced scalar action. We propagate two canonically normalised finite-time states without claiming a unique asymptotic vacuum. The reporting prescription gives the like-for-like early-plateau ratio $r_{\ell=60}=\rValClosedFRW$, with scalar-state range $[\rValClosedFRWStateMin,\rValClosedFRWStateMax]$, distinct from the observable-pivot slow-roll prediction $r\simeq3.1\times10^{-3}$. The entropy calculation gives $\mathcal I_{S\mathcal R}=\ISRAtFiducialMchi$ at $\ell=60$, with scalar-state range $[\ISRAtFiducialMchiStateMin,\ISRAtFiducialMchiStateMax]$. Here it is finite-time-state spectator power, neither $T_{\mathcal{RS}}$ nor a primordial CDM-isocurvature fraction; the required production and transfer map is not computed.

At low closed-universe harmonics the exact measure gives the two-state exploratory range $n_s^{\rm curv}\in[\nsCurvatureScaleStateMin,\nsCurvatureScaleStateMax]$; its endpoints are the instantaneous-Hamiltonian and acoustic-gradient finite-time states, respectively. Its state sensitivity and cosmic variance preclude a likelihood claim. The work establishes a fully specified finite-grid perturbative continuation through a curvature-driven bounce, not a proof of past geodesic completeness.
\end{abstract}

\section{Introduction}
\label{sec:intro}

Non-singular bouncing cosmologies \cite{Novello2008,BattefeldPeter2015,BrandenbergerPeter2017,IjjasSteinhardt2018} offer a concrete alternative to imposing an inflationary initial boundary at a singular transition \cite{Guth1981,Starobinsky1980}. The realisation studied here evolves a finite classical solution from a contracting branch through a NEC-respecting bounce and into slow-roll inflation. This removes a curvature singularity from the \emph{modelled transition}; it neither derives the contracting initial state nor establishes past geodesic completeness. The closed-FRW two-field construction was developed across earlier versions of this manuscript \cite{Kravchenko2025v1,Kravchenko2025v2,Kravchenko2025v3,Kravchenko2025v4}; the present version supersedes them and stands alone. Version~1 introduced the unregularised hyperbolic metric and curvature-driven-bounce construction. Versions~2--3 introduced the sigmoid that enforces the chosen boundedness and asymptotic-canonicity conditions, reported the convention-dependent $\sim10^{21}$ basin-expansion figure, and attempted perturbation-regularity and $\alpha$-independence analyses. Version~4 supplied the Newtonian-gauge two-field integration, an attempted wall-free shear argument, and an epistemic classification. None of those descriptions imports its conclusions: the corrections and independent v5 tests are stated below. The earlier perturbation analysis used the fiducial $\dot\chi=0$ trajectory, for which $\omega=0$ and turn-sourced transfer vanishes. A genuine dynamical transfer requires a turning trajectory and remains open; the present work instead computes the independently evolving spectator entropy power.

Curvature-driven bounces and the propagation of perturbations through them have established precedents. Gordon and Turok analysed perturbations through a closed-universe scalar-field bounce in General Relativity \cite{GordonTurok2003}; Falciano, Lilley, and Peter examined the background conditions and perturbation evolution for a classical positive-curvature bounce \cite{FalcianoLilleyPeter2008}. Martin and Peter studied scalar perturbations through a bounce in closed geometry with a single scalar field in General Relativity, finding that the bounce can modify their spectrum in a wavenumber-dependent way \cite{MartinPeter2003}. Matsui, Takahashi, and Terada exhibited NEC-respecting scalar-field bounces supported by positive curvature and a potential flat at large field, followed by slow-roll inflation \cite{MatsuiTakahashiTerada2019}; that background mechanism is not new here. Renevey, Barrau, and Martineau computed scalar spectra and subsequent CMB temperature anisotropies for curvature-bounce models, including the Starobinsky potential, and examined the influence of the vacuum choice \cite{ReneveyBarrauMartineau2022}. Other closed-universe constructions have different global histories: the emergent universe of Ellis and Maartens approaches an Einstein-static past \cite{EllisMaartens2004}, whereas the simple harmonic universe uses a negative cosmological term and matter with $-1<w<-1/3$ to support oscillations \cite{GrahamEtAl2014}. Neither global history is assumed in the finite evolution studied here.

The distinction between a local bounce and global completion is essential. The Borde--Guth--Vilenkin theorem constrains past completeness when its positive averaged-expansion condition holds along the relevant geodesic \cite{BordeGuthVilenkin2003}; a finite contracting--bounce--expanding segment establishes neither that condition on a complete past extension nor its failure. Recent work further sharpens the global question. Burwig and Easson derive obstructions to simultaneously satisfying null completeness and the averaged null energy condition (ANEC) in non-static flat or open FRW spacetimes in the regular classes they analyse, while closed geometry admits complete examples \cite{BurwigEasson2026Closed,BurwigEasson2026ANEC}. These results motivate studying $K=+1$; they do not certify completeness of an arbitrary closed solution. Easson constructs a geodesically complete closed bounce--inflation model with a single canonical scalar and a positive vacuum offset \cite{Easson2026Bounce}. Our specified two-field potential has no such offset, and we establish neither its global extension nor an affine-ANEC statement on complete geodesics. The contribution here is instead the exact finite-harmonic scalar reduction on the fiducial no-turn trajectory, two propagated canonical finite-time states, and direct constraint and cross-representation checks on a finite turning-background grid. The fiducial reduction specialises the general non-flat action of Akama and Kobayashi \cite{AkamaKobayashi2019}; its numerical implementation and validation are developed here. A generic turning-background reduced action and a late-time CMB transfer calculation remain outside the scope of this work.

\paragraph{Provenance of numerical values.}
Every load-bearing numerical value in this manuscript is either produced or evaluated within the present work, or externally sourced and cited; the sole antecedent number retained in the prose is the explicitly non-load-bearing, convention-dependent $\sim10^{21}$ basin-expansion figure, carried only to rescope that earlier claim in \S\ref{sec:numerics}. The active values fall into four transparently labelled classes. (i)~\emph{Standard slow-roll results}---$n_s=1-2/N$, $r=12/N^2$, and the Maldacena single-field-consistency $f_{\rm NL}^{\rm local}$, with the separate super-Hubble $\delta N$ contribution evaluated as a cross-check---are textbook consequences of the Starobinsky potential \cite{Starobinsky1980,MukhanovChibisov1981,Starobinsky1983,Maldacena2003,SugiyamaKomatsuFutamase2013} evaluated here from the model parameters. (ii)~\emph{Computed results}---the background, exact finite-harmonic closed-$S^3$ scalar powers, the direct scalar/tensor ratio, the $\mathcal{I}_{S\mathcal{R}}$ ratio and its $m_\chi$ scan, and the low-harmonic curvature-scale diagnostic---are recomputed here from the model parameters. (iii)~\emph{Validation diagnostics}, including the cross-representation comparison, the Einstein-constraint residuals, and independent integration-method comparisons, are computed from the same integrations and quoted directly from their numerical output. (iv)~\emph{External observational constraints and forecasts} are quoted from Planck 2018, BICEP/Keck 2018, LiteBIRD, CMB-S4, and PICO with their primary citations \cite{Planck2018,Planck2018NG,BICEPKeck2021,LiteBIRD2022,CMB-S42016,PICO2019}. No numerical value is imported from v1--v4 as a v5 result or as a normalisation target.

The present manuscript addresses that outstanding probe and develops the perturbation analysis along three structural directions. We retain the model parameters and action of the earlier work ($V_0 = 10^{-10}\,M_{\rm Pl}^4$, $\beta = \sqrt{2/3}$, $\alpha = 1$, $m_\chi = \mchiFiducial\,M_{\rm Pl}$, $\phi_0 \approx 10\,M_{\rm Pl}$ on the Starobinsky plateau); the model's complete action, field-space geometry, corrected background diagnostics, flatness estimate, and wall-free homogeneous-shear check are presented self-contained in \S\S\ref{sec:theory}--\ref{sec:bkl} below. First, we implement an independent two-basis integration of the standard field-space-covariant perturbation system \cite{SasakiStewart1996,GrootNibbelinkVanTent2002,Langlois2008}, decomposing the perturbations along and orthogonal to the background trajectory via the Gordon--Wands projection \cite{Gordon2000,Wands2002}: adiabatic mode $Q_\sigma = (\dot\phi\,\delta\phi + g_{\chi\chi}\dot\chi\,\delta\chi)/\dot\sigma$ and entropy mode $Q_s = \sqrt{g_{\chi\chi}}\,(\dot\phi\,\delta\chi - \dot\chi\,\delta\phi)/\dot\sigma$, with $\dot\sigma^2 = \dot\phi^2 + g_{\chi\chi}\dot\chi^2$ (the $\sqrt{g_{\chi\chi}}$ Jacobian on $Q_s$ is the orthonormal field-space projection of \S\ref{subsec:covariant_decomp}, Eq.~(\ref{eq:Q_sigma_Q_s})). The field-space connection $\Gamma^a_{bc}$ is retained explicitly. Two systems in the same Newtonian spacetime gauge are integrated independently in the coordinate field basis $(\delta\phi,\delta\chi,\Phi)$ and the orthonormal field basis $(Q_T,Q_N,\Phi)$; an algebraically equivalent Gordon--Wands projection supplies a third reconstruction check. Their cross-basis comparison covers a nominal $\TwoTwoTotalCells$-cell grid: $\TwoTwoIntegratedCells$ integrated bouncing cells plus $\TwoTwoControlCells$ non-integrated finite-range contracting-control cells. The maximum cross-representation residual on the integrated cells is $\TwoTwoMaxResidual$; every integrated cell satisfies the acceptance criterion $\Delta<\TwoTwoOriginalThreshold$ (\S\ref{subsec:triangulation_limitation}).

Second, on the fiducial no-turn trajectory we derive the exact finite-harmonic closed-$S^3$ reduced scalar action and propagate two explicitly specified canonical finite-time states. At $\ell=60$, the reporting state gives $P_{\mathcal R}=\PRSixtyClosedFRW$ four e-folds after geometric exit, while the two-state interval is $[\PRSixtyClosedFRWStateMin,\PRSixtyClosedFRWStateMax]$. The independently normalised tensor amplitude $P_h=\PhSixtyClosedFRW$ agrees to $\PhSameExitAgreementPct\%$ with the same-exit flat-FRW slow-roll reference $P_h^{\rm flat}=\PhSixtyFlatFRWReference$. Their direct early-plateau ratio is $r_{\ell=60}=\rValClosedFRW$, with scalar-state interval $[\rValClosedFRWStateMin,\rValClosedFRWStateMax]$. The numerical mode is not the observable pivot; the latter remains described by $16\epsilon_V(\phi_{60}^{\rm num})\approx3.08\times10^{-3}$ ($12/N^2$ at leading order).

Third, we integrate the covariant entropy mode with the raised-index mass $m_\chi^2/g_{\chi\chi}$ and form $\mathcal{I}_{S\mathcal{R}}=P_{SS}/P_{\mathcal{R}}$ directly from the entropy and exact scalar solvers at their common geometric-exit time. At $\ell=60$, the reporting scalar state gives $P_{\mathcal{R}}=\PRExitEllSixty$, $P_{SS}=\PSSExitEllSixty$, and $\mathcal{I}_{S\mathcal{R}}=\ISRAtFiducialMchi$; the scalar-state interval is $[\ISRAtFiducialMchiStateMin,\ISRAtFiducialMchiStateMax]$. The sampled-grid maximum is $\ISRSupSampledGrid$, with corresponding two-state interval $[\ISRSupSampledGridStateMin,\ISRSupSampledGridStateMax]$. Since $\omega=0$, the transfer coefficient $T_{\mathcal{RS}}$ vanishes and $\mathcal{I}_{S\mathcal{R}}$ is spectator power for the specified finite-time state rather than turn-sourced transfer. Mapping it to a late-time CDM-isocurvature fraction remains model dependent; the assumptions of a conditional comparison with the Planck density-isocurvature bound are stated in \S\ref{subsec:entropy_power_ratio}, and the numerical comparison is quantified in \S\ref{subsec:planck_comparison}.

\long\def\EarlierVersionCorrections{%
\paragraph{Corrections relative to the earlier versions.}
The curvature-formula correction applies to v2--v4; v3 repeated the same $-\alpha^2/2$ asymptote even though item (v) below names the versions containing the derivation. The BKL and flatness corrections apply to both v3 and v4: v4 repeated the BKL-stability and $3.5$-e-fold language, while its $\Sigma^2>a^4$ display omitted the factor $6M_{\rm Pl}^2$ required for a like-for-like shear/curvature ratio and was not the bounce-prevention threshold. None of those statements is inherited as a v5 result.
Because the present version replaces the earlier versions of this eprint \cite{Kravchenko2025v1,Kravchenko2025v2,Kravchenko2025v3,Kravchenko2025v4}, we record the substantive corrections it makes to their text.

(i) The second and third versions' claim that the sigmoid is ``the unique smooth solution'' of the boundary conditions is replaced by the precise statement of \S\ref{sec:theory}: Conditions~1--3 admit infinitely many regularisations, and the sigmoid is singled out by the minimal-complexity selection; the second through fourth versions treated Conditions~1--2 as physical requirements, whereas they are coordinate-frame choices and only positivity is coordinate-invariant; the accompanying geometric ``compactification'' argument is retained only as a geometric \emph{reading}, since a coordinate transformation cannot produce the regularised metric. The fourth version also classified $g_{\chi\chi}(0)=1/2$ as a mere convention; in the complete action the origin of $\phi$ is already fixed by the minimum of the Starobinsky potential, so centring the metric transition there is an additional model-building choice. (ii) The second and third versions asserted, without proof, that the gauge-invariant perturbation variables remain regular through $H = 0$; the present version integrates two regular systems across $H=0$, evaluates both Einstein constraints directly in a common immediate-bounce window, and checks tolerance refinement there, with the separate cross-basis comparison performed post-bounce (\S\ref{sec:perturbations}). (iii) The bound $r<0.036$---attributed to Planck in the third version and quoted without attribution in the fourth---is the BICEP/Keck 2018 result for $r_{0.05}$ \cite{BICEPKeck2021}; Planck 2018 alone gives $r_{0.002}<0.10$ \cite{Planck2018}. The first version's attribution to BICEP/Keck was correct. (iv) The fourth version evolved continuous flat-space Fourier modes on a closed background; the present version uses discrete $S^3$ harmonics, the curvature-shifted scalar constraint, and the standard spin-specific closed-universe scalar and tensor spectral conventions \cite{LindblomTaylorZhang2017,AkamaKobayashi2019}, with its own v5 diagnostics (\S\ref{subsec:constraint_diagnostics}). (v) The field-space curvature formula in the second and fourth versions is incorrect: the Gaussian curvature is $K_G=(\alpha^2/M_{\rm Pl}^2)(1-g_{\chi\chi})(3g_{\chi\chi}-1)$ and tends to $-\alpha^2/M_{\rm Pl}^2$, not $-\alpha^2/(2M_{\rm Pl}^2)$.

(vi) The first version's bounce figure was initialised off the Friedmann constraint surface and its contraction branch did not reverse the field momenta; the present backgrounds are constraint-initialised and verified pointwise (\S\ref{sec:numerics}). (vii) The third version's $\alpha$ scan assigned identical analytical observables to every trajectory; the present scan measures the kinematic index from each trajectory's own $N=60$ point (\S\ref{sec:observables}). (viii) The fourth version labelled the power fraction $P_{SS}/(P_{RR}+P_{SS})$ as $T_{\mathcal{RS}}$. In that version, both terms were curvature powers: $P_{RR}$ came from the adiabatic-initialised vacuum mode and $P_{SS}$ from the isocurvature-initialised vacuum mode. Thus the fraction measured the isocurvature-seeded contribution to total curvature power, not the spectator-field power used here and not the signed Gordon--Wands transfer amplitude. In the present no-turn calculation the true turn-sourced coefficient vanishes analytically because $\omega=0$, while the separately reported horizon-exit ratio $\mathcal{I}_{S\mathcal{R}}=P_{SS}/P_{\mathcal{R}}$ measures spectator field power. The v4 number is not used as transfer evidence or as a normalisation target (\S\ref{subsec:entropy_power_ratio}).

(ix) We also record transient errors corrected before the fourth version: the first version used $r=12\alpha^2/N^2$ instead of the present model's $\alpha$-independent $12/N^2$, estimated $f_{\rm NL}^{\rm local}\sim1$ from a malformed $\delta N$ expression lacking the standard squared gradient norm in the denominator and without computing the asserted turning trajectory (the supported slow-roll value is $\simeq+0.013$), and described the hyperbolic metric as a ``kinetic-energy sink'' even though at fixed canonical momentum $\pi_\chi^2/(2g_{\chi\chi})$ diverges as $g_{\chi\chi}\to0$; the second version's displayed $\chi$ equation dropped the factor of two in $2\Gamma^\chi_{\phi\chi}\dot\phi\dot\chi$, although the background equations it integrated retained the factor; and the third version gave the single-field consistency relation the wrong sign.

Finally, broader claims not re-established here are retracted or rescoped. (x) The second and third versions' Trans-Planckian claim is withdrawn. (xi) The second version's ``stable attractor'' mechanism is not invoked: standard observable-pivot predictions and the exact baseline $\alpha$-independence claim use the fiducial $\chi=\dot\chi=0$ trajectory; displaced trajectories enter only construction-consistency checks and the explicitly narrower sampled kinematic-$n_s$ stability test, with no displaced-family $f_{\rm NL}$ or perturbation-spectrum claim. $\mathcal{I}_{S\mathcal{R}}$ and the curvature-scale slope are solver diagnostics, not established late-time observables. (xii) The third version's BKL-suppression claim is reduced to the finite wall-free homogeneous-shear bounce scan of \S\ref{sec:bkl}; $\Sigma_{\rm crit}^2=8M_{\rm Pl}^6/V_0^2$ is derived only as a constant-plateau benchmark, and full Bianchi-IX/BKL dynamics are not claimed. (xiii) The third and fourth versions' ``$\sim3.5$ e-folds suffice for flatness'' figure is not retained; the third version also mislabelled $|\Omega_k|<10^{-3}$ as the Planck-2018 constraint, whereas the cited Planck inflation analysis reports late-time spatial flatness to $0.4\%$ precision at $95\%$~CL only in combination with BAO \cite{Planck2018}. The present estimate applies the full simulated inflationary dilution and a conservative post-inflationary regrowth factor only through CMB decoupling; it is not identified with the late-time $\Omega_{K,0}$ constrained by Planck+BAO (\S\ref{sec:flatness}). (xiv) The fourth version's ``phenomenological indistinguishability from Starobinsky'' and ``UV-safe'' framing is retracted because neither a complete late-time isocurvature map nor a trans-Planckian analysis is supplied here; the extra field-power diagnostic alone is not an observational distinction. (xv) The first version stated that the comoving curvature perturbation itself diverges at the bounce, and the fourth placed a ``genuine gauge singularity'' in $\mathcal R$; in the regular Newtonian description $\mathcal R=\Phi+H\delta\phi/\dot\phi$ remains finite at $H=0$ when $\dot\sigma\neq0$, while reduced variables containing inverse powers of $H$ can be singular. (xvi) The fourth version's $10^{-2232}$ estimate for a bounce-generated CMB bispectrum was obtained by squaring a linear matching hierarchy, not by a second-order calculation; it is withdrawn. The present observable-pivot $f_{\rm NL}^{\rm local}$ is the standard Maldacena single-field squeezed-limit result on the post-bounce attractor; the separate-universe $\delta N$ term is reported as a component-level cross-check, while a through-bounce bispectrum remains uncomputed.

Additional source-level corrections are required for completeness. (xvii) With the $(-,+,+,+)$ spacetime signature printed in v1, its displayed scalar action carried a plus sign in front of $G_{ab}g^{\mu\nu}\partial_\mu\phi^a\partial_\nu\phi^b/2$, inconsistent with the positive-energy Friedmann equations used later; the present action has the canonical minus sign. (xviii) In v1's constant-potential approximation, $a(t)=\omega^{-1}\cosh(\omega t)$ gives the constant Ricci scalar $R=12\omega^2$, not $6\omega^2[1+\operatorname{sech}^2(\omega t)]$; its finiteness conclusion survives, but the displayed value does not. (xix) The v2--v3 statement that finite-$k$ modes remain outside the Hubble radius for $100\%$ of the bounce is false: $k/(a|H|)\to\infty$ as $H\to0$. Their use of $Q^I=\delta\phi^I+\dot\phi^I\Psi/H$ and a claimed regular $z''/z$ therefore did not establish through-bounce conservation. The present work evolves regular Newtonian variables through $H=0$; the sampled modes satisfy $k/(a|H|)\to\infty$ at the bounce before their post-bounce exits, although this ratio alone does not define an adiabatic regime there. (xx) Boundedness of a field-coordinate metric component does not by itself establish perturbative unitarity, absence of strong coupling, or a UV completion, as claimed in v2--v3; positivity establishes the no-ghost signature of the two-derivative model, while the stronger UV claims are not made here. (xxi) The earlier versions' broad statements that the initial singularity problem was resolved, or that an eternal past contraction followed if the bounce repeated, are reduced to the finite result actually shown: a locally non-singular contracting-to-expanding transition. The origin, geodesic completion, and global past extension remain open (\S\ref{sec:conclusions}).

(xxii) The third version inferred that detecting $r\sim0.003$ would provide strong evidence for the bouncing origin. That inference is withdrawn: this is the standard Starobinsky-plateau prediction, so such a detection would test the inflationary potential but would not by itself distinguish or support the pre-inflationary bounce.

(xxiii) The second and third versions presented $A_s\approx2.1\times10^{-9}$ as an exact model prediction. It is neither exact at the retained rounded value $V_0=10^{-10}M_{\rm Pl}^4$ nor an independent prediction: the Starobinsky scalar amplitude is proportional to the freely calibrated potential scale $V_0$. The present observable claim therefore concerns the dimensionless slow-roll shape parameters; no antecedent amplitude is imported as a target. (xxiv) The fourth version's independent flat-FRW CMB benchmark initialised modes at $k/(aH)=50$ and extracted them on its requested output grid. The present benchmark does not retain those numerical outputs: it uses depth $10^3$ and evaluates the dense solution at the exact comparison epoch (\S\ref{par:flat_scalar_benchmark}); a recorded one-off sweep to depth $3\times10^3$ changed the fitted $n_s$ by approximately $1.4\times10^{-6}$.

(xxv) The second and third versions' ``$100\%$ success'' over 16 orders of magnitude and ``$100\%$ robustness'' under $20\%$ perturbations, and the fourth version's $86\%$ success figure over a 22-order coordinate-velocity grid, are not accepted as v5 basin measurements. Those rates depend on the sampled coordinate convention and on the antecedent classification criteria. The present work instead reports the three explicitly listed successful bouncing initial conditions and one finite-range contracting control; it neither estimates a basin volume nor assigns a success probability (\S\ref{sec:numerics}). (xxvi) With the convention $\Omega_k=-K/(aH)^2$, the first version's suggestion that $\Omega_k>0$ would support a closed universe had the sign reversed. The fourth version used the closed-universe sign $\Omega_k<0$ but called a future detection indirect evidence for the framework. A curvature measurement can test late-time spatial geometry; it does not by itself establish a pre-inflationary bounce, and no such inference is made here. (xxvii) The fourth version repeatedly called $V/(24\pi^2M_{\rm Pl}^4\epsilon_V)$ and its fitted slope an ``exact'' Starobinsky spectrum. It is a standard slow-roll approximation, not an exact quantum power spectrum. The present text labels the potential formula and $n_s=1-6\epsilon_V+2\eta_V$ as slow-roll/first-order quantities and uses a separately integrated, depth-converged flat-FRW mode calculation only as an independent numerical reference, never as an antecedent target. (xxviii) All four earlier versions present exponential suppression of the $\chi$ kinetic energy at $\phi<0$, where $g_{\chi\chi}\approx e^{2\alpha\phi/M_{\rm Pl}}\ll1$, as an operative ingredient of the bounce: the first version calls that suppression ``the key to achieving a low-energy bounce'', and the second through fourth head it ``Condition~1: Bounce Mechanism ($\phi\ll0$)''. Those versions also attribute the bounce itself to positive spatial curvature, as the present one does; what is withdrawn here is the suppression clause, which none of their own integrations realises. Their fiducial trajectories start at $\phi_0=+3\,M_{\rm Pl}$ (first version) and $\phi_0=+10\,M_{\rm Pl}$ (second through fourth) and remain on the positive-$\phi$ branch throughout, never entering the $\phi<0$ suppression regime; the factor is not small in either family: the first version's unregularised metric gives $g_{\chi\chi}=e^{2\alpha\phi/M_{\rm Pl}}\approx4\times10^{2}$ at its own initial point, and the regularised sigmoid of the second through fourth is saturated at $g_{\chi\chi}\simeq1$. In all four, $\phi<0$ appears only in static field-space plots, never in an integrated solution. The bounce reported here is driven by the positive spatial-curvature term of the closed Friedmann equation with $g_{\chi\chi}\to1$ throughout the integration window (\S\ref{sec:model}, \S\ref{sec:numerics}); no kinetic-suppression mechanism is invoked, and the regularisation is required for field-space regularity rather than for the bounce itself.
}

\paragraph{Corrections relative to the earlier versions.}
Because this version supersedes v1--v4, Appendix~\ref{app:earlier_corrections} records all twenty-eight substantive corrections that the author has identified. In summary, v5 repairs the action sign and field-space curvature, replaces off-shell or incomplete bounce and perturbation arguments with constraint-initialised integrations, uses discrete closed-$S^3$ harmonics, corrects observational attributions and slow-roll formulae, and retracts unsupported claims about uniqueness, unitarity, BKL stability, basin probabilities, trans-Planckian safety, late-time transfer, and global singularity resolution. No conclusion of an earlier version is used as a v5 result or numerical target.

Throughout the present work the inflaton dynamics are exactly $\alpha$-independent at the background level on the baseline $\chi = \dot\chi = 0$ trajectory: the field-space metric $g_{\chi\chi}(\alpha)$ enters the background equations of motion only through the kinetic term $\tfrac{1}{2} g_{\chi\chi} \dot\chi^2$, which vanishes identically on $\dot\chi = 0$, so the inflaton evolution is exactly insensitive to $\alpha$ on this trajectory. The standard slow-roll observables
\begin{equation}
n_s \approx \nsAnalytical,
\qquad
r \approx \rAnalytical,
\qquad
f_{\rm NL}^{\rm local} \approx \fNLCMBconsistencyLeading \quad (\text{Maldacena squeezed limit})
\label{eq:abstract_observables}
\end{equation}
hold on the slow-roll attractor at $N = 60$. The closed-FRW $K=+1$ structure additionally admits a cosmic-variance-dominated curvature-scale diagnostic at the lowest $S^3$ harmonics. Using the exact per-log-wavenumber measure, the $\ell=2\to3$ log-slope spans
\begin{equation}
n_s^{\rm curv}\in[\nsCurvatureScaleStateMin,\nsCurvatureScaleStateMax]
\quad\text{(two propagated finite-time states)},
\label{eq:abstract_curvature_scale}
\end{equation}
defined in \S\ref{subsec:curvature_scale_diagnostic}. The endpoints are $\nsCurvatureScaleInstantaneous$ for the instantaneous-Hamiltonian state and $\nsCurvatureScale$ for the acoustic-gradient state. The least-squares value over $\ell\in\{2,3,5,7,10\}$ is $\nsCurvatureScaleLS$ for the latter state. The wide two-state interval follows from propagating two finite-time canonical prescriptions through the exact reduced action; it is not a statistical error bar or a complete measure over admissible states. The scalar modes use the common exact acoustic-depth threshold $50$, while tensor and entropy modes use the separate depth $1.5\times10^3$. The result is a model-internal diagnostic with no computed late-time CMB transfer, not a second observable-pivot spectral index.

The remainder of the paper is organised as follows. \S\ref{sec:theory} states the model's coordinate-frame choices, coordinate-invariant positivity requirement, and minimal-complexity selection principle. \S\ref{sec:model} states the action, field-space geometry, and background equations. \S\ref{sec:numerics} reviews the numerical methods and background validation. \S\ref{sec:flatness} and \S\ref{sec:bkl} cover flatness resolution and the wall-free homogeneous-shear analysis. \S\ref{sec:perturbations} develops the covariant two-field framework, exact fiducial closed-$S^3$ scalar action, closed-FRW tensor solver, cross-representation comparison, and entropy-mode power ratio. \S\ref{sec:observables} separates the observable-pivot slow-roll predictions from the low-harmonic diagnostic. \S\ref{sec:conclusions} states the limitations. Appendix~\ref{app:integrated_scalar_systems} gives both complete scalar systems supplied to the turning-background integrators; Appendix~\ref{app:closed_s3_canonical} derives the exact reduced fiducial scalar action and its validation; Appendix~\ref{app:earlier_corrections} preserves the correction record for v1--v4; and Data Availability identifies the public archive.

\section{Theoretical Foundation with Epistemic Classification}
\label{sec:theory}

\begin{figure}[htbp]
    \centering
    \IfFileExists{figures/fig_sigmoid_metric.pdf}{%
        \includegraphics[width=0.82\textwidth]{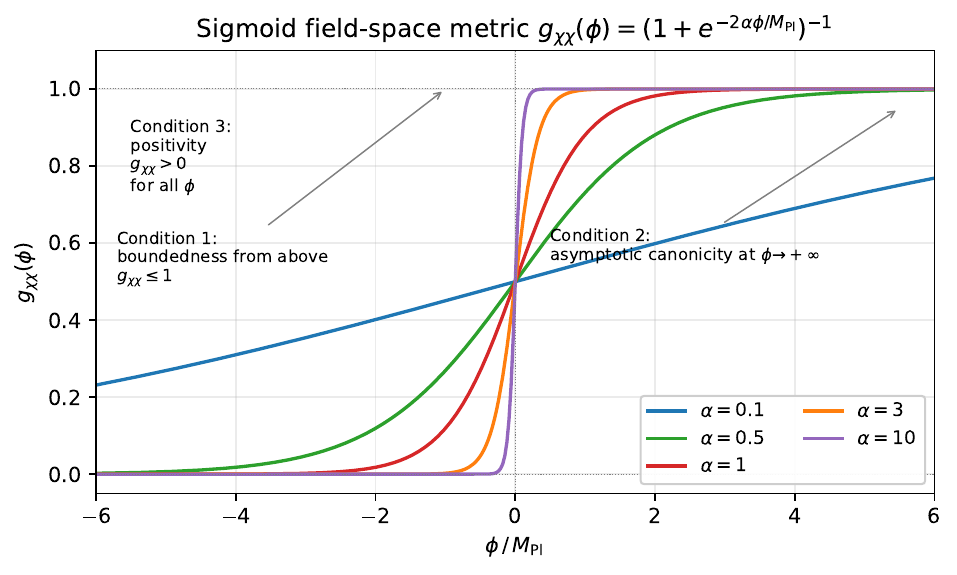}%
    }{%
        \fbox{\parbox{0.78\textwidth}{\centering
            \textbf{Figure not available in this build.}}}%
    }
    \caption{\textbf{Sigmoid field-space metric $g_{\chi\chi}(\phi) = (1 + e^{-2\alpha\phi/M_{\rm Pl}})^{-1}$.}
        Representative profiles for $\alpha \in \{0.1, 0.5, 1, 3, 10\}$. The
        sigmoid satisfies all three stated boundary conditions of \S\ref{sec:theory}:
        boundedness from above (Condition~1, $g_{\chi\chi} \leq 1$),
        asymptotic canonicity at $\phi \to +\infty$ (Condition~2,
        $g_{\chi\chi} \to 1^-$), and positivity for all $\phi$ (Condition~3,
        $g_{\chi\chi} > 0$); the regular $\phi \to -\infty$ limit
        $g_{\chi\chi} \to 0^+$ is shared by the sigmoid and the
        unregularised metric and is not itself a selection condition. The transition is centred at the potential minimum $\phi=0$ by an explicit model choice and has width $\sim M_{\rm Pl}/\alpha$;
        for the fiducial trajectory ($\alpha = 1$, $\phi_0 = 10\,M_{\rm Pl}$)
        the metric is saturated at unity throughout the perturbation window,
        with the longer simulated background giving the conservative bound
        $|1-g_{\chi\chi}|_{\max}=\gDeviationMax$. In the separate excited-
        spectator scan at the smallest $\alpha=0.1$, the measured $N=60$
        value is $g_{\chi\chi}=\gCMBminNT$, still $\mathcal{O}(1)$
        (\S\ref{sec:numerics}).}
    \label{fig:sigmoid_metric}
\end{figure}

The sigmoid field-space metric $g_{\chi\chi}(\phi) = (1 + e^{-2\alpha\phi/M_{\rm Pl}})^{-1}$ is selected using two coordinate-frame boundary choices, one coordinate-invariant positivity requirement, and a minimal-complexity selection principle (the metric is shown for representative $\alpha$ in Fig.~\ref{fig:sigmoid_metric}). The derivation in this section disentangles two distinct epistemic claims that are often conflated in the bouncing-cosmology literature: (i) the \emph{model-building} content is encoded in Conditions~1--3 below: boundedness and asymptotic canonicity are coordinate-frame regularity/normalisation choices, while positivity is the coordinate-invariant physical requirement; (ii) the \emph{mathematical} content is the existence theorem (the sigmoid is admitted by Conditions~1--3), while the minimal-complexity principle selects it as a simple representative within the stated function class. The latter is a methodological choice, not a physical prediction, and is open to alternative selection criteria; an alternative regularisation that coincides with the sigmoid on the sampled inflationary plateau can share its plateau predictions, but its derivatives and bounce-region behaviour must be reverified independently. To make this distinction explicit, each step in the derivation below is classified as a theorem (mathematical consequence), an assumption (physical input), or a minimal-complexity choice (Occam selection).

\begin{assumptionbox}
\textbf{Condition~1: boundedness from above (a coordinate-frame regularity choice).}
In the fixed field-coordinate $\chi$ adopted for the action, we require the metric component $g_{\chi\chi}(\phi)$ to remain bounded above for all $\phi$. This is a parametrisation-dependent statement about the chosen coordinate: under a field redefinition $\chi \to f(\chi)$ one has $g_{\chi\chi} \to g_{\chi\chi}/f'(\chi)^2$, so numerical boundedness of the component is not itself coordinate-invariant. It is motivated by keeping the $\chi$-kinetic normalisation controlled on the plateau. Boundedness of a metric component alone does \emph{not} guarantee bounded field-space curvature, which also depends on derivatives; bounded curvature is verified separately for the selected sigmoid below. At fixed \emph{coordinate velocity} $\dot\chi$ the $\chi$-kinetic energy $\tfrac{1}{2}\,g_{\chi\chi}\,\dot\chi^2$ scales with $g_{\chi\chi}$, so capping $g_{\chi\chi}$ caps this contribution; we note that at fixed \emph{canonical momentum} $\pi_\chi = a^3 g_{\chi\chi}\dot\chi$ the dependence inverts ($K_\chi = \pi_\chi^2/(2 a^6 g_{\chi\chi}) \propto 1/g_{\chi\chi}$), so this energetic motivation is frame-dependent rather than invariant. The unregularised hyperbolic metric $g_{\chi\chi} = e^{2\alpha\phi/M_{\rm Pl}}$ is unbounded at $\phi \to +\infty$ (the deep Starobinsky-plateau regime) and violates this condition.

\textbf{Condition~2: asymptotic canonicity at $\phi \to +\infty$ (a normalisation choice).}
At large positive $\phi$ (Starobinsky inflation regime), we normalise the metric to the canonical form $g_{\chi\chi} \to 1$ so the $\chi$ field is canonically normalised and decouples from the inflaton sector smoothly on the plateau. Like Condition~1 this fixes a chosen-coordinate value ($g_{\chi\chi} \to 1^-$) rather than a coordinate-invariant quantity; the unregularised metric $g_{\chi\chi} \to +\infty$ at $\phi \to +\infty$ violates it.

\textbf{Condition~3: positivity (the coordinate-invariant requirement).}
$g_{\chi\chi}(\phi) > 0$ for all $\phi$, so the kinetic term remains positive-definite and no ghost mode appears. This is the genuinely coordinate-invariant physical requirement (ghost-free signature); Conditions~1--2 are frame-dependent regularity/normalisation choices layered on top of it.
\end{assumptionbox}

\begin{theorembox}
\textbf{Sigmoid is admitted by the three conditions.}
The function $g_{\chi\chi}(\phi) = (1 + e^{-2\alpha\phi/M_{\rm Pl}})^{-1}$ satisfies Condition~1 (bounded above: $g_{\chi\chi} \leq 1$ for all $\phi$), Condition~2 ($\lim_{\phi \to +\infty} g_{\chi\chi} = 1^-$, asymptotically canonical), and Condition~3 (strictly positive). It is also monotone increasing, an additional property of this selected interpolation rather than part of Condition~3. The limit at $\phi \to -\infty$ is $g_{\chi\chi} \to 0^+$, smooth and bounded; both the sigmoid and the unregularised metric have this behaviour, so the $\phi \to -\infty$ limit alone does not discriminate between them. The transition scale of the sigmoid is $\phi \sim M_{\rm Pl}/\alpha$.
\end{theorembox}

\begin{occambox}
\textbf{Sigmoid is the selected minimal-complexity representative.}
Conditions~1--3 admit infinitely many functional forms; the sigmoid is selected by minimal complexity, in the operational sense that it is a one-parameter rational--exponential regularisation $g_{\chi\chi} = f(\phi/M_{\rm Pl}; \alpha)$ with no additional shape coefficients in the stated family of compositions of elementary analytic functions ($\exp$, polynomial, rational). Its limits saturate Conditions~1 and~2 without adding oscillations or higher-derivative singularities. This is a declared selection rule, not a proof of a global complexity minimum across all smooth functions. Alternative selections within the same conditions (e.g.~erf, tanh, smooth step functions) can become degenerate with the sigmoid deep on the saturated plateau, but their derivatives and their behaviour outside that regime are not fixed by the endpoint conditions and require a fresh through-bounce verification.
\end{occambox}

\paragraph{Explicit derivation of the sigmoid.}
The minimal-complexity principle becomes concrete when one looks for a simple one-parameter family of metrics satisfying Conditions~1--3. Parametrising $g_{\chi\chi}(\phi) = f(x)$ with $x = e^{2\alpha\phi/M_{\rm Pl}} \in (0, +\infty)$, the three conditions require some finite upper bound (Condition~1), $f(+\infty)=1$ (Condition~2), and $f(x)>0$ on $(0,+\infty)$ (Condition~3); they do not by themselves impose monotonicity, forbid an overshoot above unity, or fix the lower endpoint. We additionally choose the regular endpoint $f(0^+)=0^+$ and a monotone, no-overshoot interpolation as part of the stated minimal-complexity prescription. Within that selected subclass, the lowest-order homography sending $\{0,\infty\}$ to $\{0,1\}$ is the rational map
\begin{equation}
f(x) = \frac{x}{1 + x},
\label{eq:sigmoid_rational}
\end{equation}
The residual transition-location freedom is fixed by the explicit model choice $g_{\chi\chi}(0) = \tfrac{1}{2}$, aligning the metric transition with the minimum of the Starobinsky potential.  This is not merely a field-origin convention in the complete action: shifting $\phi$ in the metric alone changes its physical placement relative to $V(\phi)$.  Substituting $x = e^{2\alpha\phi/M_{\rm Pl}}$ then recovers the sigmoid $g_{\chi\chi}(\phi) = (1 + e^{-2\alpha\phi/M_{\rm Pl}})^{-1}$. Equivalently, the selected sigmoid is characterised by the autonomous first-order ODE
\begin{equation}
\frac{d g_{\chi\chi}}{d\phi} = \frac{2\alpha}{M_{\rm Pl}}\,g_{\chi\chi}(1 - g_{\chi\chi}),
\qquad g_{\chi\chi}(0) = \tfrac{1}{2},
\label{eq:logistic_ode}
\end{equation}
i.e.\ the logistic equation with growth rate $2\alpha/M_{\rm Pl}$ (inverse-growth scale $M_{\rm Pl}/(2\alpha)$): its upper fixed point $g_{\chi\chi} = 1$ realises the asymptotic canonicity of Condition~2, while the invariant interval $g_{\chi\chi} \in (0, 1)$ realises both the boundedness above of Condition~1 ($g_{\chi\chi} < 1$) and the positivity of Condition~3 ($g_{\chi\chi} > 0$); the lower fixed point $g_{\chi\chi} = 0$ is the chosen $\phi \to -\infty$ endpoint, not a selection condition. The logistic form Eq.~\eqref{eq:logistic_ode} is the lowest-degree polynomial $P(g_{\chi\chi})$ that has both endpoint fixed points. Higher-degree choices need not create additional real fixed points or oscillations, but they introduce extra unconstrained coefficients and shape freedom; rejecting that additional structure is precisely the stated minimal-complexity choice, not a theorem forced by Conditions~1--3.

A \emph{geometric reading} of the prescription is provided by the compact variable $u = (1 + e^{-2\alpha\phi/M_{\rm Pl}})^{-1}$, which maps the real line $\phi \in \mathbb{R}$ to the open interval $u \in (0, 1)$. In the unregularised hyperbolic geometry---the Poincaré half-plane $d\phi^2 + e^{2\alpha\phi/M_{\rm Pl}} d\chi^2$, a geodesically complete space of constant negative curvature $-\alpha^2/M_{\rm Pl}^2$---the metric component is $e^{2\alpha\phi/M_{\rm Pl}} = u/(1-u)$: it degenerates to zero as a coordinate effect at $u \to 0$ and grows without bound (divergent proper volume per unit $\chi$) at $u \to 1$. The sigmoid prescription $g_{\chi\chi} = u$ replaces this unbounded ratio by its bounded numerator: it retains the $\phi \to -\infty$ behaviour ($g_{\chi\chi} \to 0$ with $g_{\chi\chi}/[u/(1-u)] \to 1$, so the hyperbolic regime is reproduced where $u \ll 1$) and caps the $\phi \to +\infty$ growth at the canonical value $1$. We emphasise that this is a \emph{modification of the geometry}, not a coordinate transformation of it---no change of coordinates can alter the constant hyperbolic curvature---so the selection content resides in the boundary conditions and the minimal-complexity/model-placement choices above; $u$ is simply the natural interpolation variable in which the selected metric is the identity map. The metric sector remains $\chi$-independent, but the complete action has no exact $\chi$-translation symmetry because the potential contains $m_\chi^2\chi^2/2$.

\paragraph{Epistemic status summary.}
Condition~3 (positivity) is the coordinate-invariant physical assumption (ghost-free signature); Conditions~1--2 (boundedness and asymptotic canonicity of the chosen field-coordinate) are coordinate-frame regularity/normalisation choices rather than coordinate-invariant requirements, since a field redefinition $\chi \to f(\chi)$ rescales $g_{\chi\chi} \to g_{\chi\chi}/f'(\chi)^2$. The existence theorem (sigmoid is admitted) is a mathematical consequence of Conditions~1--3 plus the choice of family in which the simplest admitted form is sought. The minimal-complexity principle and the placement $g_{\chi\chi}(0)=1/2$ relative to the fixed Starobinsky potential are model-building choices. Together, these layers separate the coordinate-frame choices, the coordinate-invariant positivity requirement, and the methodological/model-placement selections. Plateau observables may agree for alternatives that coincide over the sampled plateau, but no through-bounce verification result is transferred without rerunning the analysis.

\begin{theorembox}
\textbf{NEC preservation.}
A closed-FRW $K=+1$ turning point at $H=0$ obeys the curvature--density balance $\rho = 3 M_{\rm Pl}^2/a^2$; the verified contracting-to-expanding solutions additionally have $\dot H>0$, as derived in \S\ref{sec:model}. The null energy condition \cite{NEC} reads $\rho + p \geq 0$, which on the two-field background is $\dot\phi^2 + g_{\chi\chi}(\phi)\dot\chi^2 \geq 0$ identically (sum of positive-definite terms). The bounce thus proceeds within standard General Relativity, without invoking ghost fields, Galileons, or any other NEC-violating mechanism.
\end{theorembox}

Under the standard Einstein--Hilbert action with the two-field matter content of this work, the bouncing geometry above provides a locally non-singular contracting-to-expanding solution in a closed $K=+1$ universe, with the sigmoid field-space metric selected from the functions admitted by Conditions~1--3 through the stated minimal-complexity and model-placement choices. Anisotropic generalisations (Bianchi~IX) require additional dynamical analysis, addressed in the wall-free homogeneous-shear truncation in \S\ref{sec:bkl}.

\section{The Complete Cosmological Model}
\label{sec:model}

\subsection{Action and Field Space Geometry}

\begin{figure}[htbp]
    \centering
    \IfFileExists{figures/fig_potential_dynamics.pdf}{%
        \includegraphics[width=\textwidth]{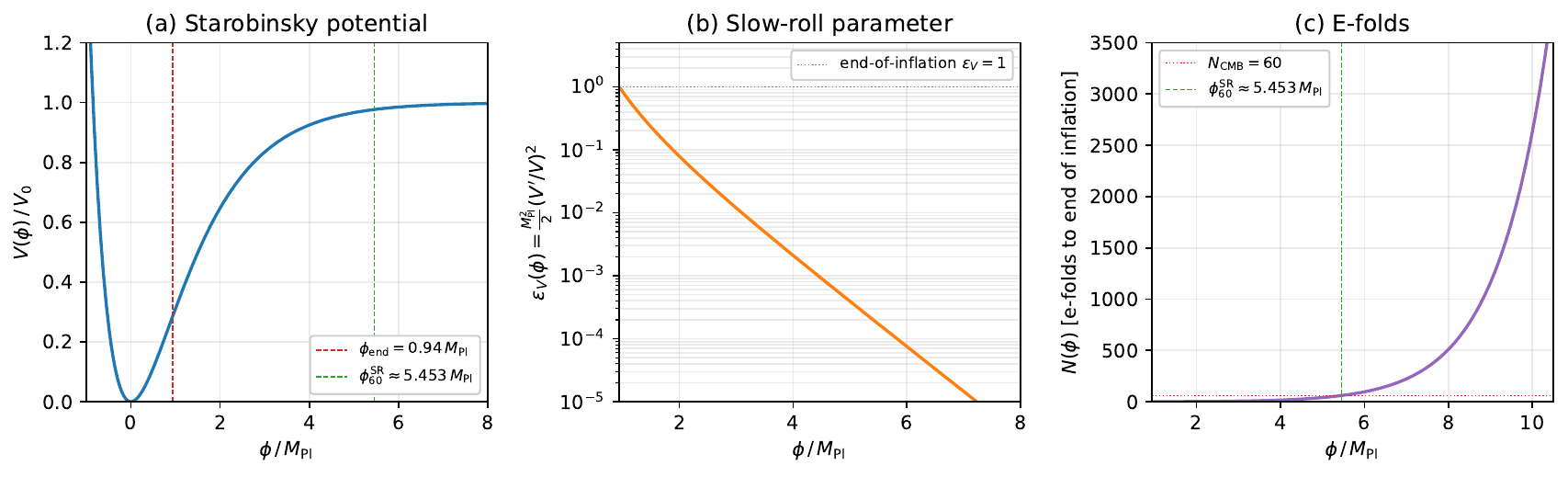}%
    }{%
        \fbox{\parbox{0.78\textwidth}{\centering
            \textbf{Figure not available in this build.}}}%
    }
    \caption{\textbf{Starobinsky inflaton dynamics.}
        (a)~Potential $V(\phi)/V_0 = (1 - e^{-\beta\phi/M_{\rm Pl}})^2$ on the
        plateau, with $\phi_{\rm end} = 0.94\,M_{\rm Pl}$ ($\epsilon_V = 1$)
        and the analytic slow-roll inversion point
        $\phi_{60}^{\rm SR} \approx 5.453\,M_{\rm Pl}$ ($N = 60$).
        (b)~Slow-roll parameter $\epsilon_V(\phi) = \tfrac{1}{2}M_{\rm Pl}^2(V'/V)^2$ on
        a log scale, illustrating the deep slow-roll regime
        ($\epsilon_V \approx 1.9 \times 10^{-4}$ at the $N = 60$ pivot,
        rising to $\approx 2.6 \times 10^{-4}$ at $N = 50$ and
        $\approx 4.0 \times 10^{-4}$ at $N = 40$) across the
        observable inflation window, and the rapid rise to $\epsilon_V = 1$
        at $\phi_{\rm end}$.
        (c)~E-fold counter $N(\phi)$ obtained from the potential slow-roll
        relation including its linear subleading term (Eq.~\ref{eq:N_of_phi}), pinpointing the CMB-pivot
        location $\phi_{60}^{\rm SR} \approx 5.453\,M_{\rm Pl}$ where $N = 60$.}
    \label{fig:potential_dynamics}
\end{figure}

The model action is
\begin{equation}
\label{eq:action}
S = \int d^4x \sqrt{-g} \left[
\frac{M_{\rm Pl}^2}{2} R
- \frac{1}{2} G_{ab}(\phi^c)\,\nabla_\mu \phi^a \nabla^\mu \phi^b
- V(\phi, \chi)
\right],
\end{equation}
with field-space metric
\begin{equation}
G_{ab} = \mathrm{diag}\bigl(1,\, g_{\chi\chi}(\phi)\bigr),
\qquad
g_{\chi\chi}(\phi) = \bigl(1 + e^{-2\alpha\phi/M_{\rm Pl}}\bigr)^{-1},
\end{equation}
and Starobinsky-class potential
\begin{equation}
V(\phi, \chi) = V_0 \bigl(1 - e^{-\beta\phi/M_{\rm Pl}}\bigr)^2 + \tfrac{1}{2} m_\chi^2 \chi^2,
\qquad
\beta = \sqrt{2/3},
\quad V_0 = 10^{-10}\,M_{\rm Pl}^4.
\end{equation}
The inflaton $\phi$ slow-rolls down the Starobinsky plateau (Fig.~\ref{fig:potential_dynamics}); the regulated direction $\chi$ couples to $\phi$ through the sigmoid kinetic term $g_{\chi\chi}(\phi)\,\dot\chi^2$ and through the explicit mass term $\tfrac{1}{2} m_\chi^2 \chi^2$. The fiducial mass is $m_\chi = \mchiFiducial\,M_{\rm Pl}$, a factor $\approx 6$ below the inflationary Hubble scale $H_{\rm inf} \approx 5.8 \times 10^{-6}\,M_{\rm Pl}$ ($m_\chi/H_{\rm inf} \approx \mchiOverHinfFiducial$); we analyse the sensitivity of perturbation results to this choice in \S\ref{subsec:entropy_power_sensitivity}.

The field-space Christoffel symbol is $\Gamma^\chi_{\phi\chi} = (\alpha/M_{\rm Pl})\,(1 - g_{\chi\chi})$, with $\dot{g}_{\chi\chi}/g_{\chi\chi} = 2\,\Gamma^\chi_{\phi\chi}\,\dot\phi$. We use covariant field-space derivatives $\nabla_a$ in the perturbation analysis of \S\ref{sec:perturbations}. The field-space geometry enters the coordinate entropy equation through the connection-induced friction term $\dot g_{\chi\chi}/g_{\chi\chi}=2\Gamma^\chi_{\phi\chi}\dot\phi$, while the inverse metric produces the raised-index potential term $V^{,\chi}=m_\chi^2\chi/g_{\chi\chi}$; the orthonormal system retains the covariant Hessian and field-space curvature explicitly.

For a two-dimensional metric $ds^2=d\phi^2+g_{\chi\chi}(\phi)d\chi^2$, the Gaussian curvature is $K_G=-(\sqrt{g_{\chi\chi}})''/\sqrt{g_{\chi\chi}}$. Substitution of the logistic derivative Eq.~\eqref{eq:logistic_ode} gives
\begin{equation}
K_G=\frac{\alpha^2}{M_{\rm Pl}^2}(1-g_{\chi\chi})(3g_{\chi\chi}-1),
\label{eq:field_space_curvature}
\end{equation}
which is bounded, tends to $-\alpha^2/M_{\rm Pl}^2$ as $\phi\to-\infty$, changes sign at $g_{\chi\chi}=1/3$, and tends to zero from above as $\phi\to+\infty$. This is the curvature formula used in v5.

Curved field spaces also enter inflationary $\alpha$-attractors \cite{KalloshLindeRoest2013}, hyperinflation \cite{Brown2018}, and geometrical destabilisation \cite{RenauxPetelTurzynski2016}. These are distinct dynamical settings, not interchangeable descriptions of the present sigmoid model. In particular, our $\alpha$ controls the sigmoid transition and is not the $\alpha$ of the $\alpha$-attractor family, which sets the constant negative curvature of that hyperbolic field space. The fiducial trajectory has no turn, and its saturated plateau has $g_{\chi\chi}>1/3$, hence $K_G\geq0$ by Eq.~\eqref{eq:field_space_curvature}; the negative-curvature contribution responsible for geometrical destabilisation is absent there. This local statement is not a stability theorem for other trajectories or for the negative-curvature end of the metric.

\subsection{Background Evolution in Closed Universe}

The Friedmann equations for $k = +1$ are
\begin{align}
H^2 &= \frac{\rho}{3 M_{\rm Pl}^2} - \frac{1}{a^2},
\label{eq:friedmann}
\\
\dot H &= -\frac{\rho + p}{2 M_{\rm Pl}^2} + \frac{1}{a^2},
\label{eq:raychaudhuri}
\end{align}
with $\rho = \tfrac{1}{2}\dot\phi^2 + \tfrac{1}{2} g_{\chi\chi}\,\dot\chi^2 + V$ and $p = \tfrac{1}{2}\dot\phi^2 + \tfrac{1}{2} g_{\chi\chi}\,\dot\chi^2 - V$. The bounce occurs when $H = 0$, i.e.\ $\rho = 3 M_{\rm Pl}^2 / a^2$, \emph{with} $\dot H > 0$ there: from Eq.~\eqref{eq:raychaudhuri} at the turning point, $\dot H = (V - \dot\phi^2 - g_{\chi\chi}\dot\chi^2)/(3 M_{\rm Pl}^2) > 0$ precisely when \emph{twice} the turning-point kinetic energy, $\dot\phi^2 + g_{\chi\chi}\dot\chi^2$, is below the potential energy. This local criterion is satisfied on the verified bouncing trajectories of \S\ref{sec:numerics}; it is not by itself a global sufficiency theorem for arbitrary initial data (a turning point with $\dot H < 0$ would instead be a recollapse maximum). Since the kinetic energy is subdominant on the slow-roll plateau ($\dot\phi^2 \ll V$), the constant-plateau approximation defines the isotropic reference radius
\begin{equation}
a_{\rm iso}\equiv\sqrt{\frac{3 M_{\rm Pl}^2}{V_0}},
\qquad
a_{\rm min}=\aMinBounce\,M_{\rm Pl}^{-1}\approx a_{\rm iso},
\label{eq:a_min}
\end{equation}
where $a_{\rm min}$ is the measured fiducial minimum and differs slightly from $a_{\rm iso}$ because the dynamical turning point samples $V(\phi_{\rm bounce})<V_0$ and nonzero kinetic energy. The inflationary Hubble scale $H_{\rm inf}=\sqrt{V_0/(3M_{\rm Pl}^2)}\approx5.8\times10^{-6}\,M_{\rm Pl}$ obeys $a_{\rm iso}H_{\rm inf}=1$ exactly and is only approximately $1/a_{\rm min}$.

\begin{assumptionbox}
\textbf{Initial conditions.}
We initialise the contracting phase with $\phi_0 \approx 10\,M_{\rm Pl}$ on the Starobinsky plateau, $\dot\phi_0 = 0$, $\chi_0 = 0$, $\dot\chi_0 = 0$, and the scale factor at $a_0 = 1.8\,a_{\rm iso}$. The origin of these initial conditions---how the universe enters a contracting phase with the inflaton on the slow-roll plateau---is not addressed here and is a common open question for curvature-driven bouncing cosmologies \cite{Novello2008,CaiEassonBrandenberger2012,BattefeldPeter2015,BrandenbergerPeter2017}. The inequality $\dot\phi_0^2 \lesssim 3\%\,V_0$ quoted in the antecedent scan is an analytic turning-point guide for the fiducial remaining data, not a v5-verified sufficiency region: the sampled family here uses $\dot\phi_0=0$ only. The brief contraction spans $\Delta N=\ln(a_0/a_{\rm min})\approx0.59$ e-folds; using the nearby reference ratio $a_0/a_{\rm iso}=1.8$ gives the heuristic homogeneous anti-friction factor $1.8^6\approx34$ and the corresponding $a^{-2}$ gradient-scaling factor $e^{2\Delta N}\approx3.2$. Those finite factors explain why the sampled near-homogeneous trajectories remain controlled, but they do not establish nonlinear stability for generic inhomogeneous initial data.
\end{assumptionbox}

On the fiducial $\chi = \dot\chi = 0$ trajectory the $g_{\chi\chi}\dot\chi^2$ contribution to $\rho + p$ vanishes identically, so the inflaton dynamics are exactly $\alpha$-independent at the background level: the spectral observables $n_s$, $r$, $f_{\rm NL}$ on this trajectory depend only on the Starobinsky potential and not on the regularisation scale $\alpha$. A displaced-family background probe of $g_{\chi\chi}(\alpha)$ via a non-zero spectator displacement is reported in \S\ref{sec:observables}; the entropy-mode equation of motion (\S\ref{sec:perturbations}) carries the explicit field-space-metric dependence structurally, through the raised-index mass $m_\chi^2/g_{\chi\chi}$. The headline absolute-power scalar, tensor, and entropy runs are evaluated on the fiducial $\alpha = 1$ trajectory, where $g_{\chi\chi}$ is saturated at unity throughout the integration window. Separately, the cross-basis consistency comparison grid integrates displaced trajectories at $\alpha\in\{1,3\}$, but tests construction-to-construction consistency rather than spectral $\alpha$-independence. The size of the $1/g_{\chi\chi}$ enhancement that would appear at smaller $\alpha$ is quantified in \S\ref{sec:perturbations}, but is not itself scanned at the perturbation level here.

\section{Numerical Validation and Robustness}
\label{sec:numerics}

\subsection{Background Integration}

\begin{figure}[htbp]
    \centering
    \IfFileExists{figures/fig_background_fiducial.pdf}{%
        \includegraphics[width=0.96\textwidth]{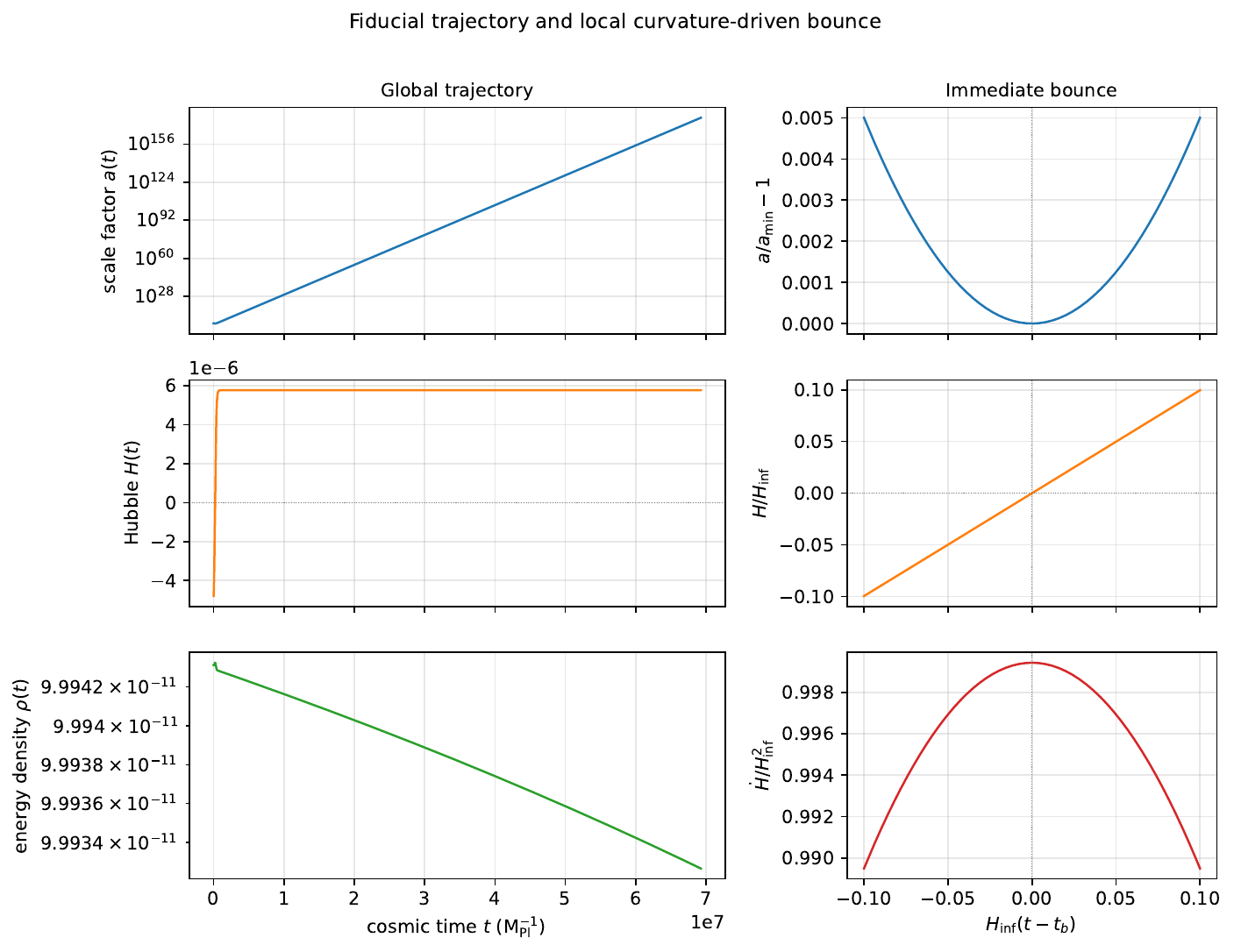}%
    }{%
        \fbox{\parbox{0.78\textwidth}{\centering
            \textbf{Figure not available in this build.}}}%
    }
    \caption{\textbf{Fiducial trajectory and immediate bounce.}
        Left: scale factor $a(t)$ (top, log scale), Hubble parameter $H(t)$
        (middle), and total energy density $\rho(t)$ (bottom, log scale) over
        the full fiducial integration. Right: the local window
        $|H_{\rm inf}(t-t_b)|\leq0.1$, showing $a/a_{\min}-1$,
        $H/H_{\rm inf}$, and $\dot H/H_{\rm inf}^2$. The scale factor reaches
        its non-singular minimum $a_{\rm min}=\aMinBounce\,M_{\rm Pl}^{-1}$
        at $H=0$, with $\dot H>0$, while the matter density remains finite.
        The displayed trajectory has $\phi_0=10\,M_{\rm Pl}$,
        $\chi_0=\dot\chi_0=0$, and $\alpha=1$ and is drawn from a fresh
        integration of the background equations.}
    \label{fig:background_fiducial}
\end{figure}

The background two-field FLRW system, Eqs.~(\ref{eq:friedmann})--(\ref{eq:raychaudhuri}) plus the Klein--Gordon equations for $\phi$ and $\chi$, is integrated with the explicit eighth-order Dormand--Prince method (DOP853), at relative tolerance $10^{-12}$ and absolute tolerance $10^{-14}$, falling back to the implicit Radau~IIA method (relative $10^{-10}$, absolute $10^{-12}$) when the explicit integrator fails to converge. Background quantities (scale factor, Hubble parameter, field values) are interpolated onto the perturbation time grid using cubic splines. The fiducial trajectory ($\phi_0 = 10\,M_{\rm Pl}$, $\chi_0 = \dot\chi_0 = 0$, $\alpha = 1$) reaches its minimum scale factor $a_{\rm min} = \aMinBounce\,M_{\rm Pl}^{-1}$ at the bounce, with Hubble parameter $H_{\rm bounce} = \HBounce$ confirming the linear $H$ zero-crossing through the bounce point (Fig.~\ref{fig:background_fiducial}). Post-bounce, the trajectory sustains $\NpostBounce$ e-folds of Starobinsky inflation within the integration window, with the inflaton still deep on the slow-roll plateau at the end of the simulated time span---the potential slow-roll parameter $\epsilon_V \equiv \tfrac{1}{2} M_{\rm Pl}^2 (V'/V)^2$ remains $\ll 1$ there. The integration window is deliberately truncated once several hundred e-folds have accumulated (already far beyond the $\sim 60$ required observationally); the analytical slow-roll inversion of \S\ref{sec:flatness} places the true end of inflation ($\epsilon_V = 1$) much later, at $N \approx 2630$ e-folds for this trajectory.

\paragraph{Independent flat-FRW scalar benchmark.}\phantomsection
\label{par:flat_scalar_benchmark}
As a separate numerical reference, we solve the single-field Mukhanov--Sasaki equation $v_k''+(k^2-z''/z)v_k=0$ on a spatially flat Starobinsky background, with $z=a\sqrt{2\epsilon_H}M_{\rm Pl}$ and conformal-time derivatives. The background starts at $\phi=6M_{\rm Pl}$ and reaches the end of inflation at $\epsilon_H=1$; $k_*=a_*H_*$ is defined 60 e-folds before that event. Five modes with $\ln(k/k_*)\in[-0.05,0.05]$ are initialised at $k/(aH)=10^3$ and evaluated half an e-fold before the end of inflation. Fitting $\ln P_{\mathcal R}$ against $\ln k$ gives $n_s=0.96774$, compared with $1-2/60=0.96667$; the difference $0.00108$ satisfies the tolerance $0.005$. A recorded one-off initial-depth sweep to $3\times10^3$ gave $n_s=0.96774314$, differing from the depth-$10^3$ value by approximately $1.4\times10^{-6}$; this sweep is not repeated by the standard verification run. This validates the independent flat calculation against a leading slow-roll reference; it is not a spectral validation of the closed two-field system and supplies none of its power normalisations.

\subsection{Sigmoid saturation diagnostic}

On the fiducial trajectory $\alpha = 1$, $\phi(t)$ remains above $5\,M_{\rm Pl}$ throughout both the perturbation window (bounce through $\sim\!65$ e-folds) and the longer simulated background window. The maximum deviation $|1-g_{\chi\chi}|_{\max}=\gDeviationMax$ over the latter occurs at its late cutoff ($\phi\approx9.80\,M_{\rm Pl}$ after $\NpostBounce$ simulated e-folds; the deviation grows monotonically as $\phi$ decreases, and at the bounce itself it is $\approx2.1\times10^{-9}$), below the perturbation thresholds. The separate $\phi_0=6M_{\rm Pl}$ excited-spectator $\alpha$ scan of \S\ref{sec:observables} reaches $g_{\chi\chi}=\gCMBminNT$ at its measured $N=60$ point for $\alpha=0.1$. This remains $\mathcal{O}(1)$, while the raised-index entropy mass carries the explicit factor $1/g_{\chi\chi}$ (about a $34\%$ increase at that point; \S\ref{sec:perturbations}).

\subsection{Friedmann constraint and NEC pointwise}

The Friedmann constraint $3 M_{\rm Pl}^2 H^2 = \rho - 3M_{\rm Pl}^2/a^2$ holds pointwise on the integrated trajectory. We measure its violation through
\begin{equation}
\Delta_F=\left|H^2-\frac{\rho}{3M_{\rm Pl}^2}+\frac{1}{a^2}\right|,
\qquad
\mathcal S=\max\!\left\{\frac{\rho}{3M_{\rm Pl}^2},\frac{1}{a^2},H_{\rm inf}^2,10^{-30}M_{\rm Pl}^2\right\},
\end{equation}
and obtain $\max(\Delta_F/\mathcal S)=\FriedmannScaledResidualMax$, against an acceptance threshold of $\FriedmannScaledResidualGate$; in the numerical $M_{\rm Pl}=1$ units, $\max\Delta_F=\FriedmannAbsoluteResidualMax$ against an independent absolute threshold $\FriedmannAbsoluteResidualGate$. The null energy condition $\rho + p = \dot\phi^2 + g_{\chi\chi}\,\dot\chi^2 \geq 0$ is satisfied identically by construction. The unreduced matter equations have luminal principal characteristics, $c_\phi^2=c_\chi^2=1$, and the Einstein--Hilbert tensor characteristic is likewise $c_T^2=1$. These facts do not by themselves determine the finite-harmonic gradient-to-kinetic ratio of the constraint-reduced curvature mode: on the fiducial no-turn trajectory it is the mode- and time-dependent $c_\ell^2=\mathcal B_\ell/\mathcal A_\ell$. This constraint-reduced finite-harmonic ratio is not a local signal speed at fixed low $\ell$; in particular, it is $\simeq\ExactScalarBounceCSqMin$ throughout the declared harmonic set at the bounce. Appendix~\ref{app:closed_s3_canonical} derives those coefficients, reports the resolved endpoints, explains the non-commuting bounce and short-wavelength limits, and verifies $\mathcal A_\ell>0$ and $\mathcal B_\ell>0$ on the declared finite grid and at every adaptive evaluation, so no ghost or gradient instability is found on that sampled fiducial domain. Separately, the same integration corroborates recovery of the Starobinsky-1980 slow-roll attractor \cite{Starobinsky1980} through the relative deviation of $\Delta N$ on the symmetric annular window $5\leq|\Delta N|\leq30$ about $\phi\approx\AttractorPhiRef\,M_{\rm Pl}$: the median is $\AttractorMedianDeviation$ (threshold $\AttractorMedianGate$) across $\AttractorPointCount$ sampled points. These statements use no antecedent numerical diagnostic.

\subsection{Finite-time leading-adiabatic states and Wronskian normalisation}

For the tensor and entropy sectors, ``Bunch--Davies'' (BD) denotes the adopted finite-time prescription with the form of a leading positive-frequency mode, where $k_{\rm phys}(t) = k_{\rm comoving}/a(t)$. Its adiabatic interpretation requires control of the canonical-frequency corrections and their time variation, not merely $k_{\rm phys}/|H| \gg 1$ (\S\ref{subsec:BD_robustness}). It is not asserted to be a unique asymptotic vacuum. A physical field mode $h_k(t)$ with a canonical conformal variable proportional to $a h_k$ satisfies
\begin{equation}
\ddot h_k + 3 H \dot h_k + (k_{\rm phys}^2 + m_{\rm eff}^2)\,h_k = 0
\label{eq:mode_function_eom}
\end{equation}
on the curved background. Rescaling to $u_k=a h_k$ and conformal time absorbs the friction exactly: $u_k''+[k_{\rm comoving}^2+a^2m_{\rm eff}^2-a''/a]u_k=0$, where $a''/a=a^2(2H^2+\dot H)$. Neglecting the mass and geometric terms requires them to be small compared with $k_{\rm comoving}^2$; the Hubble-depth condition alone does not establish this near the bounce. The resulting leading positive-frequency form specifies the finite-time initial data used here,
\begin{equation}
h_k^{\rm BD}(t) = \frac{1}{\sqrt{2\,k_{\rm phys}(t)\,a(t)^3}}\,e^{-i \int^{t} k_{\rm phys}(t')\,dt'},
\qquad
\dot h_k^{\rm BD}(t) = -\bigl(H(t)+i k_{\rm phys}(t)\bigr)\,h_k^{\rm BD}(t),
\label{eq:BD_initial_state}
\end{equation}
The real $-Hh_k$ term is the derivative of the $1/a$ amplitude obtained when the canonical conformal mode is transformed back to the physical field mode; it cancels from the antisymmetric Wronskian and therefore cannot be tested by Wronskian conservation alone. This is the finite-time leading-positive-frequency (``BD'') prescription in the unit canonical normalisation of a single scalar degree of freedom, as used for the entropy mode. In this normalisation the Wronskian
\begin{equation}
W = a^3\,(h_k\,\dot h_k^* - h_k^*\,\dot h_k) = i
\label{eq:wronskian_norm}
\end{equation}
fixes the finite-time state normalisation. The Wronskian is conserved by the exact linear evolution; its numerical drift provides a normalisation and integration diagnostic. The closed-FRW tensor solver of \S\ref{subsec:tensor_solver} uses the two-polarisation envelope normalisation fixed by the tensor action in which Eq.~\eqref{eq:BD_initial_state} is replaced by $h_k^{\rm BD} = 2 / \sqrt{a^3 k_{\rm phys}}\,e^{-i\int^t k_{\rm phys}(t')dt'}$ with $\dot h_k^{\rm BD}=-(H+i k_{\rm phys})h_k^{\rm BD}$, and the conserved Wronskian takes the value $W_{\rm init} = 8 i$; the tensor per-polarisation value is $4i$, fixed by the action coefficient in the adopted polarisation convention, and summing two polarisations gives $8i$. Relative to the unit scalar mode, the envelope amplitude is larger by $2\sqrt{2}$; this factor must be retained consistently in the tensor power and is not a freely adjustable amplitude. See \S\ref{subsec:tensor_solver} for the explicit tensor-side form.

We initialise Eq.~\eqref{eq:mode_function_eom} with the finite-time state Eq.~\eqref{eq:BD_initial_state} at the earliest pre-bounce time at which $k_{\rm phys}/|H| \geq 1.5 \times 10^3$ is satisfied, evolve through $H = 0$, and verify Wronskian conservation at a mid-contraction reference point between initialisation and the bounce (the diagnostic is restricted to the pre-bounce branch, where the oscillatory mode functions condition the Wronskian well). Both solvers---the tensor sector under the sum-over-polarisations convention $W_{\rm init} = 8 i$, and the entropy sector under the unit canonical mode convention $W_{\rm init} = i$---give Wronskian-conservation residuals $|W - W_{\rm init}|/|W_{\rm init}|$ that stay well within the imposed Wronskian tolerance ($10^{-4}$): $\lesssim 10^{-8}$ for the tensor solver and $\lesssim 10^{-7}$ for the entropic solver (rising toward the latter at the highest $\ell \sim 10^3$), consistent with the analytical conservation requirement. A separate self-contained algebraic benchmark checks the Bogoliubov acceptance identity for a free-scalar instantaneous frequency change and a constant-$G$ curved-field-space toy (maximum identity residual $\sim 10^{-13}$ vs threshold $10^{-12}$). Its helpers are defined within that benchmark; it does not exercise the numerical solvers or validate a solver-side Bogoliubov extraction. A Bogoliubov decomposition of the \emph{evolved} post-bounce mode functions into a late-time leading-adiabatic basis, $h_k^{\rm late} = \alpha_k h_k^{\rm BD} + \beta_k h_k^{{\rm BD}*}$ with $|\alpha_k|^2 - |\beta_k|^2 = 1$, is \emph{not} performed in this work; the numerical normalisation evidence is the pre-bounce Wronskian check above together with its imposed $10^{-4}$ tolerance. The fiducial adiabatic scalar mode instead uses the exact closed-space canonical momentum $\Pi_{\mathcal R,\ell}=a^3\mathcal A_\ell\dot{\mathcal R}_\ell$, with $\mathcal R_\ell\Pi_{\mathcal R,\ell}^*-\mathcal R_\ell^*\Pi_{\mathcal R,\ell}=i$; Appendix~\ref{app:closed_s3_canonical} gives the mode-dependent coefficients, two propagated finite-time state prescriptions, and the exact map to regular Newtonian initial data.

\subsection{Basin of attraction}

The sampled family probes four initial-condition points in the $(\phi_0,\dot\phi_0,\chi_0,\dot\chi_0)$ phase space at $a_0=1.8a_{\rm iso}$. The three points $(\chi_0,\dot\chi_0)\in\{(0,0),(0.3,10^{-7}),(1.0,5\times10^{-7})\}$ at $\dot\phi_0=0$ undergo a contraction-to-expansion crossing followed by more than $60$ e-folds. The contracting-control initial condition $(1.5M_{\rm Pl},2\times10^{-6})$ is instead a finite-range background control: an independent branch that enforces the Friedmann constraint algebraically reaches $a=0.1a_{\rm iso}$ without an $H^2=0$ root, while remaining contracting, with $2K/V\approx9.1\times10^5$ and increasing $\rho a^2/(3M_{\rm Pl}^2)$. It is therefore classified as a contracting control without a bounce. This establishes absence of the intended curvature bounce over the declared range; it does not prove what happens at arbitrarily smaller $a$ and, because the successful and control points differ in both $\chi_0$ and $\dot\chi_0$, it does not locate a basin boundary. All sampled points have $\dot\phi_0=0$; the $\dot\phi_0^2\lesssim3\%V_0$ condition is the analytic $\dot H>0$ turning-point criterion of \S\ref{sec:model} (probed by an antecedent scan), not a region swept here.

Within the adopted field-coordinate convention, the sigmoid regularisation provides an explicit bounded alternative to the unregularised hyperbolic metric: $0<g_{\chi\chi}\leq1$ caps the kinetic contribution of a fixed coordinate velocity $\dot\chi_0$, whereas $g_{\chi\chi}=e^{2\alpha\phi/M_{\rm Pl}}$ can assign an exponentially large kinetic weight at positive $\phi$. This is not a coordinate-invariant ordering of the two models. A static displacement contributes $m_\chi^2\chi_0^2/2\approx0.5\%V_0$ at $\chi_0=1M_{\rm Pl}$. The antecedent works \cite{Kravchenko2025v2,Kravchenko2025v3} reported a $\sim10^{21}$ expansion of the successful-bounce basin at fixed coordinate velocity, scanning $|\dot\chi_0|$ over $10^{-20}$--$10^{-4}\sqrt{V_0}$. That exponent is retained only as historical context: the exponential-metric endpoint $|\dot\chi_0|\lesssim10^{-25}\sqrt{V_0}$ was an illustrative fine-tuning threshold rather than a converged boundary, and the sigmoid endpoint was the edge of the tested range. Moreover, at $\phi_0=10M_{\rm Pl}$ the conversion between coordinate and proper velocity contributes roughly four of the quoted orders, and a fixed-canonical-momentum comparison would weight the models differently. The figure is not re-derived here. The v5 result is therefore the bounded metric plus successful bounces on the three explicitly listed bouncing initial conditions and the finite-range contracting control, not a measured basin volume or expansion factor.

\section{Flatness Problem Resolution}
\label{sec:flatness}

The closed-FRW $K=+1$ universe with $\phi_0 \approx 10\,M_{\rm Pl}$ on the Starobinsky plateau produces a long post-bounce inflationary phase that supplies the standard inflationary resolution of the flatness problem: the relative curvature contribution $\Omega_k = -1/(a H)^2$ decays exponentially during inflation and becomes negligible within the model's decoupling-era estimate. For $N(t)=\ln[a(t)/a_{\rm onset}]$, the exact kinematic relation is
\begin{equation}
\frac{|\Omega_k(t)|}{|\Omega_k|_{\rm onset}}
=e^{-2N(t)}\left[\frac{H_{\rm onset}}{H(t)}\right]^2.
\label{eq:Omega_k_exact_scaling}
\end{equation}
The familiar $e^{-2N}$ law is therefore the quasi-de Sitter plateau approximation, not an identity when $H$ evolves. We define the plateau onset by $aH=1$, so $|\Omega_k|_{\rm onset}=1$. The reciprocal reference scales $a_{\rm iso}=\sqrt{3M_{\rm Pl}^2/V_0}$ and $H_{\rm inf}=\sqrt{V_0/(3M_{\rm Pl}^2)}$ provide this normalisation, although $a=a_{\rm iso}$ and $H=H_{\rm inf}$ are not attained simultaneously: $H$ rises from zero to approximately $H_{\rm inf}$ within a fraction of an e-fold while $a$ remains near the measured $a_{\rm min}$. At the bounce point itself $H=0$ and $\Omega_k$ is not defined.

\paragraph{E-fold taxonomy.}
Three e-fold counts appear in the analysis and we keep them notationally distinct:
\begin{itemize}
\item $N_{\rm total}^{\rm ideal}(\phi_{\rm bounce})$: the analytical slow-roll inversion of the Starobinsky $N(\phi)$ relation, valid in the strict slow-roll limit. For the fiducial $\phi_{\rm bounce} = 10\,M_{\rm Pl}$ the relation
\begin{equation}
N(\phi) = \tfrac{3}{4}\,e^{\beta\phi/M_{\rm Pl}} - \tfrac{3}{4}\,e^{\beta\phi_{\rm end}/M_{\rm Pl}} - \frac{\phi - \phi_{\rm end}}{2\beta M_{\rm Pl}},
\qquad \beta = \sqrt{2/3},
\label{eq:N_of_phi}
\end{equation}
with $\phi_{\rm end} \approx 0.94\,M_{\rm Pl}$ (the $\epsilon_V = 1$ end-of-inflation point) yields $N_{\rm total}^{\rm ideal} \approx 2630$.

\item $N_{\rm total}^{\rm sim} = \NpostBounce$: the e-fold count $\ln(a_{\rm end}/a_{\rm min})$ actually accumulated by the numerical trajectory within the integration window, from the bounce to the end of the integrated time span. This is far smaller than $N_{\rm total}^{\rm ideal}$ \emph{not} because inflation ends early but because the integration window is deliberately truncated once the trajectory has accumulated several hundred e-folds---already vastly more than the observational requirement. The inflaton is still on the slow-roll plateau ($\epsilon_V \ll 1$) at the end of the window, consistent with the analytical ideal placing the true end of inflation at $N_{\rm total}^{\rm ideal} \approx 2630$. (For the shorter $\phi_0 = 6.0\,M_{\rm Pl}$ trajectory used in the baseline $\alpha$-independence scan of \S\ref{sec:observables}, where $N_{\rm total}^{\rm ideal} \approx 92$ at its bounce point fits inside the window, the integration \emph{does} run to $\epsilon_V = 1$ and the simulated count $N_{\rm total} = \NtotalInflationAlphaScan$ is the genuine end-of-inflation value.)

\item $N_{\rm CMB} = 60$: the standard CMB pivot e-folds, the conventional number of post-horizon-exit e-folds before the end of inflation at which the spectral observables are quoted. The analytic slow-roll inversion gives $\phi_{60}^{\rm SR}\approx5.453\,M_{\rm Pl}$ via Eq.~\eqref{eq:N_of_phi}; the numerically measured baseline point is defined separately as $\phi_{60}^{\rm num}\equiv\phiCMBmeasured\,M_{\rm Pl}$.
\end{itemize}
With $N_{\rm total}^{\rm sim} \approx \NpostBounce$ e-folds, direct evaluation of the exact endpoint quantity $1/(a_{\rm cutoff}H_{\rm cutoff})^2$ gives $\log_{10}|\Omega_k|_{\rm cutoff}=\OmegaKCutoffLogTen$. Thus the residual curvature at the simulated inflationary cutoff is
\begin{equation}
|\Omega_k|_{\rm cutoff}
=e^{-2 N_{\rm total}^{\rm sim}}
\left(\frac{H_{\rm onset}}{H_{\rm cutoff}}\right)^2
\simeq e^{-2 N_{\rm total}^{\rm sim}}
\lesssim 10^{\OmegaKCutoffUpperExponent},
\label{eq:Omega_k_residual}
\end{equation}
For a deliberately conservative upper estimate, suppose inflation ended immediately at that simulated cutoff and the subsequent radiation/matter history then increased $|\Omega_k|\propto(aH)^{-2}$ by a reheating-scale-dependent factor of order $10^{50}$ before CMB decoupling. This gives $\log_{10}|\Omega_k|_{\rm dec,upper}=\OmegaKDecouplingLogTen$, or $\lesssim 10^{\OmegaKDecouplingUpperExponent}$. This is a decoupling-era upper estimate, not the quantity directly constrained by Planck+BAO: the quoted $0.4\%$ precision refers to the late-time curvature parameter $\Omega_{K,0}$ \cite{Planck2018}. We therefore make no same-epoch likelihood comparison. The estimate demonstrates an enormous inflationary suppression margin, but a model-specific conversion to $\Omega_{K,0}$ requires a complete reheating and late-time background history, which is not computed here. The actual trajectory is still inflating at the cutoff, and the analytical slow-roll inversion places its end at $N^{\rm ideal}\approx2630$; those additional e-folds can only suppress curvature further. The $10^{50}$ factor is an order-of-magnitude thermal-history reference, not a computed reheating transfer or likelihood.

The substantial overshoot of the total inflationary duration ($N_{\rm total}^{\rm ideal} \approx 2630$, of which at least $N_{\rm total}^{\rm sim} = \NpostBounce$ are directly simulated) relative to the observationally required $N_{\rm CMB} = 60$ is a consequence of the chosen $\phi_{\rm bounce}$. The separate $\phi_0=6M_{\rm Pl}$ baseline scan supplies a concrete smaller-field example that still bounces and reaches $N_{\rm total}=\NtotalInflationAlphaScan$, placing its bounce-scale feature closer to observable scales; no general continuation in $\phi_0$ is claimed. The flatness resolution is therefore not a unique prediction of the model but a generic consequence of post-bounce Starobinsky inflation; the model's distinguishing feature is the finite contracting-to-expanding transition rather than the flatness solution per se.

\section{BKL Compatibility: Homogeneous-Shear Robustness in the Wall-Free Truncation}
\label{sec:bkl}

The Belinskii--Khalatnikov--Lifshitz (BKL) analysis \cite{Belinskii1970} establishes that generic anisotropic contraction in General Relativity approaches a chaotic Kasner-like singular state in which spatial shear $\Sigma^2$ dominates over matter and curvature. In a curvature-driven bounce, the question is whether the bounce can be reached before that regime develops. We address only a necessary check in the \emph{wall-free homogeneous-shear truncation}: anisotropy enters as a conserved shear-energy component on the closed-FRW $K=+1$ background. The Bianchi-IX curvature-anisotropy wall potential and the anisotropy variables $\beta_\pm$ are not integrated, and their neglect at the bounce scale is a truncation \emph{assumption}, not an established property.

\paragraph{Shear-threshold diagnostic.}
The Bianchi~IX generalisation of Eqs.~\eqref{eq:friedmann}--\eqref{eq:raychaudhuri} carries a shear-energy-density contribution $\rho_{\rm shear} = \Sigma^2/(2 a^6)$ in addition to the scalar matter content. The generalised Friedmann constraint reads
\begin{equation}
H^2 = \frac{\rho_{\rm matter} + \rho_{\rm shear}}{3 M_{\rm Pl}^2} - \frac{1}{a^2},
\qquad
\rho_{\rm shear} = \frac{\Sigma^2}{2 a^6},
\label{eq:BIX_friedmann}
\end{equation}
where $\Sigma^2$ is the conserved comoving shear amplitude (the squared shear scales as $\Sigma^2/a^6$). A useful diagnostic is the epoch at which the shear energy density overtakes the spatial-curvature term in Eq.~\eqref{eq:BIX_friedmann}:
\begin{equation}
\frac{\rho_{\rm shear}/(3 M_{\rm Pl}^2)}{1/a^2} = \frac{\Sigma^2}{6 M_{\rm Pl}^2\,a^4} > 1
\quad\Longleftrightarrow\quad \Sigma^2 > 6\,M_{\rm Pl}^2\,a^4,
\label{eq:kasner_condition}
\end{equation}
comparing the two terms of Eq.~\eqref{eq:BIX_friedmann} like for like. This shear--curvature crossover is a diagnostic, \emph{not} the bounce-prevention threshold itself. For an analytic benchmark, freeze the matter density to a constant plateau, $\rho_{\rm matter}=V_0$, and neglect kinetic evolution. Then setting $H^2 = 0$ and clearing $a^6$ gives $(V_0/3 M_{\rm Pl}^2)\,a^6 - a^4 + \Sigma^2/(6 M_{\rm Pl}^2) = 0$. Maximising the Friedmann barrier $a^4 - (V_0/3 M_{\rm Pl}^2)\,a^6$ (stationary at $a^2 = 2 M_{\rm Pl}^2/V_0$) yields the double-root value
\begin{equation}
\Sigma_{\rm crit}^2 = \frac{8 M_{\rm Pl}^6}{V_0^2} = \frac{8}{9}\,M_{\rm Pl}^2\,a_{\rm iso}^4 \approx 8 \times 10^{20}\,M_{\rm Pl}^{-2}.
\label{eq:sigma_crit}
\end{equation}
Within this constant-plateau approximation there are two positive turning-point radii for $0\leq\Sigma_0^2<\Sigma_{\rm crit}^2$, a degenerate root at equality, and none above it. The evolving two-field system is not covered by that algebraic if-and-only-if statement; Eq.~\eqref{eq:sigma_crit} is therefore used as an analytic benchmark and the actual bounce outcome is determined by the dynamical integration. The shear--curvature crossover Eq.~\eqref{eq:kasner_condition}, evaluated at the isotropic reference radius $a_{\rm iso}=\sqrt{3M_{\rm Pl}^2/V_0}$, lies a factor $6.75$ above the benchmark.

\paragraph{Dynamical evolution.}
On the closed-FRW $K=+1$ background with the matter content of \S\ref{sec:model}, we adopt the \emph{wall-free truncation as an assumption}: the curvature-anisotropy wall potential and the anisotropy variables $\beta_\pm$ are not integrated, so the wall's domain of activation is not established by the present computation. Heuristically the wall term is higher order in the anisotropy for the near-isotropic configurations considered here (shear far below threshold throughout), but every statement in this section is conditional on the truncation. Within it, the shear $\Sigma^2$ obeys the conservation theorem below.

\begin{theorembox}
\textbf{Conserved shear and the constant-plateau benchmark.}
In the wall-free homogeneous-shear truncation adopted here, $\Sigma^2$ is exactly conserved along the trajectory. If, in addition, the matter sector is frozen to $\rho_{\rm matter}=V_0$, a non-degenerate curvature turning point exists precisely for $0\leq\Sigma_0^2<\Sigma_{\rm crit}^2=8M_{\rm Pl}^6/V_0^2=(8/9)M_{\rm Pl}^2a_{\rm iso}^4$, where $a_{\rm iso}=\sqrt{3M_{\rm Pl}^2/V_0}$ is the isotropic plateau turning radius. Conservation is exact within the truncation; the critical value is an analytic constant-plateau benchmark, not a theorem for the dynamically evolving scalar system.
\end{theorembox}

The conservation follows from the truncated shear equation with vanishing anisotropic wall force: the physical shear evolves as a free $w=1$ stiff component, $\rho_{\rm shear}\propto a^{-6}$, while its comoving amplitude $\Sigma^2$ is constant.

\paragraph{Numerical verification.}
In the wall-free homogeneous-shear integration the integrated state is the FRW background $(\phi,\chi,a)$, with the conserved shear entering the generalised Friedmann constraint as an $a^{-6}$ component. The scan covers $\Sigma_0^2\in[10^{-10},10^{21}]\,M_{\rm Pl}^{-2}$ (31 orders of magnitude, eight logarithmic points) on the fiducial $\phi_0=10M_{\rm Pl}$ trajectory. All seven benchmark-subcritical points, $\Sigma_0^2\leq10^{18}$, undergo an actual contraction-to-expansion sign crossing of $\dot a$. Their physical Friedmann shear/curvature ratio, $\Sigma^2/(6M_{\rm Pl}^2a^4)$, never exceeds $1.851\times10^{-4}$; the inherited $\Sigma^2/a^4$ proxy is not used for this statement or for the bounce classification. The top point, $\Sigma_0^2=10^{21}$, lies above the constant-plateau benchmark $8\times10^{20}$ and reaches the physical shear/curvature crossover while still contracting. From that event onward the shear term exceeds the negative curvature term and the matter density remains non-negative, so a later $H=0$ crossing is impossible within the truncation; this supplies the scan's negative control without interpreting a solver failure as physics. At the isotropic reference radius its shear term is only $\approx0.19$ of the curvature term, illustrating why the crossover ($\approx5.4\times10^{21}$ there) is distinct from bounce prevention. The largest verified bouncing shear is a factor $\sim800$ below the analytic benchmark. In the integration $\Sigma^2$ is held constant by construction, so the scan tests bounce robustness under the sampled shear amplitudes rather than conservation itself. No probability measure or generic basin claim over shear initial data is made. The result is therefore limited to the sampled family and wall-free truncation; full Bianchi-IX dynamics (dynamical $\beta_\pm$ and wall potential) and the inhomogeneous problem are not computed.

\paragraph{Comparison with ekpyrotic BKL suppression.}
The standard ekpyrotic resolution \cite{KhouryOvrutSteinhardtTurok2001,EricksonWesleySteinhardtTurok2004} invokes a component with $w\gg1$ during contraction; its density grows faster than $a^{-6}$ and dynamically suppresses shear. No such attractor is demonstrated here. Instead, every successful sampled wall-free trajectory turns around while its physical shear/curvature ratio remains below $1.9\times10^{-4}$. Thus spatial curvature suffices for the finite sampled check without $w\gg1$ matter, but this is not a replacement for full Bianchi-IX or inhomogeneous BKL analysis.

\paragraph{Caveats.}
Two caveats: (i) full inhomogeneous BKL instabilities in $3+1$ numerical GR are not captured by the homogeneous wall-free shear truncation studied here, and would require dedicated numerical-relativity simulations following the programme of Ijjas--Pretorius--Steinhardt \cite{IjjasPretoriusSteinhardt2020}; (ii) the wall-free truncation is an \emph{assumption}: with $\beta_\pm$ and the wall potential absent from the integrated state, the present computation cannot establish where the wall activates; the heuristic expectation that the wall regime lies at $a \ll a_{\rm min}$ for near-isotropic configurations remains to be demonstrated dynamically. We adopt the wall-free homogeneous-shear result as a necessary but not sufficient BKL-compatibility check; the full Bianchi-IX and inhomogeneous analyses are identified as future work in \S\ref{sec:conclusions}.

\section{Covariant Two-Field Perturbations}
\label{sec:perturbations}

\begin{figure}[htbp]
    \centering
    \IfFileExists{figures/fig_perturbation_residuals.pdf}{%
        \includegraphics[width=0.82\textwidth]{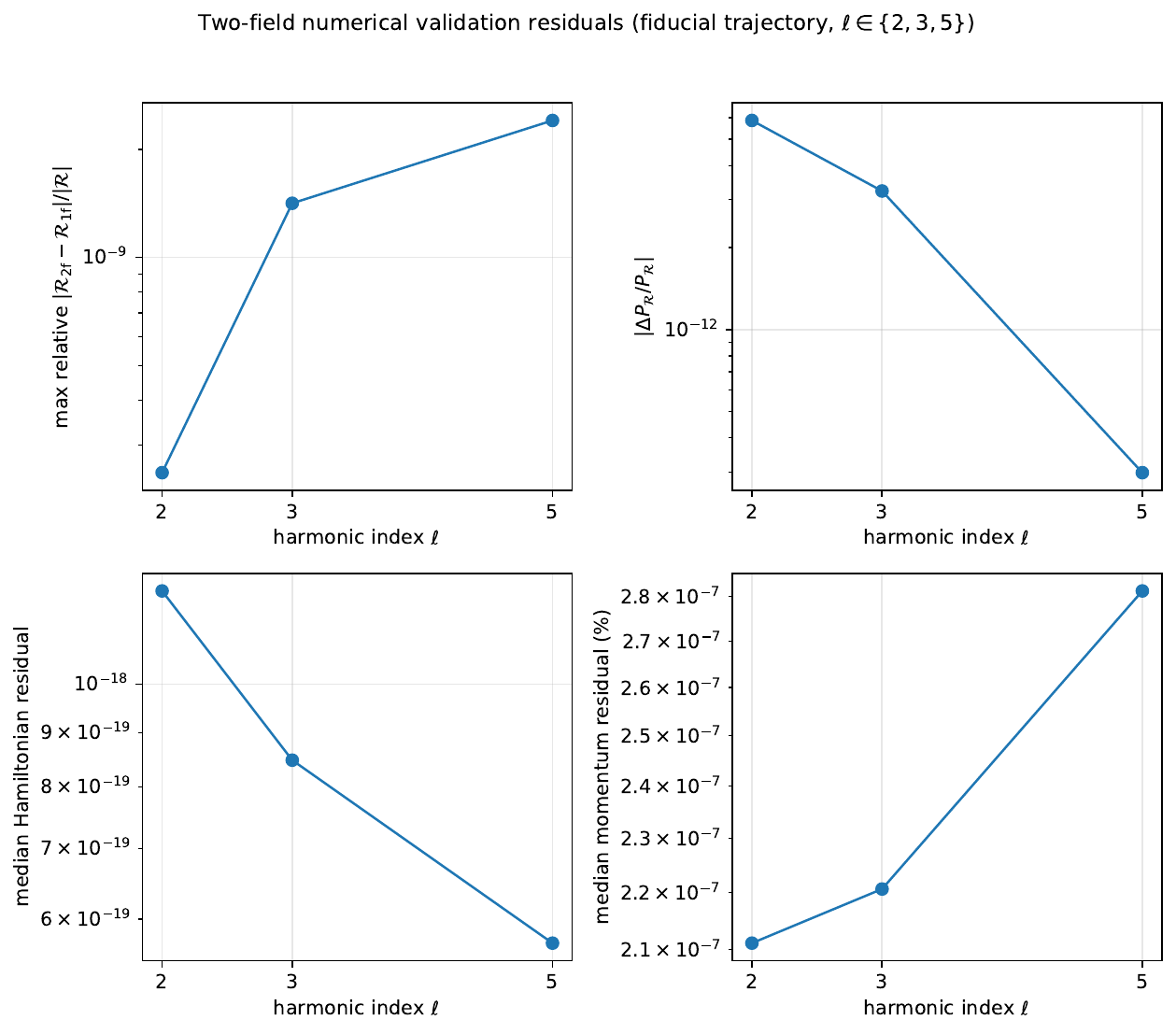}%
    }{%
        \fbox{\parbox{0.78\textwidth}{\centering
            \textbf{Figure not available in this build.}}}%
    }
    \caption{\textbf{Two-field numerical validation diagnostics on the fiducial
        trajectory.} Per-mode residuals for the harmonics
        $\ell \in \{2, 3, 5\}$ from the bounce-crossing two-field
        integration, on log axes. \emph{Top-left:} maximum relative
        cross-construction curvature residual
        $|\mathcal{R}_{\rm 2f} - \mathcal{R}_{\rm 1f}|/|\mathcal{R}|$ between the two-field
        covariant Newtonian-gauge reconstruction $\mathcal{R}_{\rm 2f}$ and the
        single-field reconstruction $\mathcal{R}_{\rm 1f}$.
        \emph{Top-right:} fractional power-spectrum drift
        $|\Delta P_{\mathcal{R}} / P_{\mathcal{R}}|$ between the covariant and single-field constructions on the post-exit tail.
        \emph{Bottom-left:} median Hamiltonian-constraint residual.
        \emph{Bottom-right:} median momentum-constraint residual (per cent).
        The maximum $\mathcal{R}$ cross-construction residual is
        $\SfRMaxResidual$, the maximum power drift is
        $\SfPowerDriftMax$, the Hamiltonian medians span
        $\SfHamiltonianMedianMin$--$\SfHamiltonianMedianMax$, and the largest
        momentum median is $\SfMomentumMedianEllFivePct\%$; together these
        confirm on the tested harmonics that the covariant two-field solver
        reproduces the single-field result and satisfies the Einstein
        constraints on their respective evaluation windows: the two upper
        panels use the post-bounce super-Hubble window
        $k_\ell/(aH)<10^{-2}$, while the two constraint medians are taken
        outside the bounce neighbourhood $|t-t_{\rm bounce}| < 2/H_{\rm inf}$
        (\S\ref{subsec:constraint_diagnostics}); neither statistic samples
        the immediate bounce. Values are those of the fiducial integration
        reported in this section.}
    \label{fig:perturbation_diagnostics}
\end{figure}

We use signature $(-,+,+,+)$ and the Newtonian-gauge scalar line element
\begin{equation}
ds^2=-(1+2\Phi)\,dt^2+a^2(t)(1-2\Phi)\,\gamma_{ij}\,dx^i dx^j,
\label{eq:newtonian_metric_convention}
\end{equation}
where vanishing anisotropic stress identifies the two scalar metric potentials.  With this convention the standard comoving curvature perturbation is $\mathcal{R}=\Phi+H Q_\sigma/\dot\sigma$ (and $\mathcal{R}=\Phi+H\,\delta\phi/\dot\phi$ in the single-field limit); the uniform-density curvature perturbation $\zeta=-\Phi-H\,\delta\rho/\dot\rho$ is a distinct quantity and is not evolved here.

The familiar flat-FRW Mukhanov--Sasaki choice $u=z\mathcal R$ with $z=a\dot\sigma/H$ is not the exact finite-harmonic canonical normalisation in a closed universe. Appendix~\ref{app:closed_s3_canonical} derives the exact mode-dependent reduced action on the fiducial $\chi=\dot\chi=0$ trajectory. Its coefficients have a finite continuation through $H=0$ whenever $\dot\phi_b\neq0$, and the corresponding canonical normalisation is $z_\ell=a(\mathcal A_\ell\mathcal B_\ell)^{1/4}$. We evolve the regular Newtonian variables directly through the bounce and use the reduced action as an independent fiducial scalar solver and state-normalisation prescription. The gauge-invariant curvature perturbation $\mathcal R=\Phi+H\delta\phi/\dot\phi$ remains finite at $H=0$ on the verified background. The same Newtonian-gauge dynamics are evolved independently in coordinate and orthonormal field-space bases, and the result is also checked with the algebraic Gordon--Wands projection; the per-mode validation residuals are summarised in Fig.~\ref{fig:perturbation_diagnostics}.

\subsection{Covariant decomposition: adiabatic and entropy projections}
\label{subsec:covariant_decomp}

On the field-space manifold $(\phi, \chi)$ with metric $G_{ab} = \mathrm{diag}(1, g_{\chi\chi}(\phi))$, the background trajectory $(\phi(t), \chi(t))$ defines a one-parameter family of tangent and normal vectors. The trajectory-tangent (adiabatic) and trajectory-normal (entropy) unit vectors are
\begin{equation}
\hat\sigma^a = \frac{1}{\dot\sigma}\,(\dot\phi,\, \dot\chi),
\qquad
\hat s^a = \frac{1}{\dot\sigma}\,(-\sqrt{g_{\chi\chi}}\,\dot\chi,\, \dot\phi/\sqrt{g_{\chi\chi}}),
\label{eq:tangent_normal}
\end{equation}
with $\dot\sigma^2 = G_{ab}\dot\phi^a \dot\phi^b = \dot\phi^2 + g_{\chi\chi}\,\dot\chi^2$. The field perturbation $\delta\phi^a$ projects onto these vectors as
\begin{equation}
Q_\sigma = G_{ab}\,\hat\sigma^a\,\delta\phi^b
        = \frac{\dot\phi\,\delta\phi + g_{\chi\chi}\,\dot\chi\,\delta\chi}{\dot\sigma},
\qquad
Q_s = G_{ab}\,\hat s^a\,\delta\phi^b
    = \frac{\sqrt{g_{\chi\chi}}\,(\dot\phi\,\delta\chi - \dot\chi\,\delta\phi)}{\dot\sigma},
\label{eq:Q_sigma_Q_s}
\end{equation}
following the Gordon--Wands convention \cite{Gordon2000,Wands2002}. In the orthonormal adiabatic--entropy basis and in the zero-turn limit, the kinetic and gradient part of the quadratic field action takes the schematic diagonal form (with mass terms and metric-constraint contributions suppressed)
\begin{equation}
\frac{\mathcal{L}_{\rm kin+grad}^{(2)}}{a^3}
= \tfrac{1}{2}\,(\dot Q_\sigma)^2 + \tfrac{1}{2}\,(\dot Q_s)^2 - \tfrac{1}{2}\,(\partial_i Q_\sigma)^2/a^2 - \tfrac{1}{2}\,(\partial_i Q_s)^2/a^2,
\end{equation}
where the vanishing-turn condition is $\omega = -V_s/\dot\sigma=0$ and $V_s = \hat s^a \nabla_a V$.  Away from that limit, rotation of the basis produces the familiar turn-rate mixing, so the displayed equation is not the complete quadratic action for the displaced bouncing trajectories.

The comoving curvature perturbation $\mathcal{R} = \Phi + H\,Q_\sigma/\dot\sigma$ inherits the kinetic mixing of the two projections through the trajectory turn rate $\omega$. On the fiducial no-turn trajectory $\chi=\dot\chi=0$ at background level and $\omega \equiv 0$ identically, so $Q_\sigma$ and $Q_s$ decouple and the standard single-field expression $\mathcal{R} = \Phi + H\,\delta\phi/\dot\phi$ holds. The displaced mildly and moderately turning trajectories used for the cross-basis consistency comparison do not satisfy this no-turn reduction.

On the fiducial $\chi=\dot\chi=0$ trajectory on which the entropy power is evaluated, the metric-source terms proportional to $\dot\chi$ and $V_\chi$ vanish. The $\ell$-harmonic coordinate perturbation $\delta\chi$---\emph{not} the canonical projection $Q_s$ of \S\ref{subsec:covariant_decomp}---then satisfies the homogeneous field-space-covariant equation
\begin{equation}
\ddot{\delta\chi} + \biggl(3 H + \frac{\dot g_{\chi\chi}}{g_{\chi\chi}}\biggr)\,\dot{\delta\chi}
+ \biggl(\frac{\ell(\ell+2)}{a^2} + \frac{V_{\chi\chi}}{g_{\chi\chi}}\biggr)\,\delta\chi = 0,
\label{eq:entropy_eom}
\end{equation}
with $V_{\chi\chi} = m_\chi^2$ for the explicit mass term in the potential. The structural feature emphasised here is the \emph{raised-index} entropy mass $V_{\chi\chi}/g_{\chi\chi} = m_\chi^2/g_{\chi\chi}$, derived directly from the field-space-covariant Klein--Gordon operator on the curved $G_{ab}$ manifold. For $\phi \gg M_{\rm Pl}/\alpha$ (the saturated sigmoid regime that holds throughout the integration window on the fiducial trajectory), $g_{\chi\chi} \to 1$ and the raised-index correction reduces to the flat-index $m_\chi^2$ that would be written in an uncorrected Newtonian-gauge formulation; for smaller $\alpha$ or smaller $\phi$, the $1/g_{\chi\chi}$ enhancement becomes numerically significant ($\sim 35\%$ at $g_{\chi\chi} = 0.74$).

The equation of motion Eq.~(\ref{eq:entropy_eom}) and the entropy Wronskian $W^{(s)}$ below are written for the coordinate entropy-field perturbation $\delta\chi$, with the field-space metric carried explicitly---in the friction term $\dot g_{\chi\chi}/g_{\chi\chi}$ above and in the Jacobian factor $g_{\chi\chi}$ of $W^{(s)}$ below. The canonically-normalised entropy projection of \S\ref{subsec:covariant_decomp}, $Q_s = \sqrt{g_{\chi\chi}}\,\delta\chi$ on the $\dot\chi = 0$ trajectory, is the variable that enters the power ratio Eq.~(\ref{eq:entropy_power_ratio}); it differs from $\delta\chi$ only by the Jacobian $\sqrt{g_{\chi\chi}}$, which equals unity to $|1 - g_{\chi\chi}| \leq \gDeviationMax$ throughout the saturated integration window, so the two are numerically interchangeable on the fiducial trajectory.

The harmonic index $\ell$ on the closed-FRW $K=+1$ background labels the $S^3$ scalar harmonic eigenvalue $\lambda_\ell = \ell(\ell+2)$ in place of the flat-FRW $k^2$, with $\ell \geq 2$ for non-trivial scalar harmonics on $S^3$. We retain $k_{\rm phys}(\ell, a) = \sqrt{\ell(\ell+2)}/a$ as the physical wavenumber entering the finite-time initialisation criteria below.

\subsection{Closed-FRW \texorpdfstring{$K=+1$}{K=+1} tensor solver}
\label{subsec:tensor_solver}

The gravitational-wave sector on closed-FRW $K=+1$ admits a complete $S^3$ harmonic decomposition into spin-2 tensor modes with transverse-traceless harmonic eigenvalue $\lambda_\ell^{\rm T} = \ell(\ell+2) - 2$ for $\ell \geq 2$ \cite{LindblomTaylorZhang2017,BongaGuptYokomizo2017Tensor}. The full mode-function equation for each tensor polarisation on the curved background combines the harmonic eigenvalue with the spatial-curvature term $+2 K$ from the closed-FRW Ricci tensor; in the $K = +1$ geometry these recombine to
\begin{equation}
\ddot h_k + 3 H\,\dot h_k + \frac{\lambda_\ell^{\rm T} + 2 K}{a^2}\,h_k
= \ddot h_k + 3 H\,\dot h_k + \frac{\ell(\ell+2)}{a^2}\,h_k = 0,
\label{eq:tensor_eom}
\end{equation}
i.e.\ the operative coefficient in the mode equation on closed-FRW $K = +1$ is $\ell(\ell+2)/a^2$, not $\lambda_\ell^{\rm T}/a^2$. We integrate Eq.~(\ref{eq:tensor_eom}) from the canonical Wronskian-normalised finite-time leading-adiabatic (``BD'') state. We retain both the harmonic eigenvalue $\lambda_\ell^{\rm T}$ (for the algebraic identity test) and the recombined coefficient $\ell(\ell+2)$ (used in the EOM). For a unit-normalised TT harmonic coefficient the action per polarisation has kinetic coefficient $(M_{\rm Pl}^2/8)a^3\dot h_{\rm pol}^2$, hence $|h_{\rm pol}^{\rm BD}|^2=2/(M_{\rm Pl}^2a^3k_{\rm phys})$. We evolve the two-polarisation power envelope, $|h_k|^2=2|h_{\rm pol}|^2$, so in Planck units its initialisation is
\begin{equation}
h_k^{\rm BD}(t) = \frac{2}{\sqrt{a(t)^3\,k_{\rm phys}(\ell, a(t))}}\,e^{-i\int^t k_{\rm phys}(t')\,dt'},
\qquad
\dot h_k^{\rm BD}(t) = -(H+i k_{\rm phys})\,h_k^{\rm BD}(t),
\label{eq:tensor_BD}
\end{equation}
which gives the standard flat-FRW slow-roll tensor power $P_h^{\rm flat\text{-}MS} = (2/\pi^2) (H/M_{\rm Pl})^2$ on the deep-inflation plateau \cite{MukhanovFeldmanBrandenberger1992}. Writing $n=\ell+1$, the standard closed-universe definition is $P_T(n)=n(n^2-3)\sum_s|h_n^{(s)}|^2/(2\pi^2)$ \cite{AkamaKobayashi2019}. The stored envelope obeys $|h_\ell|^2=\sum_s|h_\ell^{(s)}|^2$, so the production power is
\begin{equation}
P_h(\ell)=\frac{(\ell+1)\bigl[(\ell+1)^2-3\bigr]}{2\pi^2}|h_\ell|^2.
\label{eq:tensor_power_closed}
\end{equation}
The separate density-of-states construction obtained from the fixed-order per-polarisation degeneracy $(\ell-1)(\ell+3)$ \cite{LindblomTaylorZhang2017} and $d\ell/d\ln k=\ell(\ell+2)/(\ell+1)$ would instead multiply $|h_\ell|^2/(2\pi^2)$ by $[\ell(\ell+2)/(\ell+1)](\ell-1)(\ell+3)$. It is not the standard $P_T(n)$ and is not used for $P_h$, $r$, or the flat-FRW comparison; the DOS-to-standard ratio is $[(n^2-1)(n^2-4)]/[n^2(n^2-3)]$, equal to $0.999462$ at $\ell=60$ but appreciably different at the lowest harmonics. The finite-time state is imposed pre-bounce at the earliest cosmic time at which the deep sub-Hubble criterion $k_{\rm phys}/|H| \geq 1.5 \times 10^3$ is satisfied, evolved through $H = 0$, and integrated to the end of the post-bounce slow-roll window. The frozen power is evaluated at the root $\ln[a(t)/a(t_{\rm exit})]=4$, not with a constant-$H$ time-offset approximation; a background without that post-exit interval fails rather than falling back to its endpoint. Numerically Eq.~(\ref{eq:tensor_eom}) is stiff in deep sub-Hubble due to fast mode-function oscillations; the integration uses Radau (A--L stable, implicit, primary) with relative tolerance $10^{-10}$, atol $10^{-14}$, falling back to LSODA if Radau fails. The envelope Wronskian $W = a^3 (h \dot h^* - h^* \dot h)$ is checked against its conserved initial value at a mid-contraction reference point on the pre-bounce branch (where the oscillatory mode functions condition it well); the residual $|W - W_{\rm init}|/|W_{\rm init}|$ remains below $10^{-8}$ on the computed grid (the leading-adiabatic initialisation Eq.~(\ref{eq:tensor_BD}) gives $W_{\rm init} = 8 i$ for the two-polarisation envelope, twice the per-polarisation canonical value, and the conserved-quantity check verifies that this rescaled initial value is preserved by the integrator). The harmonic measure, the finite-time state, the post-exit evaluation epoch, the root finding, Wronskian normalisation, and the flat-FRW limit are each checked independently against their analytic forms.

Throughout the numerical perturbation analysis we adopt $\ell = 60$ as the \emph{pivot-proxy} harmonic, and we are explicit about what it is and is not. The reference scales $a_{\rm iso}\equiv\sqrt{3M_{\rm Pl}^2/V_0}$ and $H_{\rm inf}\equiv\sqrt{V_0/(3M_{\rm Pl}^2)}$ obey $a_{\rm iso}H_{\rm inf}=1$ exactly; the measured physical $a_{\rm min}$ differs slightly because the bounce samples $V(\phi_{\rm bounce})<V_0$ and nonzero kinetic energy. With this reference normalisation, the post-bounce comoving horizon obeys $a H \simeq e^{N}$ once the brief curvature transient has decayed ($H^2 = H_{\rm inf}^2 (1 - e^{-2N})$ on the plateau, so the relation is per-cent accurate by $N \approx 2$; at the bounce point itself $H = 0$), and a harmonic $\ell$ exits at $N_{\rm exit} \simeq \ln\sqrt{\ell(\ell+2)}$ e-folds after the bounce---$N_{\rm exit} \approx 4.1$ for $\ell = 60$, and between $\approx 1$ and $\approx 7$ for the whole sampled window $\ell \in [2, 1000]$. The mode that becomes the conventional CMB pivot $k_{\rm pivot} = 0.05\,\mathrm{Mpc}^{-1}$ instead exits $N_{\rm total}^{\rm ideal} - 60 \approx 2570$ e-folds after the bounce, at harmonic $\ell \sim 10^{1116}$ (the scale hierarchy of \S\ref{sec:observables}), far beyond direct numerical reach. Between the two exit epochs $H^2$ (and hence the leading plateau tensor amplitude) drifts by approximately $2.4\%$. This plateau-to-pivot drift is a separate extrapolation systematic and is larger than the \PhSameExitAgreementPct\% same-exit amplitude check below; the latter validates the closed solver at $\ell=60$ and does not compute the unreachable pivot mode. At the pivot-proxy harmonic the tensor power spectrum amplitude is
\begin{equation}
P_h(\ell = 60) = \PhSixtyClosedFRW
\label{eq:Ph_60}
\end{equation}
on the fiducial trajectory ($\phi_0 = 10\,M_{\rm Pl}$, $\alpha = 1$). As an amplitude-only validation, this agrees to $\PhSameExitAgreementPct\%$ with the flat-FRW slow-roll reference $P_h^{\rm flat}=2H^2(t_{\rm exit})/(\pi^2M_{\rm Pl}^2)=\PhSixtyFlatFRWReference$ \cite{MukhanovFeldmanBrandenberger1992}, evaluated at the numerical mode's own horizon-exit time. No scalar amplitude and no different-epoch $N=60$ quantity enters this comparison.

For this small residual the usual four-significant-figure powers round to the same number. Retaining additional digits gives
\begin{equation}
\begin{aligned}
H(t_{\rm exit})/M_{\rm Pl}&=\HSixtyTensorExit,\\
P_h(60)&=\PhSixtyClosedFRWDetailed,\\
P_h^{\rm flat}(t_{\rm exit})&=\PhSixtyFlatFRWDetailed.
\end{aligned}
\end{equation}
The quoted percentage is $100|P_h-P_h^{\rm flat}|/P_h^{\rm flat}$, evaluated before rounding.

The like-for-like solver ratio at this early-plateau harmonic is
\begin{equation}
r_{\ell=60} \equiv \frac{P_h(\ell=60)}{P_{\mathcal{R}}(\ell=60)}
= \frac{\PhSixtyClosedFRW}{\PRSixtyClosedFRW}
= \rValClosedFRW
\label{eq:r_ell_sixty}
\end{equation}
Here $P_{\mathcal{R}}(\ell=60)=\PRSixtyClosedFRW$ is obtained from the exact reduced action of Appendix~\ref{app:closed_s3_canonical} and evaluated, like the tensor power, exactly four e-folds after the common geometric-exit time. The reporting state and the second propagated canonical state give
\begin{equation}
P_{\mathcal R}(60)\in
[\PRSixtyClosedFRWStateMin,\PRSixtyClosedFRWStateMax],
\qquad
r_{60}\in[\rValClosedFRWStateMin,\rValClosedFRWStateMax].
\label{eq:r_ell_sixty_state_range}
\end{equation}
Thus the quoted $\rValClosedFRW$ is a specified finite-time-state convention rather than a state-independent number; the scalar-power span across the two prescriptions is $\ExactScalarStateSpanEllSixtyPct\%$. The result has the same order as the same-epoch slow-roll scale $16\epsilon_V(\text{early plateau})\approx1.7\times10^{-6}$. It must not be compared to $12/N^2=\rLythStewartReference$ at $N=60$: those refer to different exit epochs. The observable-pivot prediction remains $16\epsilon_V(\phi_{60}^{\rm num})\approx3.08\times10^{-3}$.

\subsection{Cross-basis consistency comparison of \texorpdfstring{$\mathcal{R}$}{R}}
\label{subsec:triangulation}

\begin{figure}[htbp]
    \centering
    \IfFileExists{figures/fig_triangulation_residuals.pdf}{%
        \includegraphics[width=0.82\textwidth]{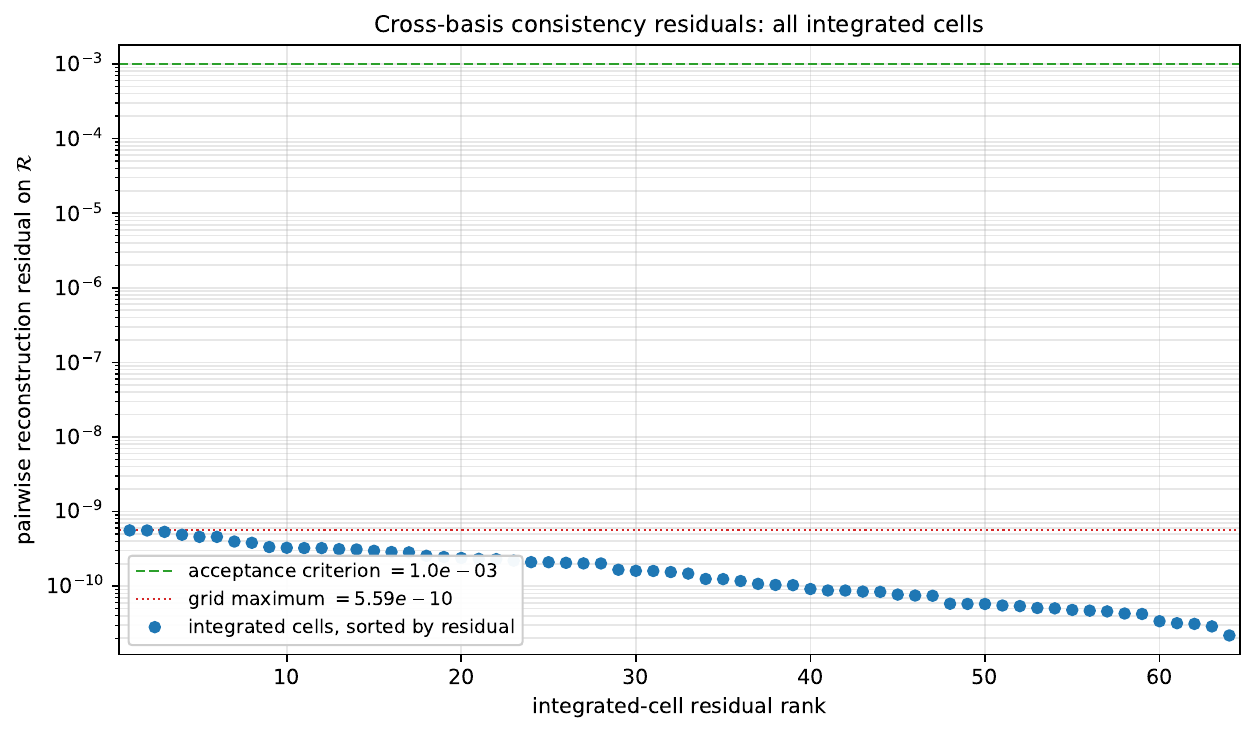}%
    }{%
        \fbox{\parbox{0.78\textwidth}{\centering
            \textbf{Figure not available in this build.}}}%
    }
    \caption{\textbf{Cross-basis consistency residuals.}
        Pairwise reconstruction residuals on $\mathcal{R}$ for all
        $\TwoTwoIntegratedCells$ integrated cells, sorted from largest to
        smallest (blue markers). The grid maximum $\TwoTwoMaxResidual$ is
        shown by the red dotted line; every integrated cell lies below the
        acceptance criterion $\Delta<\TwoTwoOriginalThreshold$ (green
        dashed). The $\TwoTwoControlCells$ contracting-control cells are a
        non-integrated finite-range background control and are not plotted.}
    \label{fig:triangulation_residuals}
\end{figure}

We compare the comoving curvature perturbation $\mathcal{R}$ across three reconstructions---two independently evolved field-basis systems in the same Newtonian spacetime gauge and one algebraic projection check. The complete systems supplied to the integrators are given in Appendix~\ref{app:integrated_scalar_systems}. Figure~\ref{fig:triangulation_residuals} displays the per-cell residual maxima.

\begin{itemize}
\item \textbf{Coordinate-field-basis construction.} Integrate the full two-field Newtonian-gauge system $(\delta\phi_k, \delta\chi_k, \Phi_k)$ using the perturbed Klein--Gordon equations for $\delta\phi_k$ and $\delta\chi_k$ on the field-space manifold and the perturbed spatial-trace Einstein equation
$\ddot\Phi + 4 H \dot\Phi + (\dot H + 3 H^2)\,\Phi = \bigl[\dot\phi\,\delta\dot\phi + g_{\chi\chi}\,\dot\chi\,\delta\dot\chi + \tfrac12 g_{\chi\chi,\phi}\dot\chi^2\delta\phi - V_{,\phi}\,\delta\phi - V_{,\chi}\,\delta\chi\bigr]/(2 M_{\rm Pl}^2)$
as the evolution equation for $\Phi_k$. The metric-variation term follows directly from $\delta[\tfrac12 g_{\chi\chi}(\phi)\dot\chi^2]$ and is retained in both the pressure source and the Hamiltonian energy-density constraint. This momentum-constraint-preserving form follows by differentiating the $(0i)$ momentum constraint: the closed-FRW spatial-trace Einstein equation carries an explicit $-K/a^2$ term, $\ddot\Phi + 4H\dot\Phi + (2\dot H + 3H^2 - K/a^2)\,\Phi = \delta p/(2M_{\rm Pl}^2)$, and folding the $-\dot\sigma^2\Phi$ lapse of the full pressure perturbation $\delta p$ through the Raychaudhuri identity $\dot\sigma^2/(2M_{\rm Pl}^2) = K/a^2 - \dot H$ collapses the coefficient to exactly $\dot H + 3H^2$ with the complete lapse-free field source (the standard longitudinal-gauge Bardeen result); we additionally verify it against the Einstein-constraint drift diagnostic of \S\ref{subsec:constraint_diagnostics}. We then reconstruct $\mathcal{R}^{\rm Newt} = \Phi_k + H\,Q_{\sigma,k}/\dot\sigma$ post-bounce.
\item \textbf{Orthonormal-field-basis evolved-$\Phi$ construction.} Use the comoving curvature variable
$\mathcal{R}_k = \Phi_k + H\,Q_{\sigma,k}/\dot\sigma$, evolving the orthonormal-basis system $(Q_{T}, Q_{N}, \Phi)_k$ from a finite-$k$ comoving initial condition: the algebraic Kodama--Sasaki $\Phi$ inversion, with coefficient $(\lambda_\ell - 3K)/a^2 - \dot\sigma^2/(2 M_{\rm Pl}^2)$, is applied at the deep-sub-Hubble initial time only (where it is well conditioned), and $\Phi_k$ is evolved thereafter via the same spatial-trace equation. The construction is regular away from $H = 0$; the algebraic inversion is therefore used only at the pre-bounce initial time, the system is then integrated through the bounce, and the cross-representation residual is evaluated post-bounce.
\item \textbf{Gordon--Wands projection.} Recompute $\mathcal{R}^{\rm GW} = \Phi + (H/\dot\sigma)\,Q_T$ by projecting the covariant-solver field perturbations onto the orthonormal adiabatic direction (the Gordon--Wands decomposition \cite{Gordon2000}); since $Q_T \equiv Q_\sigma$, this expression is algebraically equivalent to the Newtonian-gauge reconstruction, so the Newtonian--Gordon-Wands residual tests the \emph{implementation} of the orthonormal-basis projection algebra (basis construction, field-space Jacobians) rather than independent dynamics; it performs no new integration.
\end{itemize}

The coordinate and orthonormal systems are integrated separately in different dynamical field bases ($(\delta\phi, \delta\chi, \Phi)$ versus $(Q_T, Q_N, \Phi)$), and the Gordon--Wands leg is an algebraic projection check on the coordinate-basis solution. Both evolved systems use the same Newtonian spacetime gauge, background trajectory, and $\Phi$ evolution equation. The comparison therefore tests field-basis transformations, connection and turn-rate terms, and projection algebra on through-bounce integrations; it is not a comparison between spacetime gauges and not three disjoint integrations. The cross-basis consistency criterion is the pairwise residual
\begin{equation}
\frac{|\mathcal{R}^{(A)} - \mathcal{R}^{(B)}|}{\max\!\bigl(|\mathcal{R}^{(A)}|,\,|\mathcal{R}^{(B)}|,\,\mathcal{R}_{\rm floor}\bigr)} < \TwoTwoOriginalThreshold
\qquad
\forall\, (A,B) \in \{({\rm Newt}, {\rm com}), ({\rm Newt}, {\rm GW})\},
\label{eq:triangulation_gate}
\end{equation}
on every mode of the computed grid, with the denominator floor-regularised by $\mathcal{R}_{\rm floor} = \sqrt{V_0}/(2\pi M_{\rm Pl})$ and evaluated symmetrically. The $({\rm com}, {\rm GW})$ pair is omitted from the decisive comparison: since $Q_T \equiv Q_\sigma$ the Gordon--Wands leg is algebraically equivalent to the Newtonian reconstruction, so its residual against the comoving-basis construction is bounded through the triangle inequality by the two Newtonian-anchored pairs; it is retained only as a diagnostic. The acceptance criterion is $\Delta<\TwoTwoOriginalThreshold$. The grid consists of three IC trajectories (mildly turning, moderately turning, and a contracting control), two $\alpha$ values ($\alpha \in \{1, 3\}$), and the sixteen $\ell$ modes
\begin{equation}
\ell \in \{2,\,3,\,5,\,7,\,10,\,15,\,20,\,30,\,50,\,75,\,100,\,150,\,200,\,300,\,500,\,1000\},
\end{equation}
giving $\TwoTwoTotalCells$ nominal cells. The contracting-control trajectory ($\chi_0=1.5M_{\rm Pl}$, $\dot\chi_0=2\times10^{-6}$) is the finite-range contracting background control of \S\ref{sec:numerics}, which contracts without a bounce over the declared range; it is not perturbatively evolved. The cross-basis comparison is therefore integrated on the $\TwoTwoIntegratedCells$ cells of the two bouncing families (mildly and moderately turning) $\times$ two $\alpha$ $\times$ sixteen $\ell$, while the remaining $\TwoTwoControlCells$ nominal cells record the non-integrated contracting control. All $\TwoTwoStrictCells$ integrated cells satisfy it, with maximum residual $\TwoTwoMaxResidual$. This numerical agreement tests the implemented field-basis transformations and projection algebra, while a complete reduced coupled adiabatic--entropy action on generic turning backgrounds remains future work in \S\ref{sec:conclusions}; the exact no-turn reduction is given in Appendix~\ref{app:closed_s3_canonical}.

\subsection{Constraint diagnostics}
\label{subsec:constraint_diagnostics}

We assess the two Einstein constraints through three numerical-validation diagnostics. Items (a) and (b) report a-posteriori residuals on the fiducial Newtonian-gauge solution at $\ell\in\{2,3,5\}$; item (c) evaluates both evolved field-basis systems on the sixteen-harmonic turning grid and tests tolerance sensitivity around the bounce. All quoted statistics are computed in this work on the respective domains specified below.

\textit{(a) Momentum constraint.} The momentum constraint $\dot\Phi + H\Phi = \dot\phi\,\delta\phi/(2 M_{\rm Pl}^2) + g_{\chi\chi}\,\dot\chi\,\delta\chi/(2 M_{\rm Pl}^2)$ is \emph{not} used in the evolution ($\Phi$ is evolved via the spatial-trace equation, \S\ref{subsec:triangulation}); it is checked \emph{a posteriori}: the constraint's right-hand side is compared against the \emph{integrated} $\dot\Phi$ (the state variable evolved by the spatial-trace equation)---an ODE-consistency check, since the momentum constraint is not imposed during the evolution. The medians are taken over every output sample outside a symmetric window about the bounce, that is at $|t - t_{\rm bounce}| \geq 2/H_{\rm inf}$ (the plateau time for two e-folds); no sub- or super-Hubble selection is applied to this statistic. Because these harmonics are initialised only $\approx 0.05$--$0.12/H_{\rm inf}$ before the bounce, that symmetric cut removes their whole integrated contracting branch, so the medians are in effect expanding-branch statistics. On the harmonics $\ell \in \{2, 3, 5\}$ the relative residual, normalised by the constraint terms $\max(|H\Phi|,\,|\dot\phi\,\delta\phi + g_{\chi\chi}\dot\chi\,\delta\chi|/2M_{\rm Pl}^2)$, has median values $\SfMomentumMedianEllTwoPct\%$, $\SfMomentumMedianEllThreePct\%$, and $\SfMomentumMedianEllFivePct\%$, respectively (required to remain below $\SfMomentumGatePct\%$). The associated cross-solver power comparison---the covariant and single-field constructions evaluated on the last tenth of the post-bounce super-Hubble window ($t > t_{\rm bounce} + 2/H_{\rm inf}$ and $k_\ell/(aH) < 10^{-2}$)---has maximum $|\Delta P_{\mathcal{R}}/P_{\mathcal{R}}|=\SfPowerDriftMax$ on these modes. It is a solver-agreement statistic at fixed integrator settings rather than a grid- or tolerance-refinement study; the tolerance sensitivity of the through-bounce curvature solutions is tested separately by the immediate-bounce refinement in item (c) below, not for this power-agreement statistic.

\textit{(b) Hamiltonian constraint (independent check).} The perturbed $(0,0)$ Einstein equation $3 H (\dot\Phi + H\Phi) + (k^2 - 3K)\,\Phi/a^2 = -\delta\rho/(2 M_{\rm Pl}^2)$ (with the closed-FRW curvature shift $k^2 \to k^2 - 3K$, $K = +1$, in the Kodama--Sasaki form \cite{KodamaSasaki1984}; $k^2 = \lambda_\ell = \ell(\ell+2)$ on $S^3$) is likewise \emph{not} used in evolution and provides an algebraic, finite-difference-free consistency check. The Hamiltonian-constraint residual $\mathcal{H}_k = |\mathrm{LHS} - \mathrm{RHS}|/S$, normalised by the background curvature/energy scale $S = \max(\rho/3M_{\rm Pl}^2,\,1/a^2,\,H_{\rm inf}^2)$ (a finite-difference-free measure shared with the constraint-drift criterion), on the same harmonics has median values $\SfHamiltonianMedianMin$--$\SfHamiltonianMedianMax$---many orders of magnitude smaller, relative to this background scale, than the momentum-constraint relative residual above. The separately reported term-normalised diagnostic divides by $\max(|\mathrm{LHS}|,|\mathrm{RHS}|)$, which carries the small mode amplitude; its largest per-mode median is $\SfHamiltonianTermMedianMaxPct\%$. Together these residuals support numerical consistency with the Einstein constraints on the sampled window---which, as stated above, excludes the immediate bounce neighbourhood---and with the constraint-initialised background data. Behaviour across $H=0$ itself is not established by these medians, although the modes they diagnose are integrated continuously through the bounce from the declared pre-bounce finite-time test state of Appendix~\ref{app:integrated_scalar_systems}; the cross-basis agreement of such through-bounce integrations is quantified separately by the triangulation of \S\ref{subsec:triangulation}, whose residuals are likewise evaluated post-bounce.

\textit{(c) Immediate-bounce constraints and tolerance refinement.} To test the region omitted by the medians above, we evaluate the same Hamiltonian residual and the raw momentum ratio
\begin{equation}
\mathcal{C}_M=\frac{|\dot\Phi+H\Phi-B|}{\max(|H\Phi|,|B|,10^{-30})},
\qquad
B=\frac{\dot\phi\,\delta\phi+g_{\chi\chi}\dot\chi\,\delta\chi}{2M_{\rm Pl}^2},
\label{eq:bounce_momentum_residual}
\end{equation}
directly on both independently evolved systems.  The grid contains the two turning initial-condition families, $\alpha\in\{1,3\}$, all sixteen harmonics of \S\ref{subsec:triangulation}, and both evolved legs: $\BounceWindowLegCells$ leg--mode combinations, corresponding to $\BounceWindowBackgroundCells$ distinct background--mode combinations.  Every leg is evaluated from its dense solution on $\BounceWindowSamples$ equally spaced cosmic-time samples in the common fully covered interval
\begin{equation}
H_{\rm inf}|t-t_{b}|\leq \BounceWindowHalfWidth .
\label{eq:bounce_constraint_window}
\end{equation}
The global maxima are $\max\mathcal{H}_k=\BounceWindowMaxCH$ and $100\max\mathcal{C}_M=\BounceWindowMaxCMPct\%$, below the respective strict criteria $10^{-3}$ and $1.5\%$. For transparency we also evaluate each constraint against the magnitudes of its own two sides,
\begin{equation}
\mathcal{C}^{\rm term}_X
=\frac{|L_X-R_X|}{\max(|L_X|,|R_X|)},
\qquad X\in\{H,M\},
\label{eq:bounce_term_residual}
\end{equation}
with $0/0\equiv0$ and a non-zero numerator over a zero denominator represented by $+\infty$. These term-normalised quantities are diagnostics rather than acceptance criteria. Their finite-grid global maxima are $\max\mathcal{C}^{\rm term}_H=\BounceWindowMaxCHTerm$ and $100\max\mathcal{C}^{\rm term}_M=\BounceWindowMaxCMTermPct\%$. The immediate-bounce summaries, including the independent-method comparison below, are collected in Table~\ref{tab:bounce_neighbourhood_diagnostics}.

Repeating each integration with $(\mathrm{rtol},\mathrm{atol})$ tightened from $(10^{-12},10^{-20})$ to $(10^{-13},10^{-22})$, while holding the background, endpoints, initial state, method, and evaluation grid fixed, gives
\begin{equation}
\max\frac{|\mathcal{R}_{\rm ref}-\mathcal{R}_{\rm base}|}
{\max(|\mathcal{R}_{\rm ref}|,|\mathcal{R}_{\rm base}|,\mathcal{R}_{\rm floor})}
=\BounceWindowMaxRefinement<10^{-6}.
\label{eq:bounce_refinement_residual}
\end{equation}
As an independent local integration-algorithm check, the full-history DOP853 solution is evaluated at the left boundary of the compact, fully covered core $H_{\rm inf}|t-t_b|\leq\BounceWindowMethodHalfWidth$. This actual half-width is the smaller of the declared cap $0.1$ and the common immediate-bounce half-width $\BounceWindowHalfWidth$. The identical 12-component state then seeds an implicit Radau IIA solve of the same twelve-dimensional perturbation system across $H=0$. The background, equations, local endpoints, baseline tolerances, and 2001-point dense-evaluation grid are held fixed. On the preselected $\BounceWindowMethodCells$-cell subset, the symmetric DOP853--Radau curvature residual has global maximum $\BounceWindowMaxMethod<10^{-6}$; the local Radau solutions independently give $\max\mathcal H_k=\BounceWindowRadauMaxCH$ and $100\max\mathcal C_M=\BounceWindowRadauMaxCMPct\%$, below the same $10^{-3}$ and $1.5\%$ bounds. This comparison tests method sensitivity in the immediate neighbourhood of the Hubble zero; it does not claim method agreement on the wider common window or a Radau reproduction of the oscillatory pre-bounce history, refine the background interpolation, or derive an observed convergence order.

\begin{table}[htbp]
\centering
\caption{\textbf{Immediate-bounce numerical diagnostics.} The first five rows are global maxima over both evolved field-basis systems on all $\BounceWindowLegCells$ leg--mode combinations. The independent-method rows use the preselected $\BounceWindowMethodCells$-cell subset spanning both turning families, both $\alpha$ values, both evolved legs, and $\ell\in\{2,50,1000\}$. The term-normalised rows are reported as diagnostics; all other rows have pre-committed acceptance thresholds.}
\label{tab:bounce_neighbourhood_diagnostics}
\begin{tabular}{@{}ll@{}}
\toprule
Quantity & Global maximum \\
\midrule
Background-scale Hamiltonian residual $\mathcal{H}_k$ & $\BounceWindowMaxCH$ \\
Floor-regularised momentum residual $100\mathcal{C}_M$ & $\BounceWindowMaxCMPct\%$ \\
Term-normalised Hamiltonian residual $\mathcal{C}^{\rm term}_H$ & $\BounceWindowMaxCHTerm$ \\
Term-normalised momentum residual $100\mathcal{C}^{\rm term}_M$ & $\BounceWindowMaxCMTermPct\%$ \\
Tolerance-refinement ratio & $\BounceWindowMaxRefinement$ \\
Local DOP853--Radau $\mathcal R$ ratio & $\BounceWindowMaxMethod$ \\
Local Radau background-scale Hamiltonian residual $\mathcal H_k$ & $\BounceWindowRadauMaxCH$ \\
Local Radau floor-regularised momentum residual $100\mathcal C_M$ & $\BounceWindowRadauMaxCMPct\%$ \\
\bottomrule
\end{tabular}
\end{table}

Background conditioning quantities are also recorded but do not enter the acceptance decision. This finite-grid result establishes direct numerical consistency across $H=0$ on the stated turning trajectories; it is not a proof for untested initial data or a derivation of the reduced coupled adiabatic--entropy action on generic turning backgrounds.

\subsection{Direct entropy-mode power ratio}
\label{subsec:entropy_power_ratio}

The $\ell$-harmonic entropy-field perturbation $\delta\chi$ satisfies the equation of motion Eq.~(\ref{eq:entropy_eom}) with the raised-index entropy mass $m_\chi^2/g_{\chi\chi}$; its canonical field-space projection $Q_s = \sqrt{g_{\chi\chi}}\,\delta\chi$ (\S\ref{subsec:covariant_decomp}) is what enters the power ratio Eq.~(\ref{eq:entropy_power_ratio}) below. We integrate this equation directly from the canonical Wronskian-normalised finite-time leading-adiabatic (``BD'') state at deep sub-Hubble scales (the same $k_{\rm phys}/|H| \geq 1.5 \times 10^3$ criterion as the tensor solver of \S\ref{subsec:tensor_solver}). The canonical initial state, the horizon-exit extraction from the dense output, the normalisation, Wronskian conservation, and the power-ratio identity are each checked independently against their analytic forms. Transforming the canonical variable $q=a\sqrt{g_{\chi\chi}}\,\delta\chi$ back to the coordinate perturbation gives $\delta\chi_{\rm BD}=1/\sqrt{2g_{\chi\chi}a^3k_{\rm phys}}$ and $\dot{\delta\chi}_{\rm BD}=-(H+\dot g_{\chi\chi}/2g_{\chi\chi}+ik_{\rm phys})\delta\chi_{\rm BD}$ at zeroth adiabatic order. The Wronskian
\begin{equation}
W^{(s)} = a^3\,g_{\chi\chi}\,(\delta\chi\,\dot{\delta\chi}^* - \delta\chi^*\,\dot{\delta\chi})
\end{equation}
is conserved within the imposed numerical Wronskian tolerance, $|W^{(s)} - i|/|i|<10^{-4}$, on every successfully reported mode.

The per-mode entropy-to-curvature field-power ratio at horizon exit is defined as
\begin{equation}
\mathcal{I}_{S\mathcal{R}}(\ell) = \frac{P_{SS}(\ell)}{P_{\mathcal{R}}(\ell)}
            = \frac{|\mathcal{S}_{\ell}^{\rm exit}|^2}{|\mathcal{R}_{\ell}^{\rm exit}|^2},
\label{eq:entropy_power_ratio}
\end{equation}
The entropy ratio uses the scalar power at horizon exit, whereas Eq.~(\ref{eq:r_ell_sixty}) uses the scalar power four e-folds later. Numerically,
\begin{equation}
\frac{\PRExitEllSixty}{\PRSixtyClosedFRW}\simeq1.99.
\end{equation}
This is horizon-crossing-to-frozen evolution, not a normalisation mismatch; neither same-epoch ratio mixes the two evaluation times.
Here $\mathcal{S}=(H/\dot\sigma)Q_s$ and $\mathcal{R}=\Phi+(H/\dot\sigma)Q_\sigma$ are the entropy and curvature perturbations; the equality is not an identification of $\mathcal{R}$ with $(H/\dot\sigma)Q_\sigma$ in Newtonian gauge. The scalar and entropy modes are evaluated at the same numerically solved geometric-exit time, and each amplitude is converted with the same exact per-log-wavenumber closed-$S^3$ prefactor $\ell(\ell+1)(\ell+2)/(2\pi^2)$; this prefactor therefore cancels in the ratio. On the fiducial $\chi=\dot\chi=0$ trajectory, $Q_\sigma\to\delta\phi$ and $Q_s\to\sqrt{g_{\chi\chi}}\,\delta\chi$. At $\ell=60$ the reporting scalar state gives $P_{\mathcal{R}}(t^*)=\PRExitEllSixty$ and the fixed entropy prescription gives $P_{SS}(t^*)=\PSSExitEllSixty$, hence $\mathcal{I}_{S\mathcal{R}}=\ISRAtFiducialMchi$. Propagating the second declared scalar state changes the ratio to the interval $[\ISRAtFiducialMchiStateMin,\ISRAtFiducialMchiStateMax]$. This interval varies the scalar state only; it is not a complete joint finite-time-state comparison for the entropy and adiabatic sectors. No observational scalar amplitude is used as an input.

The notation is deliberately separated from the Gordon--Wands transfer matrix \cite{Wands2002}. In that formalism $T_{\mathcal{RS}}$ is the off-diagonal coefficient mapping an initial entropy perturbation into a later curvature perturbation. It is identically zero here because the no-turn background has $\omega=0$. By contrast, Eq.~(\ref{eq:entropy_power_ratio}) compares the powers of two independently quantised modes at the same time; for light fields their horizon-exit powers are expected to be comparable. The condition $T_{\mathcal{RS}}^{\rm sup}<10^{-2}$ therefore applies to the transfer coefficient, not to $\mathcal{I}_{S\mathcal{R}}$; the corresponding turn-sourced test is not exercised on a trajectory with $\omega\equiv0$.

On the fiducial no-turn trajectory ($\phi_0=10\,M_{\rm Pl}$, $\chi_0=\dot\chi_0=0$, $\alpha=1$, $m_\chi=\mchiFiducial\,M_{\rm Pl}$), the per-$\ell$ ratios at the seven sampled harmonics are reported in Table~\ref{tab:entropy_power_ratio_per_ell}.

\begin{table}[htbp]
\centering
\caption{\textbf{Per-mode entropy-to-curvature power ratio $\mathcal{I}_{S\mathcal{R}}(\ell)$ on the fiducial no-turn trajectory.}
Both powers are evaluated at the same geometric-exit epoch with the exact per-log-wavenumber closed-$S^3$ measure. The second numerical column uses the specified reporting scalar state; the third propagates the two declared scalar states while holding the entropy prescription fixed. The reporting-state sampled-grid maximum is $\ISRSupSampledGrid$, and the corresponding two-state range is $[\ISRSupSampledGridStateMin,\ISRSupSampledGridStateMax]$.}
\label{tab:entropy_power_ratio_per_ell}
\begin{tabular}{ccc}
\toprule
$\ell$ & reporting state & scalar-state range \\
\midrule
2     & \ISRAtEllTwo & $[\ISRAtEllTwoStateMin,\ISRAtEllTwoStateMax]$ \\
5     & \ISRAtEllFive & $[\ISRAtEllFiveStateMin,\ISRAtEllFiveStateMax]$ \\
60    & \ISRAtEllSixty & $[\ISRAtEllSixtyStateMin,\ISRAtEllSixtyStateMax]$ \\
100   & \ISRAtEllHundred & $[\ISRAtEllHundredStateMin,\ISRAtEllHundredStateMax]$ \\
200   & \ISRAtEllTwoHundred & $[\ISRAtEllTwoHundredStateMin,\ISRAtEllTwoHundredStateMax]$ \\
500   & \ISRAtEllFiveHundred & $[\ISRAtEllFiveHundredStateMin,\ISRAtEllFiveHundredStateMax]$ \\
1000  & \ISRAtEllThousand & $[\ISRAtEllThousandStateMin,\ISRAtEllThousandStateMax]$ \\
\bottomrule
\end{tabular}
\end{table}

The table is a direct solver-side comparison, not an externally normalised proxy. Its low-$\ell$ variation probes the closed-FRW curvature-scale regime, while the intermediate-to-high harmonics test the light-spectator plateau.

\emph{Interpretation.} Since $\omega\equiv0$ on the no-turn trajectory, Eq.~(\ref{eq:entropy_eom}) is homogeneous: $\mathcal{I}_{S\mathcal{R}}$ measures spectator entropy power, while the true turn-sourced coefficient $T_{\mathcal{RS}}$ vanishes analytically. This statement is independent of the fourth version, whose quantity named $T_{RS}$ was the fraction of total curvature power contributed by its isocurvature-initialised vacuum mode, $P_{SS}/(P_{RR}+P_{SS})$. Its $P_{SS}$ denoted curvature output, not the entropy-field power denoted by $P_{SS}$ here; that nonnegative fraction was not a signed Gordon--Wands transfer amplitude \cite{Kravchenko2025v4}. Extrapolating the directly computed $\ell=60$ ratio to the unreachable observable pivot is an estimate. More importantly, Planck constrains the late-time density-isocurvature fraction $\beta_{\rm iso}=P_{II}/(P_{RR}+P_{II})$, whereas this work computes a conventionally normalised horizon-exit field ratio. The remaining inflationary evolution, reheating, and the mechanism by which $\chi$ could become CDM are not modelled \cite{Wands2002}; on the $\chi=0$ background, $\delta\rho_\chi=m_\chi^2\chi\,\delta\chi$ vanishes at linear order. Thus the factor-$25$ comparison quantified in \S\ref{subsec:planck_comparison} is a hypothetical mapping exercise conditional on both an order-unity field-to-density transfer and the plateau-to-pivot extrapolation, not a model exclusion.

\subsection{Sensitivity of \texorpdfstring{$\mathcal{I}_{S\mathcal{R}}$}{I-S-R} to \texorpdfstring{$m_\chi$}{m-chi}}
\label{subsec:entropy_power_sensitivity}

The substantive power ratio reported in Table~\ref{tab:entropy_power_ratio_per_ell} is conditional on the fiducial entropic mass $m_\chi = \mchiFiducial\,M_{\rm Pl}$, a factor $\approx 6$ below the inflationary Hubble scale ($m_\chi/H_{\rm inf} \approx \mchiOverHinfFiducial$). To probe the sensitivity of the finding to this choice, we perform a scan of the entropy-mode equation of motion~(\ref{eq:entropy_eom}) over five values of $m_\chi$ at the pivot-proxy harmonic $\ell = 60$ (\S\ref{subsec:tensor_solver}), holding all other parameters at their fiducial values; the results are collected in Table~\ref{tab:entropy_power_sensitivity}.

\begin{table}[htbp]
\centering
\caption{\textbf{Sensitivity of $\mathcal{I}_{S\mathcal{R}}(\ell = 60)$ to the entropic mass $m_\chi$.}
Each row corresponds to a fixed $m_\chi$ with all other parameters held at fiducial values. The reporting-state value and the interval across the two propagated scalar states use the same entropy-mode numerator. The log-log fit slope of the reporting-state column is $\ISRLoglogSlope$, with strong departures from a single power law at the $m_\chi \gtrsim H_{\rm inf}$ heavy-mass freeze-out boundary.}
\label{tab:entropy_power_sensitivity}
\begin{tabular}{cccc}
\toprule
$m_\chi/M_{\rm Pl}$ & $m_\chi/H_{\rm inf}$ & reporting state & scalar-state range \\
\midrule
$10^{-8}$  & $1.7 \times 10^{-3}$ & $\ISRAtMchiEightMinus$ & $[\ISRAtMchiEightMinusStateMin,\ISRAtMchiEightMinusStateMax]$ \\
$10^{-7}$  & $1.7 \times 10^{-2}$ & $\ISRAtMchiSevenMinus$ & $[\ISRAtMchiSevenMinusStateMin,\ISRAtMchiSevenMinusStateMax]$ \\
$\mchiFiducial$ (fiducial) & $\mchiOverHinfFiducial$  & $\ISRAtMchiSixMinus$ & $[\ISRAtFiducialMchiStateMin,\ISRAtFiducialMchiStateMax]$ \\
$10^{-5}$  & $1.7$  & $\ISRAtMchiFiveMinus$ & $[\ISRAtMchiFiveMinusStateMin,\ISRAtMchiFiveMinusStateMax]$ \\
$10^{-4}$  & $17$   & $\ISRAtMchiFourMinus$ & $[\ISRAtMchiFourMinusStateMin,\ISRAtMchiFourMinusStateMax]$ \\
\bottomrule
\end{tabular}
\end{table}

The scan reveals two regimes. For $m_\chi\lesssim H_{\rm inf}$ the entropy mode is effectively massless at exit and $\mathcal{I}_{S\mathcal{R}}$ saturates near $\ISRAtMchiEightMinus$. For $m_\chi\gtrsim10H_{\rm inf}$ it is genuinely heavy and the ratio falls to $\ISRAtMchiFourMinus$; the $m_\chi/H_{\rm inf}=1.7$ point lies in the crossover.

The full-range log-log slope $\ISRLoglogSlope$ is only a summary of this non-power-law crossover. The fiducial result is a model-internal horizon-exit diagnostic at one parameter point, not a universal prediction of the bounce. In the light regime it conditionally exceeds the Planck CDI bound under the transfer and pivot assumptions above. No late-time conclusion follows from the heavy endpoint either, because its super-Hubble mode decays and the same unmodelled transfer controls the final fraction.

The former external-baseline normalisation caveat is closed: every value in both tables is now a same-epoch ratio of the fixed entropy solution to a scalar solution of the exact reduced action. The reporting column uses the acoustic-gradient scalar state and the interval column propagates the two declared scalar states. The remaining caveats concern the fixed entropy-state prescription, the non-exhaustive scalar-state comparison, physical transfer, and the unreachable observable pivot---not an arbitrary external scalar normalisation.

\section{Observational Predictions and Parameter Independence}
\label{sec:observables}

\subsection{Standard slow-roll predictions at the observable pivot}

The Starobinsky slow-roll formulae deliver
\begin{align}
n_s &= 1 - \frac{2}{N} \approx \nsAnalytical \quad (N = 60),
\\
r &= \frac{12}{N^2} \approx \rAnalytical,
\\
f_{\rm NL}^{\rm local} &= \tfrac{5}{12}(1 - n_s) \approx \fNLCMBconsistencyLeading,
\end{align}
applicable in the Starobinsky inflation regime that the post-bounce trajectory of \S\ref{sec:numerics} settles into. The covariant framework of \S\ref{sec:perturbations} corroborates the tensor sector of these predictions on the slow-roll attractor: a closed-FRW tensor amplitude agreeing with the same-exit flat-FRW benchmark to $\PhSameExitAgreementPct\%$, and the direct like-for-like early-plateau ratio $r_{\ell=60}=\rValClosedFRW$ from independently normalised tensor and exact reduced-action scalar solvers (\S\ref{subsec:tensor_solver}). Propagating the two declared scalar states gives $r_{\ell=60}\in[\rValClosedFRWStateMin,\rValClosedFRWStateMax]$; this finite-state range is not a statistical error bar. The numerical ratio is not compared with the different-epoch $N=60$ value. The observable-pivot $n_s$ is \emph{not} obtained from a spectral slope of the covariant solvers: the pivot harmonic is numerically unreachable (\S\ref{subsec:tensor_solver}), so the quoted value is the analytic slow-roll expression evaluated at the measured pivot $\phi_{60}^{\rm num}$. Two other numerical spectral slopes occur in this work, but neither supplies that observable-pivot value. The independent flat-FRW Mukhanov--Sasaki benchmark described in \S\ref{par:flat_scalar_benchmark} extracts $n_s$ from $d\ln P_{\mathcal R}/d\ln k$ and checks that flat calculation against the leading slow-roll reference; it is not a closed-space spectral cross-check. The curvature-scale log-slope of \S\ref{subsec:curvature_scale_diagnostic} is instead a model-internal diagnostic at the lowest harmonics. The observable-pivot prediction at first order in slow roll is $16\,\epsilon_V \approx 3.08 \times 10^{-3}$, with matching first-order $n_s = 1 - 6\epsilon_V + 2\eta_V \approx 0.9672$ agreeing with the unrounded measured scan value (quoted as $\nsAlphaMean$) to $\sim 10^{-5}$; the complete squeezed-limit $f_{\rm NL}^{\rm local}$ follows from Maldacena's consistency relation on the same attractor.

\paragraph{Non-Gaussianity: squeezed-limit result and the $\delta N$ contribution.}
For Gaussian field fluctuations, the separate-universe formula \cite{SugiyamaKomatsuFutamase2013} gives the super-Hubble contribution
\begin{equation}
f_{\rm NL}^{\delta N}(\phi_*) = \frac{5}{6}\,\frac{N''(\phi_*)}{[N'(\phi_*)]^2},
\qquad
N(\phi) = \tfrac{3}{4}\,e^{\beta\phi/M_{\rm Pl}} - \tfrac{3}{4}\,e^{\beta\phi_{\rm end}/M_{\rm Pl}} - \frac{\phi - \phi_{\rm end}}{2\beta M_{\rm Pl}},
\label{eq:fNL_deltaN}
\end{equation}
giving $f_{\rm NL}^{\delta N}=\fNLCMBdeltaN$ at the analytic $N(\phi_*)=60$ inversion point $\phi_*\approx5.453\,M_{\rm Pl}$ of Eq.~\eqref{eq:N_of_phi}. This is not by itself the complete primordial bispectrum: the intrinsic horizon-crossing non-Gaussianity of the field fluctuation supplies the remaining slow-roll contribution. The complete single-field squeezed-limit prediction is Maldacena's consistency relation \cite{Maldacena2003}, $f_{\rm NL}^{\rm local}=\tfrac{5}{12}(1-n_s)=\fNLCMBconsistency$ with the first-order Starobinsky $n_s$ at the same point; the difference $|f_{\rm NL}^{\rm local}-f_{\rm NL}^{\delta N}|=\fNLCMBdelta$ is the expected component omitted by the Gaussian-field $\delta N$ term. No second-order perturbation system or in-in bispectrum is evolved through the bounce, so this standard post-bounce result does not bound a separate bounce-generated contribution at the observable pivot.

\subsection{\texorpdfstring{$\alpha$}{alpha}-universality on the baseline trajectory}

The $\alpha$-universality of $(n_s, r, f_{\rm NL})$ holds on the kinematically trivial $\chi = \dot\chi = 0$ trajectory: $g_{\chi\chi}(\alpha)$ enters the background equations only through the inactive $\chi$-kinetic term, so the inflaton dynamics are exactly $\alpha$-independent. A scan over $\nAlphaPoints$ logarithmically-spaced values $\alpha \in [0.1, 10]$ at $\phi_0 = 6.0\,M_{\rm Pl}$ (giving $N_{\rm total} = \NtotalInflationAlphaScan$), run through the background integrator of \S\ref{sec:numerics}, yields $n_s^{\rm kin} = \nsAlphaMean$ at the actual $N = 60$ point of each trajectory, with $\sigma(n_s) < 10^{-6}$ across the scan---exact $\alpha$-independence at the precision of the integrator. This numerical scan evaluates $n_s^{\rm kin}$ only; baseline $r$ and $f_{\rm NL}$ inherit their $\alpha$-independence analytically from the same decoupled slow-roll solution rather than from separate scanned estimators. (This scan and the excited-spectator scan below use the same background integration and inputs stated here.)

\paragraph{Excited-spectator kinematic-stability test.}
To test one background statistic outside the $\chi = \dot\chi = 0$ initial-data choice, we repeat the scan with an excited spectator $\chi_0 = 1\,M_{\rm Pl}$, $\dot\chi_0 = 0$. The spectator is slow-rolling during inflation ($m_\chi < H_{\rm inf}$), contributing $K_\chi = \tfrac{1}{2} g_{\chi\chi}\,\dot\chi^2$ up to $K_\chi/K_{\rm tot} = \KchiMaxNT\%$ at each trajectory's measured point $\phi_{60}^{\rm num}(\alpha)$, with $g_{\chi\chi}(\phi_{60}^{\rm num}(\alpha))$ varying from $\gCMBminNT$ ($\alpha = 0.1$) to $\gCMBmaxNT$ ($\alpha \geq 1$). Despite this $\sim 25\%$ variation in the active field-space metric and the non-negligible $\chi$-kinetic fraction, the scan yields $n_s = \nsAlphaMeanNT$ with $\sigma(n_s) = \nsAlphaStdNT$ across $\nAlphaPointsNT$ values of $\alpha$. This records stability of the sampled kinematic statistic on the displaced-background family; it does not identify or establish the ``stable attractor'' mechanism claimed in v2. The same scan also records a background-level kinematic $r=16\epsilon$ for each row, but the displaced-family summary promoted here is the $n_s$ statistic; no displaced-family $f_{\rm NL}$ is evaluated. We note the scope of this scan carefully: these are background-level slow-roll estimators at the measured $N = 60$ point of each trajectory, not integrated multifield spectra. What the covariant computed grid of \S\ref{subsec:triangulation} adds on spectator-displaced trajectories (mildly and moderately turning, $\chi_0$ up to $1\,M_{\rm Pl}$, at $\alpha \in \{1, 3\}$) is the \emph{integrability and cross-construction consistency} of the full two-field perturbation system there---not the $\alpha$-independence of the physical spectra. The scalar absolute-normalisation bridge is now closed, but the perturbation-level test of $\alpha$-independence remains open: $\mathcal{I}_{S\mathcal{R}}$ is not the transfer coefficient $T_{\mathcal{RS}}$, and the remaining perturbation-level spectral $\alpha$ scan has not been performed. The only displaced-family $\alpha$ statement promoted in this work is therefore the background-level stability of the sampled kinematic $n_s$ statistic.

One solver-level observation is compatible with (but does not test) this universality: the cross-basis residuals (\S\ref{subsec:triangulation}) at $\alpha = 1$ and $\alpha = 3$ are comparable in magnitude across the computed grid---a consistency statement about the agreement among $\mathcal{R}$ reconstructions, not a statistical test or a measurement of the $\alpha$-dependence of absolutely-normalised spectra. A genuine perturbation-level test (spectra over $\alpha$, including $\mathcal{I}_{S\mathcal{R}}$ and a transfer calculation at fixed $\chi_0 \neq 0$) is identified as future work in \S\ref{sec:conclusions}.

\subsection{Closed-FRW \texorpdfstring{$K=+1$}{K=+1} low-harmonic curvature-scale diagnostic}
\label{subsec:curvature_scale_diagnostic}

The closed-FRW $K=+1$ geometry admits a low-harmonic diagnostic at $\ell=2,3$ that has no flat-continuum counterpart. The scalar harmonics have $\lambda_\ell=\ell(\ell+2)$ and physical reduced modes begin at $\ell=2$. We define the diagnostic as the finite-difference log-slope of the exact dimensionless scalar power, using for each harmonic the arithmetic mean of the saved power samples in the final e-fold of the integrated expanding branch, rather than the power at geometric exit:
\begin{equation}
n_s^{\rm curv} - 1 = \frac{d \ln P_{\mathcal{R}}}{d \ln k}, \qquad
P_{\mathcal{R}}(\ell) = \frac{\ell(\ell+1)(\ell+2)}{2\pi^2}\,|\mathcal{R}_\ell|^2, \qquad
k(\ell) = \sqrt{\ell(\ell+2)},
\label{eq:ns_curvature_def}
\end{equation}
with $|\mathcal R_\ell|^2$ obtained from the exact reduced action of Appendix~\ref{app:closed_s3_canonical}. The factor $\ell(\ell+1)(\ell+2)$ is fixed by the scalar-harmonic degeneracy and the Jacobian from harmonic order to logarithmic physical wavenumber; approximate per-log-$\ell$ variants are not averaged with it and are not treated as an uncertainty distribution. Across the two propagated finite-time canonical states, the result is
\begin{equation}
n_s^{\rm curv}(2\to3)\in
[\nsCurvatureScaleStateMin,\nsCurvatureScaleStateMax].
\label{eq:ns_curvature_scale}
\end{equation}
The endpoints are state resolved rather than interpreted as an error bar:
\begin{equation}
n_{s,\,\mathrm{inst.H}}^{\rm curv}(2\to3)
=\nsCurvatureScaleInstantaneous,
\qquad
n_{s,\,\mathrm{ac.grad.}}^{\rm curv}(2\to3)
=\nsCurvatureScale,
\qquad
n_{s,\,\mathrm{LS,ac.grad.}}^{\rm curv}(2,3,5,7,10)
=\nsCurvatureScaleLS.
\label{eq:ns_curvature_scale_state_resolved}
\end{equation}
Here ``inst.H'' denotes the instantaneous-Hamiltonian state and ``ac.grad.'' the acoustic-gradient state of Appendix~\ref{app:closed_s3_canonical}. The range is not a statistical error bar and the two states do not exhaust all admissible finite-time prescriptions. Its width shows that the lowest-harmonic slope is not a state-robust prediction. The least-squares value is reported only to expose the additional fit dependence across five discrete harmonics; neither estimator is substituted for the observable-pivot $n_s$.

\paragraph{Effective (cosmic-variance-limited) unobservability.}
The $\ell = 2, 3$ harmonics on the closed $K=+1$ universe carry only a few independent amplitudes each: the $S^3$ harmonic $\ell$ has degeneracy $(\ell+1)^2$, i.e.\ $9$ modes at $\ell = 2$ and $16$ at $\ell = 3$, so a single closed universe furnishes at most that many realisations per harmonic. Any power-spectrum estimate at these harmonics is therefore cosmic-variance-dominated, with fractional standard deviation $\sqrt{2/(\ell+1)^2} \approx 47\%$ and $35\%$ respectively---the closed-geometry analogue, and at the lowest harmonics the structural floor, of the familiar $\ell = 2$ CMB cosmic-variance limit \cite{Planck2018}. Moreover, connecting $n_s^{\rm curv}$ to actual CMB or large-scale-structure observables would require the late-time transfer through the post-inflationary history, which is not computed here. The curvature-scale diagnostic Eq.~(\ref{eq:ns_curvature_scale}) is therefore reported as a \emph{model-internal} low-harmonic diagnostic of the closed bouncing model: severely cosmic-variance-limited at its native harmonics and not connected to data in this work, rather than strictly unobservable in principle.

\subsection{Finite-time state-prescription robustness}
\label{subsec:BD_robustness}

At a finite numerical start time, a positive-frequency prescription specifies initial data; the name ``Bunch--Davies'' alone establishes neither an asymptotic state nor an adiabatic regime. The dependence on canonical variables and vacuum prescription, including in curved inflationary universes, is established in the literature \cite{Handley2019,AgocsEtAl2020,LeteyEtAl2024}; the contribution here is the explicit propagation of two specified states on this bounce background. The exact fiducial scalar action makes the required choices explicit: the canonical pair $(\mathcal R_\ell,\Pi_{\mathcal R,\ell})$, a common initial slice at acoustic depth $\min_\ell[c_\ell\sqrt{\Lambda_\ell}/(a|H|)]=50$, and a positive-frequency prescription. We propagate both the acoustic-gradient state and the instantaneous-Hamiltonian state defined in Appendix~\ref{app:closed_s3_canonical}. Their difference is reported for the scalar powers, tensor-to-scalar and entropy-to-curvature ratios, and the two-harmonic slope; the five-harmonic least-squares diagnostic is quoted only for the acoustic-gradient state. However, two states do not define a probability distribution or exhaust admissible finite-time states; in particular, higher-order adiabatic accuracy has not been established for these two prescriptions. The turning-grid cross-basis calculation separately selects each start by the first-acceptance condition $\sqrt{\lambda_\ell}/(a|H|)\geq50$ and maps one declared finite-time test state exactly between the two field-space bases before the independent evolutions; it does not supply a vacuum-uniqueness or state-insensitivity test for the coupled turning system.

\begin{assumptionbox}[title=\textbf{Scope of the Initial-Depth Condition}]
\textbf{Hubble-rate parameter, not a bound on state uncertainty.}
For the tensor and entropy solvers the initial-depth condition $k_{\rm phys}/|H|\geq1.5\times10^3$ bounds only the following dimensionless Hubble-rate parameter:
\begin{equation}
\delta_{H,\,T/S}^{\rm init}\equiv\Bigl(\tfrac{|H|}{k_{\rm phys}}\Bigr)^2_{\rm init}\leq(1.5\times10^3)^{-2}\simeq4.4\times10^{-7}.
\end{equation}
The criterion is reached only near the bounce on the contracting branch, so this is not an asymptotic-past construction. This small parameter does not bound differences between initial states, their higher-order adiabatic corrections, or propagated powers. The depth condition does not control $\dot H/k_{\rm phys}^2$, $K/(a^2k_{\rm phys}^2)$, or other time derivatives in the canonical frequency. For the tensor gradient scale $k_{\rm phys}^2=\lambda_\ell/a^2$, the curvature ratio is $K/\lambda_\ell$, independently of the Hubble depth: it equals $1/8=12.5\%$ at $\ell=2$ and $1/35\simeq2.9\%$ at $\ell=5$. These ratios are not estimates of the propagated power error. The tensor and entropy calculations each fix one leading prescription; alternatives in those sectors are not propagated here. Wronskian conservation preserves the norm of each chosen state but does not bound the power difference between distinct states.
\end{assumptionbox}

The Hubble-rate parameter is small at the selected tensor and entropy initial times, but no quantitative state-insensitivity bound follows from it. The independent flat-FRW scalar benchmark uses depth $10^3$; its separate recorded one-off depth sweep is described in \S\ref{par:flat_scalar_benchmark} and does not normalise the closed-space scalar result. For the exact closed scalar sector, the relevant evidence is instead the directly propagated two-state interval: at $\ell=60$ the frozen scalar power spans $\ExactScalarStateSpanEllSixtyPct\%$, while the low-harmonic slope spans $[\nsCurvatureScaleStateMin,\nsCurvatureScaleStateMax]$. These are finite comparisons between two named states, not a universal state-prescription bound. No observable-pivot headline relies on them because $(n_s,r)$ at the pivot are evaluated from the post-bounce slow-roll attractor.

The entropy numerator retains one specified canonical state, so the intervals in Tables~\ref{tab:entropy_power_ratio_per_ell} and \ref{tab:entropy_power_sensitivity} propagate only the two exact scalar denominators. A joint entropy-state comparison remains future work. The model-internal $m_\chi$ dependence is characterised separately in \S\ref{subsec:entropy_power_sensitivity}.

\subsection{Comparison to Planck 2018}
\label{subsec:planck_comparison}

\begin{figure}[htbp]
    \centering
    \IfFileExists{figures/fig_ns_r_plane.pdf}{%
        \includegraphics[width=0.82\textwidth]{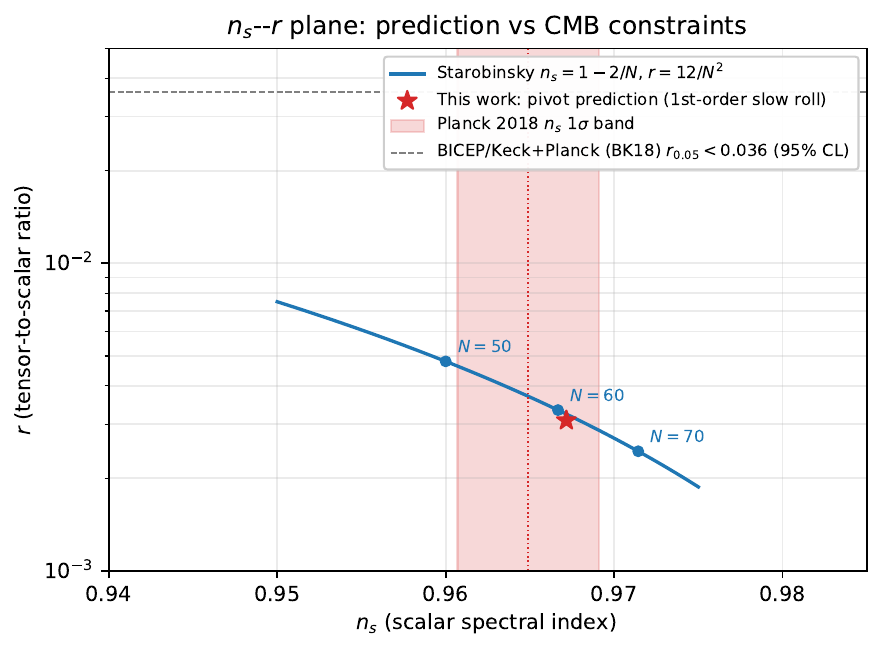}%
    }{%
        \fbox{\parbox{0.78\textwidth}{\centering
            \textbf{Figure not available in this build.}}}%
    }
    \caption{\textbf{$n_s$--$r$ plane: this work vs CMB constraints.}
        Solid blue curve: Starobinsky slow-roll prediction
        $(n_s, r) = (1 - 2/N,\ 12/N^2)$ for $N \in [40, 80]$, with markers at
        $N \in \{50, 60, 70\}$. Red star: this work's observable-pivot prediction as the
        \emph{consistent first-order slow-roll pair}
        $(n_s, r) = (1 - 6\epsilon_V + 2\eta_V,\ 16\epsilon_V) \approx
        (0.9672,\ 3.08 \times 10^{-3})$, both evaluated at the measured
        pivot point $\phi_{60}^{\rm num} = \phiCMBmeasured\,M_{\rm Pl}$ (the
        $n_s$ agrees with the unrounded measured scan value, quoted as
        $\nsAlphaMean$, to $\sim 10^{-5}$); it
        lies slightly off the leading-order $(1 - 2/N,\ 12/N^2)$ curve. The direct closed-FRW early-plateau ratio
        $r_{\ell=60}=\rValClosedFRW$ (\S\ref{subsec:tensor_solver}) is a physical
        same-epoch ratio at a different exit epoch, not an observable-pivot
        prediction, and is deliberately not shown on this plane. Red shaded band: Planck 2018 $1\sigma$
        constraint on $n_s = 0.9649 \pm 0.0042$~\cite{Planck2018}. Grey dashed
        line: BICEP/Keck+Planck (BK18) upper bound $r_{0.05} < 0.036$ at $95\%$~CL
        \cite{BICEPKeck2021}. The model's
        observable-pivot prediction sits squarely within the Planck $1\sigma$
        band for $n_s$ and well below the $r$ upper bound, accessible to
        next-generation experiments LiteBIRD~\cite{LiteBIRD2022},
        CMB-S4~\cite{CMB-S42016}, and PICO~\cite{PICO2019}.}
    \label{fig:ns_r_plane}
\end{figure}

Combining the standard slow-roll predictions \cite{StewartLyth1993} and the closed-FRW $K=+1$ refinements:
\begin{itemize}
\item $n_s = \nsAnalytical$ at the observable pivot at leading order, and $n_s = 1 - 6\epsilon_V + 2\eta_V \approx 0.9672$ at first order at the measured pivot point (agreeing with the unrounded measured scan value, quoted as $\nsAlphaMean$, to $\sim 10^{-5}$)---$0.4\sigma$ at leading order and $0.5\sigma$ at first order from the Planck-2018 central value $0.9649 \pm 0.0042$ \cite{Planck2018}; the corresponding $(n_s, r)$ pairs are $(\nsAnalytical,\ \rLythStewartReference)$ at leading order and $(0.9672,\ 3.08 \times 10^{-3})$ at first order;
\item $r = 16\,\epsilon_V(\phi_{60}^{\rm num}) \approx 3.08 \times 10^{-3}$ at the observable pivot, at first order in slow roll ($12/N^2 = \rLythStewartReference$ at leading order). At the numerically accessible early-plateau harmonic the closed-FRW tensor and exact scalar solvers independently give the reporting-state value $r_{\ell=60}=\rValClosedFRW$ and the two-scalar-state interval $[\rValClosedFRWStateMin,\rValClosedFRWStateMax]$, while the tensor amplitude agrees with the same-exit flat-FRW benchmark to $\PhSameExitAgreementPct\%$ (\S\ref{subsec:tensor_solver}); the direct early-plateau ratio and the observable-pivot prediction refer to different exit epochs and are not equated. The observable value is well below the cited $95\%$~CL upper bound $r_{0.05} < 0.036$ from BICEP/Keck 2018 data combined with Planck \cite{BICEPKeck2021} (Planck 2018 alone gives $r_{0.002} < 0.10$ \cite{Planck2018}). Forecasts depend on foreground and delensing assumptions: the cited LiteBIRD study requires total uncertainty $\delta r<0.001$ \cite{LiteBIRD2022}, the CMB-S4 Science Book lists the aspirational target $\sigma(r)=5\times10^{-4}$ \cite{CMB-S42016}, and the cited PICO study forecasts a $5\sigma$ detection at $r=5\times10^{-4}$ \cite{PICO2019};
\item $f_{\rm NL}^{\rm local} = \fNLCMBconsistency$ from the single-field squeezed-limit consistency relation (with $f_{\rm NL}^{\delta N}=\fNLCMBdeltaN$ the separately reported super-Hubble contribution). Unlike the first-order $n_s$ and $r$ above, which are evaluated at the measured pivot $\phi_{60}^{\rm num}$, both $f_{\rm NL}$ values are evaluated at the analytic inversion point $\phi_{60}^{\rm SR}$ of Eq.~\eqref{eq:N_of_phi}; the two field values differ by $0.024\,M_{\rm Pl}$ and the corresponding $f_{\rm NL}$ by about $2\%$, far below the observational precision. The result is well within the Planck 2018 constraint $f_{\rm NL}^{\rm local} = -0.9 \pm 5.1$ ($68\%$~CL) \cite{Planck2018NG}.
\end{itemize}
A detection of $r\sim0.003$ would test the Starobinsky inflationary plateau; it is not a bounce-specific discriminator and would not by itself evidence the pre-inflationary transition.
The observable-pivot $(n_s, r, f_{\rm NL})$ triple is consistent with the cited Planck-2018/BK18 constraints (Fig.~\ref{fig:ns_r_plane}): $n_s$ is within $1\sigma$, $r$ is well below the $95\%$~CL bound, and $f_{\rm NL}$ is well within the non-Gaussianity constraint. This sector is testable by next-generation CMB experiments. The closed-FRW $K=+1$ curvature-scale diagnostic Eq.~(\ref{eq:ns_curvature_scale}) is effectively unobservable---cosmic-variance-dominated at its $(\ell+1)^2$-fold-degenerate native harmonics, with no late-time transfer computed (\S\ref{subsec:curvature_scale_diagnostic})---and is not subjected to a Planck likelihood here. The horizon-exit field diagnostic $\mathcal{I}_{S\mathcal{R}}$ (\S\ref{subsec:entropy_power_ratio}--\S\ref{subsec:entropy_power_sensitivity}) is likewise not part of the standard $(n_s,r,f_{\rm NL})$ consistency statement. Its reporting-state values are $\ISRAtFiducialMchi$ at $\ell=60$ and a sampled-grid maximum $\ISRSupSampledGrid$; varying the two declared exact scalar states while holding the entropy state fixed gives the respective ranges $[\ISRAtFiducialMchiStateMin,\ISRAtFiducialMchiStateMax]$ and $[\ISRSupSampledGridStateMin,\ISRSupSampledGridStateMax]$. The distinct turn-sourced coefficient is $T_{\mathcal{RS}}=0$ on the no-turn background. Planck defines the primordial density-isocurvature fraction as $\beta_{\rm iso}=P_{II}/(P_{RR}+P_{II})$ and obtains $\beta_{\rm iso}(k_{\rm mid}=0.05\,\mathrm{Mpc}^{-1})<0.038$ at $95\%$~CL for $n_{II}=1$ uncorrelated CDI using Planck TT,TE,EE+lowE+lensing \cite{Planck2018}; identifying the present field-space $P_{SS}$ with $P_{II}$ would require a production mechanism, post-horizon transfer, and the $\ell=60$-to-pivot extrapolation. If all three were imposed with order-unity efficiency, the comparison is ratio to ratio: the bound $\beta_{\rm iso}<0.038$ corresponds to $P_{II}/P_{RR}<\beta_{\rm iso}/(1-\beta_{\rm iso})=0.0395$, which the reporting-state $\mathcal{I}_{S\mathcal{R}}=\ISRAtFiducialMchi$ would exceed by about a factor $25$. Since none is computed, this is a conditional sensitivity estimate, not an observational exclusion; no late-time statement follows from the heaviest scanned value $\mathcal{I}_{S\mathcal{R}}=\ISRAtMchiFourMinus$ either.

\section{Discussion and Conclusions}
\label{sec:conclusions}

\subsection{Summary of results}

We have presented a covariant two-field perturbation framework for a finite, locally non-singular contracting-to-expanding solution in a closed $K=+1$ universe with a regularised two-field sigma-model metric. Building on the earlier closed-FRW two-field model \cite{Kravchenko2025v1,Kravchenko2025v2,Kravchenko2025v3} and its gauge-invariant Newtonian-gauge perturbation integration through the bounce \cite{Kravchenko2025v4}, this work provides three concrete calculations: (i) an independent coordinate- and orthonormal-field-basis integration of the standard covariant system \cite{GrootNibbelinkVanTent2002,Langlois2008} through a curvature bounce in the same Newtonian spacetime gauge plus a Gordon--Wands projection check on $\TwoTwoIntegratedCells$ integrated bouncing cells; the nominal grid also contains $\TwoTwoControlCells$ non-integrated finite-range contracting-control cells (maximum integrated-cell residual $\TwoTwoMaxResidual$, with every integrated cell satisfying $\Delta<\TwoTwoOriginalThreshold$; \S\ref{subsec:triangulation}), supplemented by direct immediate-bounce constraints, DOP853 tolerance refinement, and an independent local Radau comparison on the preselected $\BounceWindowMethodCells$-cell subset (\S\ref{subsec:constraint_diagnostics}); (ii) derivation and integration of the exact finite-harmonic closed-$S^3$ reduced scalar action on the fiducial no-turn trajectory, including two explicitly specified Wronskian-normalised finite-time states, together with an independently normalised tensor solver; these give the reporting-state early-plateau ratio $r_{\ell=60}=\rValClosedFRW$ and the two-state interval $[\rValClosedFRWStateMin,\rValClosedFRWStateMax]$ while keeping the distinct observable-pivot prediction $16\,\epsilon_V\approx3.08\times10^{-3}$ (\S\ref{subsec:tensor_solver}, Appendix~\ref{app:closed_s3_canonical}); (iii) a direct calculation of the covariant entropy-mode power spectrum, with the raised-index entropy mass $m_\chi^2/g_{\chi\chi}$ derived from the field-space-covariant equation of motion, reporting the direct horizon-exit field-power ratio $\mathcal{I}_{S\mathcal{R}}$ of the turn-decoupled spectator entropy mode (\S\ref{subsec:entropy_power_ratio}).

The new quantitative entropy-sector result is the spectator power relative to the exact scalar curvature power on the fiducial trajectory at the fiducial $m_\chi=\mchiFiducial\,M_{\rm Pl}$. With the fixed entropy state, the reporting scalar state gives $\mathcal{I}_{S\mathcal{R}}=\ISRAtFiducialMchi$ at $\ell=60$ and sampled-grid maximum $\mathcal{I}_{S\mathcal{R}}^{\max}=\ISRSupSampledGrid$ over $\ell\in\{2,5,60,100,200,500,1000\}$; the two declared scalar states give the respective intervals $[\ISRAtFiducialMchiStateMin,\ISRAtFiducialMchiStateMax]$ and $[\ISRSupSampledGridStateMin,\ISRSupSampledGridStateMax]$. No external scalar baseline or analytic conversion enters those values. On the $\dot\chi=0$ trajectory the turn rate vanishes, so $T_{\mathcal{RS}}=0$ and the measured quantity is not dynamical isocurvature-to-curvature transfer: it is the power of an independently evolving spectator entropy mode in the specified finite-time state, with raised-index mass $m_\chi^2/g_{\chi\chi}$, relative to the curvature power. Its near-unity light-mass value is expected for two light, independently quantised modes and is not by itself a viability failure. Translating it into a late-time density-isocurvature statement requires an uncomputed production mechanism, post-horizon transfer, species normalisation, and observable-pivot matching; moreover, $\delta\rho_\chi$ vanishes linearly on the $\chi=0$ background without an additional production or misalignment mechanism. The $m_\chi$ scan documents a light-mass plateau and heavy-mass suppression to $\mathcal{I}_{S\mathcal{R}}=\ISRAtMchiFourMinus$ at $m_\chi=10^{-4}M_{\rm Pl}$ for the reporting scalar state, but does not establish a late-time viability boundary.

The standard slow-roll observables $n_s$, $r$, $f_{\rm NL}$ at the observable pivot are recovered on the slow-roll attractor and remain consistent with the cited Planck-2018/BK18 constraints. The exact closed-FRW scalar calculation also yields the model-internal two-state interval $n_s^{\rm curv}\in[\nsCurvatureScaleStateMin,\nsCurvatureScaleStateMax]$ at $\ell=2,3$ (\S\ref{subsec:curvature_scale_diagnostic}), with the instantaneous-Hamiltonian and acoustic-gradient states giving the lower and upper endpoints, respectively. The interval is not a statistical uncertainty, and the diagnostic is cosmic-variance-dominated with no computed late-time transfer. Initial-depth statements are sector-specific (exact closed scalar $50$; turning-grid cross-basis test $\sqrt{\lambda_\ell}/(a|H|)\geq50$ under the first-acceptance rule; flat-FRW scalar benchmark $10^3$; tensor/entropy $1.5\times10^3$). Two exact scalar states are propagated to the reporting epochs, whereas the tensor and entropy sectors retain one specified state each and an exhaustive finite-time-state comparison is not claimed (\S\ref{subsec:BD_robustness}). The horizon-exit $\mathcal{I}_{S\mathcal{R}}$ diagnostic is likewise separate from both the standard observable sector and the transfer coefficient. Its translation into primordial density isocurvature remains open, so this work neither claims full phenomenological validation nor asserts an exclusion of the fiducial model.

\subsection{Limitations and open questions}

For transparency we first state plainly what this work does \emph{not} claim. Each item is expanded either in the paragraphs below or in the section it cites. The standard observable-pivot results ($n_s$, $r$, $f_{\rm NL}$), the closed-FRW tensor-sector validation, and the direct solver-side entropy-power ratio do not rely on an external scalar normalisation. The remaining limitations concern the perturbation-level spectral $\alpha$ scan and the turn-sourced transfer test, a through-bounce bispectrum, higher-order perturbative control, finite-time-state-prescription propagation, harmonic completeness, anisotropic dynamics, parameter coverage, the physical mapping from horizon-exit field power to late-time density isocurvature, and the model-internal status of the curvature-scale log-slope.
\begin{enumerate}
\item \textbf{No exhaustive finite-time-state comparison.} The exact fiducial scalar calculation propagates two explicitly specified canonical states to geometric exit and the frozen reporting epoch, but those two states do not exhaust higher-order adiabatic choices. The tensor and entropy sectors retain one specified state each; alternatives in those sectors are not propagated, and no quantitative bound on their differences is established here. The initial-depth condition bounds only the Hubble-rate parameter $(|H|/k_{\rm phys})^2\leq(1.5\times10^3)^{-2}$ (\S\ref{subsec:BD_robustness}).
\item \textbf{No background-discretisation convergence order.} The immediate-bounce tests include full-history DOP853 tolerance refinement on the common window $H_{\rm inf}|t-t_b|\leq\BounceWindowHalfWidth$ and an independent local Radau IIA propagation across the compact fully covered core $H_{\rm inf}|t-t_b|\leq\BounceWindowMethodHalfWidth$ on a preselected 24-cell subset. They establish tolerance and local integration-method stability on the stated finite grids, but do not compare methods on the wider window or full pre-bounce history, supply a controlled maximum-step sequence, refine the background interpolation, or measure a convergence order. Those remain future numerical checks.
\item \textbf{The perturbation-level $\alpha$-independence and transfer tests remain open.} Canonical scalar normalisation is closed, but the $T_{\mathcal{RS}}<10^{-2}$ condition concerns turn-sourced transfer, not the measured entropy power ratio $\mathcal{I}_{S\mathcal{R}}$. The no-turn trajectory has $T_{\mathcal{RS}}=0$ analytically; the spectral $\alpha$ scan and controlled observable-pivot mapping required for that combined statement have not been completed.
\item \textbf{No through-bounce bispectrum.} The quoted $f_{\rm NL}^{\rm local}$ is the standard slow-roll Maldacena squeezed-limit value on the post-bounce attractor; the $\delta N$ expression is only its separately reported super-Hubble component. Linear matching cannot determine a second-order three-point function, and no bounce-generated bispectrum is computed or bounded here.
\item \textbf{No exact reduced two-field action on generic turning backgrounds.} On the fiducial no-turn trajectory the finite-harmonic closed-$S^3$ scalar action is derived and integrated exactly (Appendix~\ref{app:closed_s3_canonical}). The turning-background grid instead evolves the full metric--matter perturbation equations in two field-space bases and checks their constraints and mutual agreement; it does not derive a reduced canonical action for the coupled adiabatic--entropy system on those backgrounds. Extending the exact action-level reduction and state construction to generic turn rate remains open. The tensor sector is separate and uses the spin-2 transverse-traceless eigenvalue and standard closed-universe spectral convention; vector modes are not evolved.
\item \textbf{No cubic-action or strong-coupling analysis.} Positivity of the sampled quadratic coefficients excludes the corresponding ghost and gradient-sign instabilities on the declared finite domain, but it does not establish perturbative control of higher-order interactions near the bounce. The cubic action and associated strong-coupling scale are not computed. In particular, the large positive finite-harmonic ratio $c_{\ell,b}^2\simeq\ExactScalarBounceCSqMin$ is neither a diagnosis of strong coupling nor evidence that it is absent.
\item \textbf{No inhomogeneous (numerical-relativity) BKL analysis, and no full Bianchi-IX dynamics.} BKL compatibility is assessed only on the sampled family in the wall-free homogeneous-shear truncation (conserved $\Sigma^2$ entering the Friedmann constraint as a parameter; the anisotropy variables $\beta_\pm$ and the curvature-anisotropy wall potential are not integrated).
\item \textbf{No complete $(m_\chi, \alpha, \chi_0)$ viability landscape.} The entropy sector is characterised only along the scanned $m_\chi$ ray at fixed $(\alpha, \chi_0)$, and only at the horizon-exit level. The fiducial light-mass ratio is $\mathcal{I}_{S\mathcal{R}}\simeq1$, the heaviest scanned point has $\mathcal{I}_{S\mathcal{R}}=\ISRAtMchiFourMinus$, and no late-time viability boundary is located. A CDM-isocurvature interpretation would require an explicit production and transfer model.
\item \textbf{No post-horizon-exit field-to-density isocurvature map.} The mapping from the horizon-exit field ratio to the late-time primordial CDM-isocurvature fraction that Planck bounds---through the remaining inflation, reheating, CDM formation, and species normalisation---is not computed; on the $\chi = 0$ background the linear $\delta\rho_\chi$ vanishes identically, so a linear CDM-isocurvature reading additionally requires a misalignment or production mechanism beyond this work. The factor-$25$ sensitivity estimate assumes all of those steps plus the extrapolation of the directly computed $\ell = 60$ ratio to the numerically unreachable observable pivot (\S\ref{subsec:planck_comparison}; the mapping assumptions are discussed in \S\ref{subsec:entropy_power_ratio}).

\item \textbf{The curvature-scale log-slope is a model-internal diagnostic, not a measurable spectral index.} $n_s^{\rm curv}$ is obtained from the exact dimensionless scalar power across the few lowest $S^3$ harmonics, where the spectrum is not a power law. The per-log-wavenumber measure is fixed, not varied as an uncertainty model. The two state-resolved $\ell=2\to3$ values demonstrate strong finite-time-state sensitivity, while the five-harmonic least-squares value separately exposes estimator dependence. Neither is a statistical error bar or a complete distribution over states. The diagnostic carries no late-time transfer, is cosmic-variance limited at its native harmonics, and is not part of the observable-pivot $(n_s,r,f_{\rm NL})$ set.
\end{enumerate}

\paragraph{Cross-basis consistency scope.}
\label{subsec:triangulation_limitation}
The cross-basis consistency criterion (\S\ref{subsec:triangulation}) compares $\mathcal{R}$ from two evolved field-basis systems and one algebraic projection on the $\TwoTwoIntegratedCells$ perturbatively integrated cells. The nominal grid includes another $\TwoTwoControlCells$ contracting-control cells, but those are a finite-range contracting background control and are not perturbatively evolved. All $\TwoTwoStrictCells$ integrated cells satisfy $\Delta<\TwoTwoOriginalThreshold$, with maximum residual $\TwoTwoMaxResidual$. The direct constraint/refinement check separately covers the immediate-bounce interval on both evolved legs of the same two turning families. Together they establish numerical consistency on the declared finite turning-background grid, not an exact reduced action for the coupled adiabatic--entropy system, all harmonics, or all initial data. That generic turning-background action-level extension remains future work; the exact no-turn scalar reduction is given separately in Appendix~\ref{app:closed_s3_canonical}.

\paragraph{\texorpdfstring{$m_\chi$}{m-chi} sensitivity of the \texorpdfstring{$\mathcal{I}_{S\mathcal{R}}$}{I-S-R} diagnostic.}
The direct perturbation-level ratio ($\mathcal{I}_{S\mathcal{R}}^{\max}=\ISRSupSampledGrid$ on the sampled grid; \S\ref{subsec:entropy_power_ratio}) is conditional on the fiducial entropic mass $m_\chi=\mchiFiducial\,M_{\rm Pl}$. A scan over $m_\chi\in[10^{-8},10^{-4}]\,M_{\rm Pl}$ at $\ell=60$ documents a light-mass plateau at $m_\chi\lesssim H_{\rm inf}$ and heavy-mass suppression at $m_\chi\gtrsim10H_{\rm inf}$, where the ratio falls to $\ISRAtMchiFourMinus$. The magnitude is model-internal and parameter-dependent, not a universal prediction of the bounce. A complete $(m_\chi,\alpha,\chi_0)$ perturbation-level landscape and a physical post-horizon field-to-density map remain future work.

\paragraph{Closed scalar normalisation and the remaining pivot extrapolation.}
The canonical scalar-normalisation problem is closed on the fiducial no-turn trajectory: $P_{\mathcal R}$ follows from the exact finite-harmonic reduced action, two canonical scalar states satisfy the exact Wronskian and are propagated, and $P_{\mathcal R}$ and $P_{SS}$ use the same exact per-log-wavenumber closed-$S^3$ measure. The fixed entropy state satisfies its own Wronskian, and $\mathcal{I}_{S\mathcal R}$ is formed directly at a common geometric-exit time. What remains inaccessible numerically is the conventional observable-pivot harmonic, $\ell\sim10^{1116}$. Carrying the measured $\ell=60$ entropy ratio to that epoch is therefore an explicit plateau extrapolation, not a solver result. A direct high-harmonic asymptotic treatment or an analytically controlled matching calculation is future work.

\paragraph{Origin of initial conditions and past completion.}
For the explicitly sampled initial-data family, the model demonstrates that entry into a short contraction phase with $\phi$ on the Starobinsky plateau can be followed by a robust bounce and subsequent inflation. It does not establish a generic basin for all subdominant-kinetic initial data, explain the turnaround from expansion to contraction, make the contraction phase an attractor in phase space, or establish a past-eternal/geodesically complete extension of the finite numerical trajectory.  The result is therefore a locally non-singular contracting-to-expanding solution, not a theorem that every cosmological past singularity has been removed.  The origin and past completion of the contracting branch are principal conceptual limitations and are identified as future work via a cyclic extension or a quantum-cosmological construction.

\paragraph{Inhomogeneous BKL analysis.}
The wall-free homogeneous-shear analysis of \S\ref{sec:bkl} establishes bounce robustness only for the sampled conserved-shear amplitudes of the fiducial family within that truncation; the full Bianchi-IX dynamics ($\beta_\pm$ with the wall potential active) is not integrated. Inhomogeneous BKL instabilities in 3+1 numerical general relativity are likewise not captured by the homogeneous truncation and would require dedicated numerical-relativity simulations following the program of Ijjas--Pretorius--Steinhardt \cite{IjjasPretoriusSteinhardt2020}.

\paragraph{Cyclic extension.}
A cyclic extension connecting multiple bounce-expansion cycles requires a mechanism for turnaround and re-contraction; it is identified as future work.

\paragraph{Quantum fluctuations through the bounce.}
Evolution through $H=0$ can mix positive- and negative-frequency components relative to a chosen post-bounce adiabatic basis. In the exact fiducial scalar sector, two finite-time canonical prescriptions are propagated through the bounce to geometric exit and the frozen reporting epoch, producing the explicitly reported state intervals. This comparison is finite and non-exhaustive; it is not a Bogoliubov decomposition with respect to a specified post-bounce adiabatic basis. The tensor and entropy sectors each retain one initial prescription, so their alternative-state propagation is also open. Computing Bogoliubov coefficients between explicitly specified pre- and post-bounce adiabatic bases remains future work.

\paragraph{Embedding in fundamental theory.}
Embedding the sigmoid field-space metric in string theory or loop quantum gravity remains an open challenge.

\subsection{Robustness}

On the baseline $\chi=\dot\chi=0$ trajectory, the background equations and hence the standard slow-roll $n_s$, $r$, and $f_{\rm NL}$ are analytically independent of $\alpha$; the numerical baseline scan directly checks only the kinematic $n_s$ statistic over $\nAlphaPoints$ logarithmically-spaced values $\alpha\in[0.1,10]$, finding $\sigma(n_s)<10^{-6}$. The spectator-displacement scan supports the narrower statement that the sampled kinematic $n_s$ remains stable when $g_{\chi\chi}\dot\chi^2$ reaches $\sim\KchiMaxNT\%$ of total kinetic energy ($\sigma(n_s)=\nsAlphaStdNT$ across $\nAlphaPointsNT$ values at each trajectory's measured $N=60$ point, \S\ref{sec:observables}); it records kinematic $r=16\epsilon$ rows but does not compute displaced-family $f_{\rm NL}$ or perturbation spectra. The computed grid establishes perturbation-level integrability and cross-construction consistency on displaced trajectories, not spectral $\alpha$-independence. The closed-FRW tensor amplitude agrees with the same-exit flat-FRW benchmark to $\PhSameExitAgreementPct\%$; its direct early-plateau ratio uses the exact scalar action and spans $[\rValClosedFRWStateMin,\rValClosedFRWStateMax]$ across the two propagated scalar states. Those states are initialised at exact acoustic depth $50$. The turning-grid cross-basis test separately uses the first-acceptance threshold $\sqrt{\lambda_\ell}/(a|H|)\geq50$ for the common test state mapped between its independently evolved representations. The tensor and entropy sectors use depth $1.5\times10^3$ and one specified state each; the independent flat-FRW scalar benchmark uses depth $10^3$. These choices support the stated finite comparisons but do not constitute an exhaustive finite-time-state analysis.

The perturbation-level $\mathcal{I}_{S\mathcal{R}}$ diagnostic is sensitive to $m_\chi$, the finite-time state choices, and the uncomputed post-horizon field-to-density map, but no longer to an arbitrary external $P_{\mathcal{R}}$ normalisation. Future refinements---an exact reduced coupled action on generic turning backgrounds, a joint scalar--entropy state study, controlled observable-pivot matching, a genuine transfer calculation, and the $(m_\chi,\alpha,\chi_0)$ landscape---are required before assigning it phenomenological status.

\appendix
\section{Integrated two-field scalar systems}
\label{app:integrated_scalar_systems}

This appendix states the complete linear scalar systems supplied to the two numerical integrators. Both use the Newtonian metric convention of Eq.~(\ref{eq:newtonian_metric_convention}), the same background, and the closed-space scalar eigenvalue $\lambda_\ell=\ell(\ell+2)$. Their dynamical variables are expressed in different field-space bases. Each displayed complex equation is integrated as coupled equations for its real and imaginary parts.

Write $g\equiv g_{\chi\chi}(\phi)$, $g_\phi\equiv \partial_\phi g$, and $g_{\phi\phi}\equiv\partial_\phi^2g$. In the coordinate field basis, the evolved state is $(\Phi,\dot\Phi,\delta\phi,\delta\dot\phi,\delta\chi,\delta\dot\chi)$ and obeys
\begin{align}
\ddot\Phi+4H\dot\Phi+(\dot H+3H^2)\Phi
&=\frac{1}{2M_{\rm Pl}^2}\biggl[
\dot\phi\,\delta\dot\phi+g\dot\chi\,\delta\dot\chi
+\frac{g_\phi}{2}\dot\chi^2\delta\phi
-V_{,\phi}\delta\phi-V_{,\chi}\delta\chi\biggr],
\label{eq:appendix_phi_evolution}\\
\delta\ddot\phi+3H\delta\dot\phi-g_\phi\dot\chi\,\delta\dot\chi
&+\biggl(\frac{\lambda_\ell}{a^2}+V_{,\phi\phi}
-\frac{g_{\phi\phi}}{2}\dot\chi^2\biggr)\delta\phi
+V_{,\phi\chi}\delta\chi
-4\dot\phi\dot\Phi+2V_{,\phi}\Phi=0,
\label{eq:appendix_coordinate_phi}\\
\delta\ddot\chi+\biggl(3H+\frac{g_\phi}{g}\dot\phi\biggr)\delta\dot\chi
&+\frac{g_\phi}{g}\dot\chi\,\delta\dot\phi
+\biggl(\frac{\lambda_\ell}{a^2}+\frac{V_{,\chi\chi}}{g}\biggr)\delta\chi
-4\dot\chi\dot\Phi+2\frac{V_{,\chi}}{g}\Phi
\nonumber\\
&+\biggl[
\frac{V_{,\phi\chi}}{g}-\frac{g_\phi V_{,\chi}}{g^2}
+\frac{g g_{\phi\phi}-g_\phi^2}{g^2}\dot\phi\dot\chi
\biggr]\delta\phi=0.
\label{eq:appendix_coordinate_chi}
\end{align}
The metric-kinetic term $g_\phi\dot\chi^2\delta\phi/2$ in Eq.~(\ref{eq:appendix_phi_evolution}) is retained. The momentum and Hamiltonian constraints are not substituted during evolution; they initialise the metric mode and are then evaluated independently:
\begin{align}
\dot\Phi+H\Phi
&=\frac{\dot\phi\,\delta\phi+g\dot\chi\,\delta\chi}{2M_{\rm Pl}^2},
\label{eq:appendix_momentum_constraint}\\
3H(\dot\Phi+H\Phi)+\frac{\lambda_\ell-3K}{a^2}\Phi
&=-\frac{\delta\rho}{2M_{\rm Pl}^2},
\label{eq:appendix_hamiltonian_constraint}\\
\delta\rho
&=\dot\phi\,\delta\dot\phi+g\dot\chi\,\delta\dot\chi
+\frac{g_\phi}{2}\dot\chi^2\delta\phi
-(\dot\phi^2+g\dot\chi^2)\Phi
+V_{,\phi}\delta\phi+V_{,\chi}\delta\chi .
\end{align}

For the independently evolved orthonormal field-basis system, let $T^a=\dot\phi^a/\dot\sigma$ and choose $N^a$ as in Eq.~(\ref{eq:tangent_normal}), so that
\begin{equation}
D_tT^a=\omega N^a,
\qquad D_tN^a=-\omega T^a,
\qquad
\omega=\frac{g\dot\chi V_{,\phi}-\dot\phi V_{,\chi}}
{\sqrt g\,\dot\sigma^2}.
\label{eq:appendix_turn_rate}
\end{equation}
The coefficient $\dot\omega$ is evaluated analytically from the background equations and is never finite-differenced. Explicitly, defining
\begin{equation}
A=g\dot\chi V_{,\phi}-\dot\phi V_{,\chi},
\qquad B=\sqrt g\,\dot\sigma^2,
\qquad \dot\omega=\frac{\dot A B-A\dot B}{B^2},
\label{eq:appendix_turn_rate_derivative}
\end{equation}
the derivatives $\dot A$ and $\dot B$ are formed from the background Klein--Gordon equations, $\dot g=g_{\phi}\dot\phi$, $\dot V_{,a}=V_{,ab}\dot\phi^b$, and $d(\dot\sigma^2)/dt=2\dot\phi\ddot\phi+g_{\phi}\dot\phi\dot\chi^2+2g\dot\chi\ddot\chi$.
The non-zero Christoffel symbols are $\Gamma^\phi_{\chi\chi}=-g_\phi/2$ and $\Gamma^\chi_{\phi\chi}=\Gamma^\chi_{\chi\phi}=g_\phi/(2g)$. The covariant Hessian used by the integrator is
\begin{equation}
V_{;\phi\phi}=V_{,\phi\phi},\qquad
V_{;\phi\chi}=V_{,\phi\chi}-\frac{g_\phi}{2g}V_{,\chi},\qquad
V_{;\chi\chi}=V_{,\chi\chi}+\frac{g_\phi}{2}V_{,\phi},
\end{equation}
with $M_{AB}=e_A^ae_B^bV_{;ab}$ for $e_T^a=T^a$ and $e_N^a=N^a$. The Gaussian field-space curvature is
\begin{equation}
K_G=\frac{\alpha^2}{M_{\rm Pl}^2}(1-g)(3g-1).
\end{equation}
With $V_\parallel=T^aV_{,a}$, the evolved state $(Q_T,\dot Q_T,Q_N,\dot Q_N,\Phi,\dot\Phi)$ obeys
\begin{align}
\ddot Q_T+3H\dot Q_T
&+\biggl(\frac{\lambda_\ell}{a^2}+M_{TT}-\omega^2\biggr)Q_T
-2\omega\dot Q_N
+(M_{TN}-\dot\omega-3H\omega)Q_N
\nonumber\\
&-4\dot\sigma\dot\Phi+2V_\parallel\Phi=0,
\label{eq:appendix_orthonormal_T}\\
\ddot Q_N+3H\dot Q_N
&+\biggl(\frac{\lambda_\ell}{a^2}+M_{NN}-\omega^2
+K_G\dot\sigma^2\biggr)Q_N
+2\omega\dot Q_T
\nonumber\\
&+(M_{NT}+\dot\omega+3H\omega)Q_T
-2\dot\sigma\omega\Phi=0.
\label{eq:appendix_orthonormal_N}
\end{align}
The same metric equation, Eq.~(\ref{eq:appendix_phi_evolution}), closes this system after the orthonormal state is mapped to coordinate perturbations by
\begin{align}
\delta\phi^a&=T^aQ_T+N^aQ_N,\\
D_t\delta\phi^a
&=T^a(\dot Q_T-\omega Q_N)+N^a(\dot Q_N+\omega Q_T),\\
\delta\dot\phi^a
&=D_t\delta\phi^a-\Gamma^a{}_{bc}\dot\phi^b\delta\phi^c.
\label{eq:appendix_inverse_basis_map}
\end{align}
Thus the $-\omega^2$ diagonal terms, the $\pm2\omega$ derivative couplings, and the $\dot\omega\pm3H\omega$ mixed terms are active on the turning trajectories rather than being inferred from the no-turn reduction. The two systems share their spacetime gauge and metric equation but reach the same $\mathcal R=\Phi+HQ_T/\dot\sigma$ through independently integrated field-basis states.

\paragraph{Turning-grid initial state.}
For each turning background and harmonic, the initial slice $t_i$ is selected on the pre-bounce branch by the first-acceptance rule $\sqrt{\lambda_\ell}/(a_i|H_i|)\geq50$: an already admissible earliest endpoint is retained, while an interior first crossing is root-refined to equality.  Both evolved representations are initialised from the same declared finite-time test state, defined from the instantaneous symmetric matrix below and mapped exactly between the field-space bases; only their subsequent evolution implementations are independent.  Let $\mathbf e_c=(e_T,e_N)^T$ be the unit eigenvector with the largest tangent overlap of that matrix,
\begin{equation}
\Omega_{i,\ell}^2=
\frac{\lambda_\ell}{a_i^2}\mathbf 1+
\begin{pmatrix}
M_{TT}-\omega^2 & M_{TN}\\
M_{TN} & M_{NN}-\omega^2+K_G\dot\sigma^2
\end{pmatrix}_{\!i},
\qquad
\Omega_{i,\ell}^2\mathbf e_c=\omega_c^2\mathbf e_c,
\label{eq:turning_grid_frequency_matrix}
\end{equation}
where the selected eigenvalue must be positive.  For the adiabatic branch used on the turning grid, the otherwise arbitrary global sign is fixed by $\operatorname{sgn}(e_T)=\operatorname{sgn}(T_i^{\phi})$ (with the positive convention if either entry vanishes), so the tangent-axis contribution to $\delta\phi_i$ is non-negative.  The antisymmetric gyroscopic derivative coupling is not folded into this instantaneous symmetric matrix.  The state in Eq.~\eqref{eq:turning_grid_matter_initial_state} is therefore a declared finite-time test-state prescription for the cross-basis consistency calculation; it is not claimed to be the unique or a higher-order adiabatic vacuum of the coupled turning system.  Choosing zero initial phase gives
\begin{equation}
\begin{pmatrix}Q_T\\Q_N\end{pmatrix}_{\!i}
=\frac{\mathbf e_c}{\sqrt{2a_i^3\omega_c}},
\qquad
\begin{pmatrix}\dot Q_T\\\dot Q_N\end{pmatrix}_{\!i}
=-(H_i+\mathrm{i}\omega_c)
\begin{pmatrix}Q_T\\Q_N\end{pmatrix}_{\!i}.
\label{eq:turning_grid_matter_initial_state}
\end{equation}
The coordinate perturbations and their ordinary derivatives follow from Eq.~\eqref{eq:appendix_inverse_basis_map}.  Defining
\begin{align}
\mathcal J_i&=\dot\phi_i\delta\phi_i+g_i\dot\chi_i\delta\chi_i,\\
\mathcal E_i&=\dot\phi_i\delta\dot\phi_i
+g_i\dot\chi_i\delta\dot\chi_i
+\frac{(g_{,\phi})_i}{2}\dot\chi_i^2\delta\phi_i
+(V_{,\phi})_i\delta\phi_i+(V_{,\chi})_i\delta\chi_i,\\
\mathcal C_i&=\frac{\lambda_\ell-3K}{a_i^2}
-\frac{\dot\sigma_i^2}{2M_{\rm Pl}^2},
\end{align}
the Hamiltonian and momentum constraints fix the two metric entries,
\begin{equation}
\Phi_i=-\frac{\mathcal E_i+3H_i\mathcal J_i}
{2M_{\rm Pl}^2\mathcal C_i},
\qquad
\dot\Phi_i=\frac{\mathcal J_i}{2M_{\rm Pl}^2}-H_i\Phi_i.
\label{eq:turning_grid_metric_initial_state}
\end{equation}
The two state orderings are
\[
\begin{aligned}
\text{coordinate:}\quad
  &(\Phi,\dot\Phi,\delta\phi,\delta\dot\phi,\delta\chi,\delta\dot\chi)_i,\\
\text{orthonormal:}\quad
  &(Q_T,\dot Q_T,Q_N,\dot Q_N,\Phi,\dot\Phi)_i.
\end{aligned}
\]
In each case the six complex entries are integrated as a twelve-real state.  The algebraic metric inversion is used only on this initial slice, and $\Phi$ is evolved thereafter.

\section{\texorpdfstring{Exact finite-harmonic closed-$S^3$ scalar action on the fiducial trajectory}{Exact finite-harmonic closed-S3 scalar action on the fiducial trajectory}}
\label{app:closed_s3_canonical}

This appendix supplies the canonical scalar-harmonic completion used for every absolute scalar-power statement in the main text. Its scope is the fiducial background $\chi=\dot\chi=0$. On that trajectory the quadratic action factorises exactly into a gravity--inflaton scalar mode and a spectator entropy oscillator. This is sufficient for the reported scalar, tensor-to-scalar, and entropy-to-scalar powers; it is not a reduced canonical action for a generic turning two-field background. The turning-background calculations use the complete regular classical systems in Appendix~\ref{app:integrated_scalar_systems}.

\subsection{Harmonics and exact factorisation}

Let $\gamma_{ij}$ be the unit-three-sphere metric. Orthonormal scalar harmonics obey
\begin{equation}
D^2Q_{\ell\mu}=-\lambda_\ell Q_{\ell\mu},\qquad
\lambda_\ell=\ell(\ell+2),\qquad
\mu=1,\ldots,(\ell+1)^2.
\label{eq:s3_scalar_harmonics_exact}
\end{equation}
The scalar-derived traceless tensor harmonic
\begin{equation}
Q^{(S)}_{ij}=\lambda_\ell^{-1}D_iD_jQ_{\ell\mu}
+\tfrac13\gamma_{ij}Q_{\ell\mu}
\end{equation}
has norm $\tfrac23(\lambda_\ell-3K)/\lambda_\ell$. Hence the physical inhomogeneous scalar sector begins at $\ell=2$, and we define
\begin{equation}
\Lambda_\ell\equiv\lambda_\ell-3K>0.
\label{eq:Lambda_ell_def}
\end{equation}
This curvature-shifted eigenvalue follows from eliminating the lapse and scalar shift in the non-flat scalar action \cite{AkamaKobayashi2019,BongaGuptYokomizo2016,MonteseYokomizo2024}.

For $\chi=\dot\chi=0$, one has $V_{,\chi}=V_{,\phi\chi}=0$ and no linear metric source proportional to $\dot\chi$. The term $g_{\chi\chi,\phi}\,\delta\phi(\partial\delta\chi)^2$ first appears at cubic order. Therefore
\begin{equation}
S^{(2)}_{\rm scalar,fid}=S^{(2)}_{\mathcal R}+S^{(2)}_\chi
\label{eq:fiducial_scalar_split}
\end{equation}
exactly at quadratic order.

\subsection{Reduced curvature action and regular bounce limit}

Akama and Kobayashi write the unitary-gauge spatial metric as
$g_{ij}=a^2e^{2\zeta_{\rm AK}}\gamma_{ij}$ \cite{AkamaKobayashi2019}. On the
comoving slice, the present Newtonian convention gives
$g_{ij}=a^2(1-2\mathcal R)\gamma_{ij}$ at linear order, with
$\mathcal R=\Phi+H\delta\phi/\dot\phi$. Hence
$\zeta_{\rm AK}=-\mathcal R$. This overall sign cancels from the quadratic
action and does not alter either coefficient. In the present convention the
exact mode action is
\begin{equation}
S^{(2)}_{\mathcal R}=\frac12
\sum_{\ell=2}^{\infty}\sum_{\mu=1}^{(\ell+1)^2}
\int dt\,a^3\left[
\mathcal A_\ell|\dot{\mathcal R}_{\ell\mu}|^2
-\mathcal B_\ell\frac{\Lambda_\ell}{a^2}
|\mathcal R_{\ell\mu}|^2\right].
\label{eq:exact_closed_R_action}
\end{equation}
Away from $H=0$, define
\begin{equation}
\epsilon_\phi=\frac{\dot\phi^2}{2M_{\rm Pl}^2H^2},\qquad
\Xi_\ell=\frac{\Lambda_\ell}{\Lambda_\ell+K\epsilon_\phi}.
\end{equation}
For a canonical scalar minimally coupled to Einstein gravity, the quantities
in the general non-flat action specialise to
\begin{equation}
\mathcal G_T=\mathcal F_T=M_{\rm Pl}^2,\qquad
\Theta=M_{\rm Pl}^2H,\qquad
\Sigma=\frac12\dot\phi^2-3M_{\rm Pl}^2H^2,
\label{eq:AK_canonical_specialisation}
\end{equation}
so that $\mathcal G_T\Sigma/\Theta^2=\epsilon_\phi-3$. Acting on one scalar
harmonic, the operator ratio in that action becomes
\begin{align}
\frac{D^2+3K}
{D^2-\mathcal G_T\Sigma K/\Theta^2}Q_{\ell\mu}
&=\frac{-\lambda_\ell+3K}
{-\lambda_\ell-K(\epsilon_\phi-3)}Q_{\ell\mu}\notag\\
&=\frac{\Lambda_\ell}{\Lambda_\ell+K\epsilon_\phi}Q_{\ell\mu}
=\Xi_\ell Q_{\ell\mu}.
\label{eq:AK_operator_ratio}
\end{align}
The cited convention writes the scalar action without the overall factor
$1/2$ used in Eq.~\eqref{eq:exact_closed_R_action}. Consequently
$\mathcal A_\ell=2\mathcal G_S$ and
$\mathcal B_\ell=2\mathcal F_S$. Substitution into the general coefficients
then gives
\begin{align}
\mathcal A_\ell&=2M_{\rm Pl}^2\epsilon_\phi\Xi_\ell,\label{eq:Aell_compact}\\
\mathcal B_\ell&=2M_{\rm Pl}^2
\left(\Xi_\ell-1+\epsilon_\phi\Xi_\ell+\frac{\dot\Xi_\ell}{H}\right).
\label{eq:Bell_compact}
\end{align}
The $\Xi_\ell-1$ term is essential. For numerical evaluation through $H=0$, introduce
\begin{equation}
\kappa_\phi=\frac{\dot\phi^2}{2M_{\rm Pl}^2},\qquad
\Delta_\ell=\Lambda_\ell H^2+K\kappa_\phi.
\end{equation}
Algebraic reduction using $\dot H=K/a^2-\kappa_\phi$ yields the regular forms
\begin{align}
\mathcal A_\ell&=\frac{2M_{\rm Pl}^2\Lambda_\ell\kappa_\phi}{\Delta_\ell},
\label{eq:Aell_regular}\\
\mathcal B_\ell&=\frac{2M_{\rm Pl}^2}{\Delta_\ell^2}
\left[(\Lambda_\ell-K)\kappa_\phi\Delta_\ell
+\Lambda_\ell K(2\kappa_\phi\dot H-H\dot\kappa_\phi)\right],
\label{eq:Bell_regular}
\end{align}
where
\begin{equation}
\dot\kappa_\phi=-6H\kappa_\phi
-\frac{\dot\phi V_{,\phi}}{M_{\rm Pl}^2}.
\end{equation}
No numerical derivative is used. If $\dot\phi_b\neq0$, the exact bounce limits are finite:
\begin{align}
\mathcal A_{\ell,b}&=\frac{2M_{\rm Pl}^2\Lambda_\ell}{K},\\
\mathcal B_{\ell,b}&=\frac{2M_{\rm Pl}^2}{K}
\left(\Lambda_\ell-K+
\frac{2\Lambda_\ell\dot H_b}{\kappa_{\phi,b}}\right).
\label{eq:AB_bounce_limits}
\end{align}
Thus the divergence of the flat-space quantity $a\dot\phi/H$ is not a divergence of the finite-harmonic closed-space action.

\subsection{Canonical evolution and two finite-time states}

The derivative-free Hamiltonian system is
\begin{equation}
\Pi_{\mathcal R,\ell}=a^3\mathcal A_\ell\dot{\mathcal R}_\ell,\qquad
\dot{\mathcal R}_\ell=\frac{\Pi_{\mathcal R,\ell}}{a^3\mathcal A_\ell},\qquad
\dot\Pi_{\mathcal R,\ell}=-a\mathcal B_\ell\Lambda_\ell\mathcal R_\ell,
\label{eq:R_first_order_system}
\end{equation}
with canonical normalisation
\begin{equation}
\mathcal R_\ell\Pi_{\mathcal R,\ell}^*
-\mathcal R_\ell^*\Pi_{\mathcal R,\ell}=i.
\label{eq:R_canonical_wronskian}
\end{equation}
Define the mode-dependent acoustic time and canonical field (the closed-space analogue of the curved-universe canonical variable of \cite{Handley2019}) by
\begin{equation}
dy_\ell=\frac{c_\ell}{a}dt,\qquad
c_\ell^2=\frac{\mathcal B_\ell}{\mathcal A_\ell},\qquad
u_\ell=z_\ell\mathcal R_\ell,\qquad
z_\ell=a(\mathcal A_\ell\mathcal B_\ell)^{1/4}.
\label{eq:closed_canonical_variables}
\end{equation}
Then
\begin{equation}
u_\ell''+\left(\Lambda_\ell-\frac{z_\ell''}{z_\ell}\right)u_\ell=0,
\qquad
u_\ell u_\ell^{*\prime}-u_\ell'u_\ell^*=i,
\label{eq:closed_u_equation}
\end{equation}
where primes denote $d/dy_\ell$.

\paragraph{Initial depth is not an adiabaticity criterion.}
The acoustic-depth condition below compares a mode frequency with $|H|$, not with the time variation of the canonical normalisation. It therefore does not imply $|z_\ell''/z_\ell|\ll\Lambda_\ell$. For example, as $H\to0$ the depth diverges when $c_\ell/a$ remains finite and nonzero, irrespective of the size of $z_\ell''/z_\ell$. A WKB interpretation of Eq.~\eqref{eq:closed_u_equation} would instead require a real positive $\Omega_\ell^2=\Lambda_\ell-z_\ell''/z_\ell$ and control of its acoustic-time variation, including $|\Omega_\ell'/\Omega_\ell^2|\ll1$ at leading order. No such adiabatic regime is inferred from depth alone. The two prescriptions below are declared finite-time states, not an established adiabatic vacuum for every sampled harmonic.

All ten harmonics $\ell\in\{2,3,5,7,10,60,100,200,500,1000\}$ share the last pre-bounce slice on which
\begin{equation}
\min_\ell\frac{c_\ell\sqrt{\Lambda_\ell}}{a|H|}=50.
\label{eq:exact_common_depth}
\end{equation}
At that finite time we propagate two normalised states, neither asserted to be unique. The acoustic-gradient reporting state uses
\begin{equation}
u_i=\frac{1}{\sqrt{2W_\ell}},\qquad
u_i'=-iW_\ell u_i,\qquad W_\ell=\sqrt{\Lambda_\ell},\qquad
\mathcal R_i=\frac{u_i}{z_i},\qquad
\Pi_i=z_i u_i'-z_i'u_i.
\label{eq:acoustic_gradient_state}
\end{equation}
The time-dependent $-z_i'u_i$ term is part of the canonical transformation. The comparison state diagonalises the instantaneous $(\mathcal R,\Pi_{\mathcal R})$ Hamiltonian,
\begin{equation}
\omega_{\ell,i}=\frac{c_{\ell,i}\sqrt{\Lambda_\ell}}{a_i},\qquad
\mathcal R_i=\frac{1}{\sqrt{2a_i^3\mathcal A_{\ell,i}\omega_{\ell,i}}},\qquad
\Pi_i=-ia_i^3\mathcal A_{\ell,i}\omega_{\ell,i}\mathcal R_i.
\label{eq:instantaneous_hamiltonian_state}
\end{equation}
The states have equal initial $|\mathcal R_i|$ and differ in canonical momentum. Both are integrated by DOP853 over all ten harmonics with relative tolerance $10^{-12}$ and absolute tolerance $10^{-20}$. The independent Radau comparison uses the same inputs and tolerances on the subset $\ell\in\{2,3,60\}$ for both states, giving six comparison cells. Powers are extracted at the common geometric exit $aH=\sqrt{\lambda_\ell}$, four e-folds later, and over the final post-bounce e-fold. Their propagated range is reported rather than interpreted as a statistical uncertainty.

\subsection{Exact map to regular Newtonian variables}

The reduced action provides an independent scalar evolution and a constraint-satisfying initial map for the regular Newtonian systems. With $\dot{\mathcal R}=\Pi_{\mathcal R}/(a^3\mathcal A_\ell)$,
\begin{align}
\Phi_\ell&=\frac{\kappa_\phi}{\Delta_\ell}
\left(K\mathcal R_\ell-a^2H\dot{\mathcal R}_\ell\right),
\label{eq:R_to_Phi_exact}\\
\delta\phi_\ell&=\frac{\dot\phi}{\Delta_\ell}
\left(\Lambda_\ell H\mathcal R_\ell
+\kappa_\phi a^2\dot{\mathcal R}_\ell\right),
\label{eq:R_to_deltaphi_exact}\\
S_\ell&\equiv\dot\Phi_\ell+H\Phi_\ell
=\frac{\dot\phi\,\delta\phi_\ell}{2M_{\rm Pl}^2},
\qquad \dot\Phi_\ell=S_\ell-H\Phi_\ell,\\
\delta\dot\phi_\ell&=\frac{1}{\dot\phi}
\left[-2M_{\rm Pl}^2\left(3HS_\ell+\frac{\Lambda_\ell}{a^2}\Phi_\ell\right)
+\dot\phi^2\Phi_\ell-V_{,\phi}\delta\phi_\ell\right].
\label{eq:R_to_deltaphidot_exact}
\end{align}
These expressions satisfy $\mathcal R=\Phi+H\delta\phi/\dot\phi$ and both linear Einstein constraints at the initial slice. They remain finite at $H=0$ when $\dot\phi\neq0$, where $\Phi_{\ell,b}=\mathcal R_{\ell,b}$. Subsequent $\Phi$ and $\dot\Phi$ are evolved, not reconstructed by reinverting a constraint.

\subsection{Spectator action and power convention}

The exactly decoupled spectator action is
\begin{equation}
S^{(2)}_\chi=\frac12\sum_{\ell,\mu}\int dt\,a^3g_{\chi\chi}
\left[|\delta\dot\chi_{\ell\mu}|^2
-\left(\frac{\lambda_\ell}{a^2}+\frac{m_\chi^2}{g_{\chi\chi}}\right)
|\delta\chi_{\ell\mu}|^2\right].
\label{eq:exact_entropy_action_s3}
\end{equation}
In conformal time, $q_{s,\ell}=a\sqrt{g_{\chi\chi}}\,\delta\chi_\ell$ obeys
\begin{equation}
q_{s,\ell}''+\left[\lambda_\ell+
\frac{a^2m_\chi^2}{g_{\chi\chi}}
-\frac{(a\sqrt{g_{\chi\chi}})''}{a\sqrt{g_{\chi\chi}}}\right]q_{s,\ell}=0.
\label{eq:exact_entropy_canonical_s3}
\end{equation}
For the harmonic normalisation above, the exact per-log-wavenumber powers are
\begin{equation}
\mathcal P_{\mathcal R}(\ell)=
\frac{\ell(\ell+1)(\ell+2)}{2\pi^2}|\mathcal R_\ell|^2,
\qquad
\mathcal P_{SS}(\ell)=
\frac{\ell(\ell+1)(\ell+2)}{2\pi^2}|\mathcal S_\ell|^2.
\label{eq:exact_s3_power_measure}
\end{equation}
This fixes the measure used in \S\ref{subsec:curvature_scale_diagnostic}; alternative low-harmonic approximations are not averaged into an error bar.

\subsection{Independent numerical validation}

Table~\ref{tab:exact_closed_s3_validation} distinguishes the full ten-harmonic set from the cross-check subset $\ell\in\{2,3,60\}$; both propagated states are included in each applicable check. Coefficient positivity is checked on a 20,001-point time grid for each harmonic and at every adaptive evaluation reached by the mode integrations; this is numerical finite-grid evidence, not an analytic continuum proof. At the exactly inserted bounce, and over the fixed coefficient grid, respectively, the recorded scales are
\begin{equation}
\begin{aligned}
\frac{\kappa_{\phi,b}}{M_{\rm Pl}^2}
&=\ExactScalarBounceKappaOverMPlSq,
&
\min_{\ell,t}\frac{\mathcal A_\ell}{2M_{\rm Pl}^2}
&=\ExactScalarMinAOverTwoMPlSq,\\
\min_{\ell,t}\frac{\mathcal B_\ell}{2M_{\rm Pl}^2}
&=\ExactScalarMinBOverTwoMPlSq,
&
\min_{\ell,t}c_\ell^2
&=\ExactScalarMinCSq.
\end{aligned}
\label{eq:exact_scalar_positive_minima}
\end{equation}
At the bounce the same regular coefficients give
\begin{equation}
\begin{gathered}
c_{\ell,b}^2
=1-\frac{K}{\Lambda_\ell}
+\frac{2\dot H_b}{\kappa_{\phi,b}},\\
c_{2,b}^2=\ExactScalarBounceCSqMin,
\qquad
c_{1000,b}^2=\ExactScalarBounceCSqMax,
\qquad
c_{1000,b}^2-c_{2,b}^2=K\left(\Lambda_2^{-1}-\Lambda_{1000}^{-1}\right)\simeq0.2000.
\end{gathered}
\label{eq:exact_scalar_bounce_c2}
\end{equation}
The large positive bounce value does not represent a superluminal matter characteristic. It is the finite-harmonic gradient-to-kinetic ratio after the elliptic gravitational constraints have been eliminated. The canonical-GR short-wavelength principal-symbol result is obtained by taking $\mathcal D^2\gg K$ before the $H\to0$ reduction \cite{AkamaKobayashi2019}; at fixed harmonic the short-wavelength and bounce limits do not commute. The three minima in Eq.~\eqref{eq:exact_scalar_positive_minima} are global over the ten declared harmonics and their fixed reporting grids. Adaptive evaluations are additionally required to remain positive, but are not mixed into the quoted fixed-grid minima. The raw late-time determinant in Eq.~\eqref{eq:R_canonical_wronskian} becomes ill-conditioned after inflationary squeezing. Its quoted acceptance bound is therefore restricted to samples where the binary64 subtraction resolves a unit Wronskian, including the initial slice and exact bounce. The full late-time amplitude is checked independently by the nonlinear amplitude--phase representation
\begin{equation}
\mathcal R=\rho e^{-i\theta},\qquad
\Pi_{\mathcal R}=\left(p_\rho-\frac{i}{2\rho}\right)e^{-i\theta},
\end{equation}
which enforces the Wronskian algebraically and is integrated for all twenty harmonic--state cells.

\begin{table}[htbp]
\centering
\caption{\textbf{Validation of the exact fiducial closed-$S^3$ scalar action.}
All inequalities are strict. Rows 1--4 cover all ten harmonics (and both states where applicable); rows 5--10 cover only $\ell\in\{2,3,60\}$ and both states, six cells. The mapped constraints use both Newtonian representations and only the bounce window $|t-t_b|\leq t_b-t_i$, where $t_i$ is the common initial time and $t_f$ the end of the reporting integration. Evolution and method comparisons use the full reporting interval $[t_i,t_f]$.}
\label{tab:exact_closed_s3_validation}
\begin{tabular}{lcc}
\toprule
diagnostic & maximum residual & threshold \\
\midrule
compact vs regular coefficients & $\ExactScalarCompactResidual$ & $10^{-10}$ \\
analytic bounce limits & $\ExactScalarBounceLimitResidual$ & $10^{-10}$ \\
condition-resolved canonical Wronskian & $\ExactScalarWronskianResidual$ & $10^{-8}$ \\
direct vs amplitude--phase evolution & $\ExactScalarAmplitudePhaseResidual$ & $10^{-6}$ \\
canonical-to-Newtonian initial map & $\ExactScalarInitialMapResidual$ & $10^{-10}$ \\
reduced action vs coordinate Newtonian & $\ExactScalarCoordinateResidual$ & $10^{-6}$ \\
reduced action vs orthonormal Newtonian & $\ExactScalarOrthonormalResidual$ & $10^{-6}$ \\
mapped Hamiltonian constraint (bounce window) & $\ExactScalarHamiltonianResidual$ & $10^{-3}$ \\
mapped momentum constraint (bounce window) & $\ExactScalarMomentumResidual$ & $1.5\times10^{-2}$ \\
DOP853 vs Radau & $\ExactScalarMethodResidual$ & $10^{-6}$ \\
\bottomrule
\end{tabular}
\end{table}

All listed criteria pass on their respective domains. In particular, the two mapped Einstein-constraint maxima apply only to the six cross-check cells and the symmetric bounce window $|t-t_b|\leq t_b-t_i$, whose duration is approximately $0.54\%$ of $t_f-t_i$. They do not bound either constraint over the full evolution. The initial-map check is at $t_i$, while the cross-representation and DOP853--Radau comparisons cover the full reporting interval on those six cells. The amplitude--phase comparison covers all twenty harmonic--state cells. At $\ell=60$ the scalar-power span between the two propagated states is $\ExactScalarStateSpanEllSixtyPct\%$ four e-folds after exit. That dependence is carried into every scalar-normalised ratio in the main text. A generic turning-background reduced action and finite-time state prescription remain outside this appendix's scope.

\section{Corrections to the superseded versions}
\label{app:earlier_corrections}

\EarlierVersionCorrections

\section*{Data Availability}
\label{sec:data}

The analysis code, the numerical implementations of every equation solved here, the validation suite, the generated macro file \texttt{results\_macros.tex}, the fingerprinted JSON artefacts, the machine-readable Stage-3 verification report \texttt{verify\_solver\_report.json} from which the quoted values are taken (with its source and criteria digests), the figure scripts, and the manuscript sources are publicly available in the project repository \cite{Kravchenko2026code}:\\[2pt]
\url{https://github.com/OkMathOrg/bouncing-cosmology}, release \texttt{5.0.0},\\[2pt]
archived at Zenodo under the concept DOI \href{https://doi.org/10.5281/zenodo.17684433}{10.5281/zenodo.17684433}, which resolves to the latest archived release. The numerical results generated in this work can be reproduced from the archived code, the documented inputs, and the numerical prescriptions stated above; externally quoted observational constraints and the one retained antecedent number are cited to their sources rather than recomputed.

\bibliographystyle{plain}
\bibliography{v5}

\end{document}